# Cybersecurity for Modern Smart Grid against Emerging Threats

Daisuke Mashima[1], Yao Chen[3], Muhammad M. Roomi[1], Subhash Lakshminarayana[4] and Deming Chen[1,2]

[1] *Illinois Advanced Research Center at Singapore, Singapore; {daisuke.m, roomi.s}@iarcs-create.edu.sg*
[2] *University of Illinois at Urbana-Champaign, IL, USA; dchen@illinois.edu*
[3] *National University of Singapore, Singapore; yaochen@nus.edu.sg*
[4] *University of Warwick, UK; Subhash.Lakshminarayana@warwick.ac.uk*

ABSTRACT

Smart Grid is a power grid system that uses digital communication technologies. By deploying intelligent devices throughout the power grid infrastructure, from power generation to consumption, and enabling communication among them, it revolutionizes the modern power grid industry with increased efficiency, reliability, and availability. However, reliance on information and communication technologies has also made the smart grids exposed to new vulnerabilities and complications that may negatively impact the availability and stability of electricity services, which are vital for people's daily lives. The purpose of this book is to provide an up-to-date and comprehensive survey and tutorial on the cybersecurity aspect of smart grids. The book focuses on the sources of the cybersecurity issues, the taxonomy of threats, and the survey of various approaches to overcome or mitigate such threats. It covers the state-of-the-art research results in recent years, along with remaining open challenges. We hope that this book can be used both as learning materials for beginners who are embarking on research in this area and as a useful reference for established researchers in this field.

# 1

# Introduction

## 1.1 What and Why

Modernization of power grid systems has brought us a number of benefits for efficient and reliable electricity services, speedy recovery from system failures and natural disasters, and increased penetration of intermittent renewable energy resources. The key enabler of the smart grid is real-time situation awareness and remote control enabled by a number of intelligent devices, which have communication and computation capabilities, such as intelligent electronic devices (IEDs), programmable logic controllers (PLCs), and remote terminal units (RTUs). Such devices are deployed throughout the power grid infrastructure and communicate with one another as well as with the control center by using standardized communication protocols, such as IEC 61850 and IEC 60870-5-104.

Traditionally, cybersecurity was not a primary concern or design consideration in many industrial control systems, including power grid systems, because of the closed, isolated nature of the system infrastructure. However, owing to modernization and digitalization, now such industrial control systems are increasingly connected to other systems and networks, such as enterprise IT systems and even the Internet. Therefore, the “security by air-gap” assumption is no longer valid, and the smart grid systems are under an increasing amount of attacks in recent years.

Cybersecurity solutions for industrial control systems have some fundamental differences from the ones for enterprise IT systems. Thus, it is not straightforward to apply established cybersecurity solutions for IT systems to industrial control systems, or it is even not practical or possible to do so. Cyber attacks are typically discussed from the perspective of CIA, which are confidentiality, integrity, and availability, as defined below:

- **Confidentiality**: Confidentiality in industrial control systems (ICS) means protecting sensitive information from unauthorized access or disclosure. Industrial control systems often handle sensitive data, such as production process details, intellectual property, and other proprietary information. Ensuring confidentiality helps protect an organization's competitive advantage and prevents unauthorized personnel or adversaries from tampering with or compromising the ICS. Attacks violating confidentiality could be adversaries illegally accessing unauthorized resources by eavesdropping, security mechanism bypass, illegal escalation of privileges, identity fabricating, etc. Different from IT systems, protecting user-specific consumption data in smart grid is crucial. Unauthorized access can lead to privacy breaches where malicious actors can infer users' behavior patterns (like when they're home or away). Additionally, grid operational data should remain confidential to prevent potential sabotage or malicious grid manipulation.

- **Integrity**: Integrity refers to maintaining the consistency, accuracy, and trustworthiness of data and system components within the ICS. Ensuring integrity means that the data and commands transmitted between the system components are accurate and have not been tampered with. This is crucial in ICS, as data corruption or unauthorized modification can lead to incorrect decision-making, system malfunction, or even catastrophic events in critical infrastructure. Attacks violating integrity will damage the consistency of data. Adversaries could illegally tamper or destroy the original stored or transmitted information to cause di-rect damages or hide their illegal behaviors from future intrusions. Ensuring the integrity of commands and data for the smart grid is vital. Altered data could lead to grid instability, incorrect billing, or even catastrophic failures. For instance, a tampered command could shut down parts of the grid or redirect power flows, causing outages.

- **Availability**: Availability is about ensuring that the ICS is accessible and operational when required by authorized users. In an industrial environment, maintaining system availability is essential to preventing downtime, loss of production, or potential safety hazards. This includes protecting the system from various threats, such as hardware failures, software bugs, cyberattacks, and natural disasters, as well as implementing redundancy, backup, and disaster recovery strategies to minimize downtime. Attacks violating availability will reject the regular usage of resources by legitimate users. Adversaries could illegally consume the computing or communication resources of the target system so that it is unable to respond to the normal request of legitimate users. In addition, adversaries could also intercept the normal request to make the target service appears to be unavailable. The availability of the smart grid and its data is of paramount importance. Power needs to be available 24/7, and control commands should always reach their intended components. Any downtime can result in significant economic and societal disruptions.

Adversaries may try to exploit vulnerabilities in power systems to gain unauthorized access to private information or to cause damage to power grid facilities. This can be done for various reasons, such as economic gain or disruption of services. It is important to implement security measures to prevent and detect these types of attacks on power systems. The attacks can be categorized by their impact on the CIA requirements. For example, False Data Injection (FDI), Global Positioning System (GPS) spoofing, and Time synchronization attacks violate the integrity of information while Impersonation Attack camouflages the legitimate parties in a system or network protocol, and a Denial-of-Service (DoS) attack violates the availability of data.

Because of such distinctions, dedicated efforts have been devoted to developing security technologies for protecting our critical infrastructure. This book elaborates on representative state-of-the-art solutions of different kinds.

One universal challenge for designing and developing effective cybersecurity solutions for smart grid systems is the lack of an evaluation environment. While it would be ideal to use the real smart grid infrastructure for conducting empirical evaluation, it would never be an option for the fear that such experiments would cause a negative and potentially severe impact on the stability and availability of power grid operations. Thus, the research community has been devoting efforts to creating dedicated testbeds, which are isolated from the production environment. Such testbeds are categorized into 3 types: hardware-based testbeds, virtual, software-based testbeds (also called digital twin or cyber range), and hybrid testbeds, which are positioned in between. Digital twin has become a promising technology to address this challenge in various industry and network scenarios since it acts as a virtual representation of the real-world entity or system [Gartner, n.d.]. It is initially proposed by Grieves in 2003 for product manufacturing process [Grieves, 2015]. Until recently, its development has received extensive attention.

## 1.2 Smart Grid Technology Overview

The smart grid is a collection of energy control and monitoring devices, together with software, networking, and communications infrastructure that are installed in homes, businesses, and throughout the electricity distribution grid. This collective system provides the ability to monitor and control energy consumption comprehensively in real time.

Smart grid technology is poised to revolutionize the modern industry with powerful solutions that improve the efficiency of traditional power grids. The smart grid is an energy supply network that uses digital communication technology. Rapid advances in Information and Communication Technologies (ICT) are increasingly integrated into several infrastructure layers covering all aspects of the electricity grid and its associated operations. In addition, intelligent networked devices are emerging whose Internet of Things (IoT) interactions create new capabilities in the monitoring and management of the electricity grid and the interaction between its stakeholders. IT-empowered innovations integrated with the electricity network and the stakeholders' interactions have paved the way towards a "Smart Grid" that takes advantage of sophisticated bidirectional interactions.

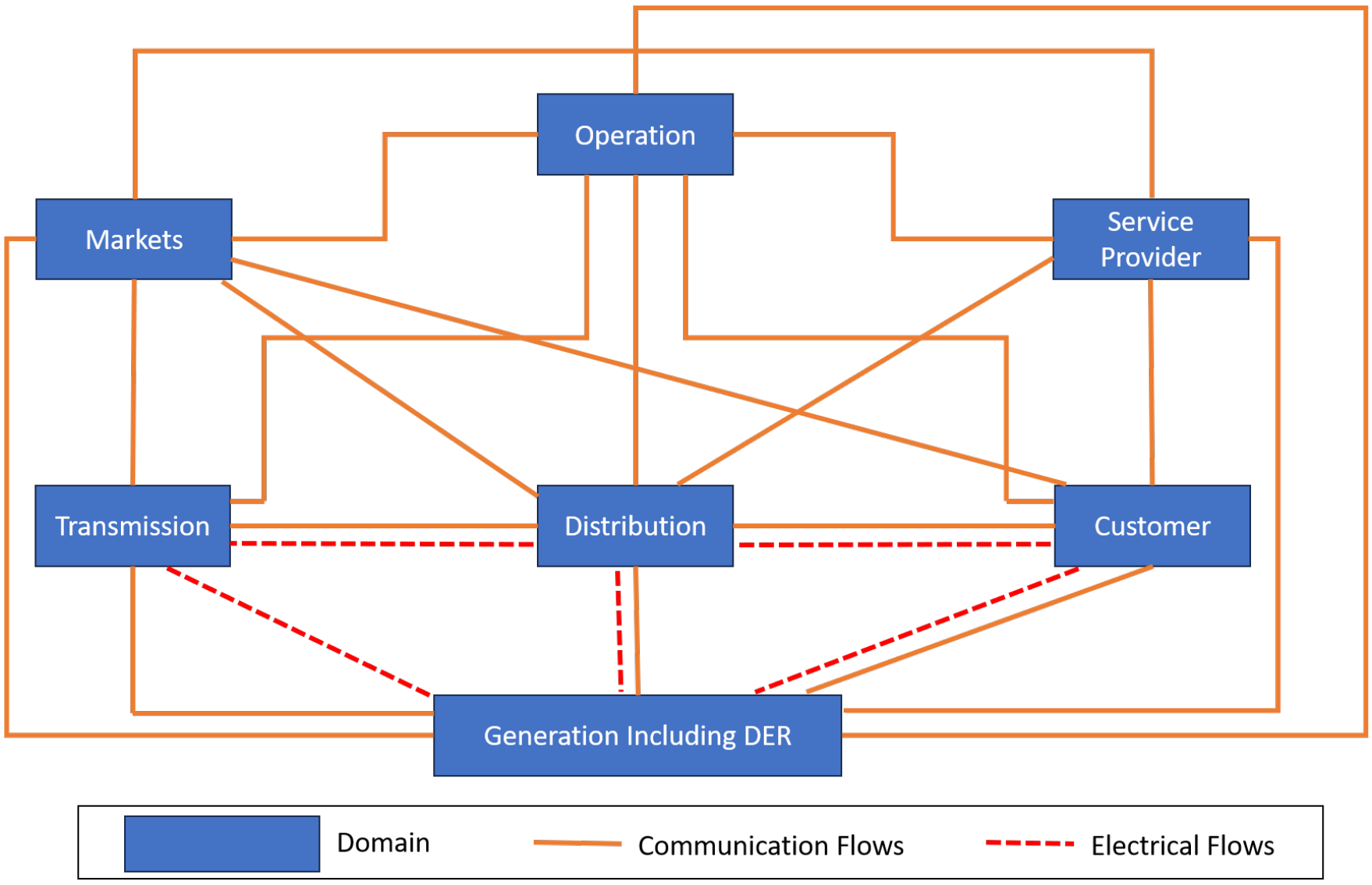

**Figure 1.1:** The Smart Grid Conceptual Model by NIST [Gopstein *et al.*, n.d.].

The US National Institute of Standards and Technology (NIST) Smart Grid Conceptual Model [Gopstein *et al.*, n.d.] defines a framework that outlines seven domains: Bulk Generation, Transmission, Distribution, Customers, Operations, Markets, and Service Providers, as shown in Figure 1.1. Complementary to that, the IEEE views the Smart Grid as "a large System of Systems" where each NIST Smart Grid domain is expanded into three Smart Grid foundational layers: (i) the Power and Energy Layer, (ii) the Communication Layer, and (iii) the IT/Computer Layer. The interplay of these layers via a highly sophisticated ICT infrastructure brings intelligence to the grid and enables it to provide new added-value services to its stakeholders [Karnouskos, 2011].

## 1.3 Motivation of Cybersecurity

The Smart Grid provides quick and enhanced services for the customers, with a reduced amount of response time delay, where energy savings can be achieved by implementing the system efficiently. However, the Smart Grid technology comes with vulnerabilities and complications, with the largest challenge being to secure the information that is the most vital asset. Cybersecurity in Smart Grid is of great importance because numerous devices are connected via a series of networks to communicate and deliver the information to the desired destinations with various techniques [Maglaras *et al.*, 2018]. The system will frequently exchange information that could be sensitive information and need protection.

International standards such as IEC 60870-5-101/104 and IEC 61850 are derived to communicate the information between the devices from different manufacturers effectively. Though these standards define the protection mechanism to be implemented in smart grids, they lack security design. Hence, even with the integration of these protection mechanisms, without adequate cyber protection, such systems are vulnerable to a wide range of attacks, including impersonation, false command injection, false data injection, and tampering with messages. There are numerous real-world attacks that have taken place in the past decade.

- *Stuxnet.* Stuxnet is a malicious computer worm targeting Supervisory Control and Data Acquisition (SCADA). It is known to be developed since at least 2005 and was first uncovered in 2010. It is able to manipulate PLCs, which are responsible for controlling the electromechanical processes of machines. The Stuxnet computer worm uses the Microsoft Windows operating system and networks. Stuxnet managed to infiltrate Iran's Nuclear Power Plant control system and was able to manipulate the PLCs, which control the centrifuges and separate nuclear materials by triggering them to spin faster in resolves to tear the equipment apart. The impact of the Stuxnet computer worm attacking Iran's Nuclear Power plant shows how fatal a computer security breach can be.

- *Trojanhorse malware BlackEnergy.* It took place during a civil war context on 25th December 2015. An electrical power station

in Ukraine's Ivano-Frankivsk city was targeted for a cyber-attack which affected a colossal impact on eighty thousand (80,000) people by putting them in the dark, whereas one million, four hundred thousand (1.4 million) people were affected. It was done using the spear-phishing email and a Trojan horse malware known as the "BlackEnergy3". This powerful malware was capable of opening backdoor to enable command and control (C2) communication with the remote attackers. Thus it enabled information gathering to identify VPN servers and login credentials. It also had component called KillDisk, which had capability of deleting the data, and destroy hard drives to leave affected systems unavailable, slowing down the recovery processes. This situation became very critical because the cyber-attack not just stole the information but also destabilized a country's critical infrastructure.

- *Intrusion into the utility company system in the US.* In 2018, it is reported that state-sponsored Russian hackers' intrusion broke into the network of a utility company in the US. The attack was said to be mounted via third-party vendors' system. The attack procedure is similar to the one used for the Ukrainian power plant attack in 2015, and they started with spear-phishing to the targeted employees to get access to the vendors' network. Then, attackers collected information and credentials associated with the utility company's system. While the severe consequence was not caused, it is said that the hackers were close to the position of causing a massive outage. Along with the Ukrainian case, this is seen as evidence that committed hackers are, in reality, interested in targeting critical infrastructure.
- *Ransomware attack.* WannaCry was known to be launched on the 12th of May 2017 and infected more than 200,000 computers - demanding a ransom to unlock the victim's files. The result is a global cyber-attack on well-known organizations, including Renault, and FedEx. The hackers demanded cryptocurrency bitcoins worth three hundred dollars ($300) to be sent to a specific address to decrypt the whole system, and if victims didn't pay in the time given, then their whole system files were permanently

deleted. The attack has caused the biggest impact by hacking the National Health Service (NHS) hospital computers, and more than nineteen thousand (19,000) appointments were canceled, and the WannaCry Ransomware has put many lives in jeopardy. The total cost of restoring the NHS's IT system worth twenty million pounds (£20m) and a further seventy-two million pounds (£72m) on ensuring cleaning up and upgrading the IT systems. The Ransomware attack was spread using various methods, which include outdated systems without up-to-date security patches on Microsoft as well as phishing emails. More recently, ransomware attack is also reported in the energy sector, such as the Newalker attack against K-Electric in Pakistan in 2020 and the Colonial Pipeline attack in the US in 2021. K-Electric is the largest power supplier in Pakistan, and as a result of the ransomware attack, customer accounts on the system became inaccessible. Fortunately, the power grid operation itself was not affected by this incident. On the other hand, the attack against Colonial Pipeline caused the outage of their pipeline operation for a few days and a shortage of gas supply in multiple states.

- *Supply chain attack.* SolarWinds incident reported in 2020 is a concrete example of a supply chain attack. SolarWinds is a vendor that offers IT management systems and remote monitoring tools, and their systems have been widely used in major companies as well as government agencies. An attacker penetrated into the SolarWinds corporate network and injected a malicious code into their product. Then, without knowing the compromise, SolarWinds distributed the software update to the customers, resulting in planting a backdoor for more than 18,000 customers' systems. While the impact on the smart grid operators was not reported, it is definitely possible that a similar attack may affect the smart grid systems. Because many of the industrial control systems and SCADA systems rely on periodic firmware/software updates and also because the global market of industrial control systems is mostly dominated by a relatively small number of vendors (e.g., Siemens, ABB, etc.), the impact of the similar attack could be massive.

Cybersecurity forms a critical pillar of the trustworthiness of smart grid systems due to the importance of these systems in managing the generation, distribution, and consumption of electricity. Smart grids involve a complex, interconnected network of systems, devices, and stakeholders that must work together seamlessly and reliably. To ensure the cybersecurity for modern smart grid systems, we introduce the technical highlights from the following aspects:

- Formulation. Formulating cyber threats in the context of the smart grid involves identifying potential points of vulnerability and imagining how they might be exploited. The formulation of cyber threats provides a blueprint of potential attacks, which can guide the development and implementation of robust cybersecurity measures in smart grids. It's a vital step in ensuring the resilience, reliability, and security of these critical infrastructures.

- Deterrence. In the context of cybersecurity for smart grids, deterrence refers to strategies designed to prevent cyber attacks by discouraging potential attackers. This is typically achieved by increasing the perceived difficulty or cost of launching a successful attack and by enhancing the likelihood of identifying and prosecuting attackers. By focusing on deterrence, smart grid operators can aim to prevent attacks from happening in the first place instead of merely reacting to them. This forms a crucial aspect of a comprehensive cybersecurity strategy.

- Prevention. Prevention in the context of cybersecurity for smart grids refers to measures taken to raise the bar for attackers to be successful in damaging or negatively impacting the system. Prevention measures step in after attackers get a footprint in the system. It is a critical component of a comprehensive cybersecurity strategy for smart grids, aiming to stop cyberattacks before they can cause harm.

- Detection. Detection, in the context of cybersecurity for smart grids, refers to the processes and systems in place to identify when

a cyber attack is occurring or has occurred. This is a critical aspect of maintaining the security of a smart grid. Once a potential cyber attack has been detected, quick action can be taken to contain it and limit its impact. This makes detection a key element of a comprehensive cybersecurity strategy for smart grids. Timely detection of attack further enables us to trigger measures not only to contain the impact of attack to maintain the stability of the power grid but also to recover/restore the system.

Without trust in the smart grid system, users may be hesitant to adopt the technology and its benefits, such as improved energy efficiency, cost savings, and better grid management. In summary, cybersecurity is of critical importance for ensuring trustworthiness for smart grid systems. This, in turn, helps maintain public trust in the energy sector and ensures the delivery of safe, reliable, and efficient electricity services.

## 1.4 Summary of this book

This mini-book provides an overview of the challenges and solutions for the cybersecurity of smart grid systems.

We first formulate the cyber threats, attack tactics, and procedures using the widely-recognized framework, namely MITRE ATT&CK Matrix for ICS (industrial control systems) in Chapter 2. Then we introduce representative cybersecurity solutions that were developed by the research community and map them into the MITRE ATT&CK Matrix to illustrate where each technology can contribute to counter the cyber attacks in Chapter 3, Chapter 4 and Chapter 5.

Secondly, we shed light on the evaluation of such cybersecurity solutions on smart grid, featuring state-of-the-art smart grid testbeds available now in Chapter 6. We pick implementations of different kinds, namely hardware-based and software-based (i.e., cyber range or digital twin), and make comparisons to provide guidance for the readers. We also provide a cyberattack case study using such a testbed.

Overall, the book offers a comprehensive and practical guide to ensuring the cybersecurity of smart grids. It provides valuable insights and practical recommendations for organizations and individuals that are interested in the cybersecurity of smart grid systems.

# 2

# Formulation of Cyber Threats against Smart Grid Systems

Cyber attacks against smart grid systems are conducted through multiple stages. Formulation of cyber attack procedures against industrial control systems (ICS) has been attempted in ICS Cyber Kill Chain by SANS Institute [Assante and Lee, 2015], which models what the adversaries must complete to execute cyber attacks against industrial control systems. As shown in Figure 2.1 and Figure 2.2, the model defines 2 stages, namely intrusion and ICS attack.

The intrusion stage is further divided into phases that an attacker should undertake to gain access to the target ICS. The first phase is reconnaissance, which aims at collecting information about the target system through various means and aims at identifying system vulnerability. The collected knowledge can be utilized, then, to prepare attack vectors, such as weaponized files, as well as to identify victims to exploit. For instance, in the Ukraine incident in 2015, an attacker compromised a virtual private network (VPN) interface based on the information collected. What follows next is a cyber intrusion, which attempts to deliver the attack vectors (e.g., weaponized files as an email attachment), which are then exploited by an attacker to install back doors, for instance. Such a backdoor would be utilized to establish a command

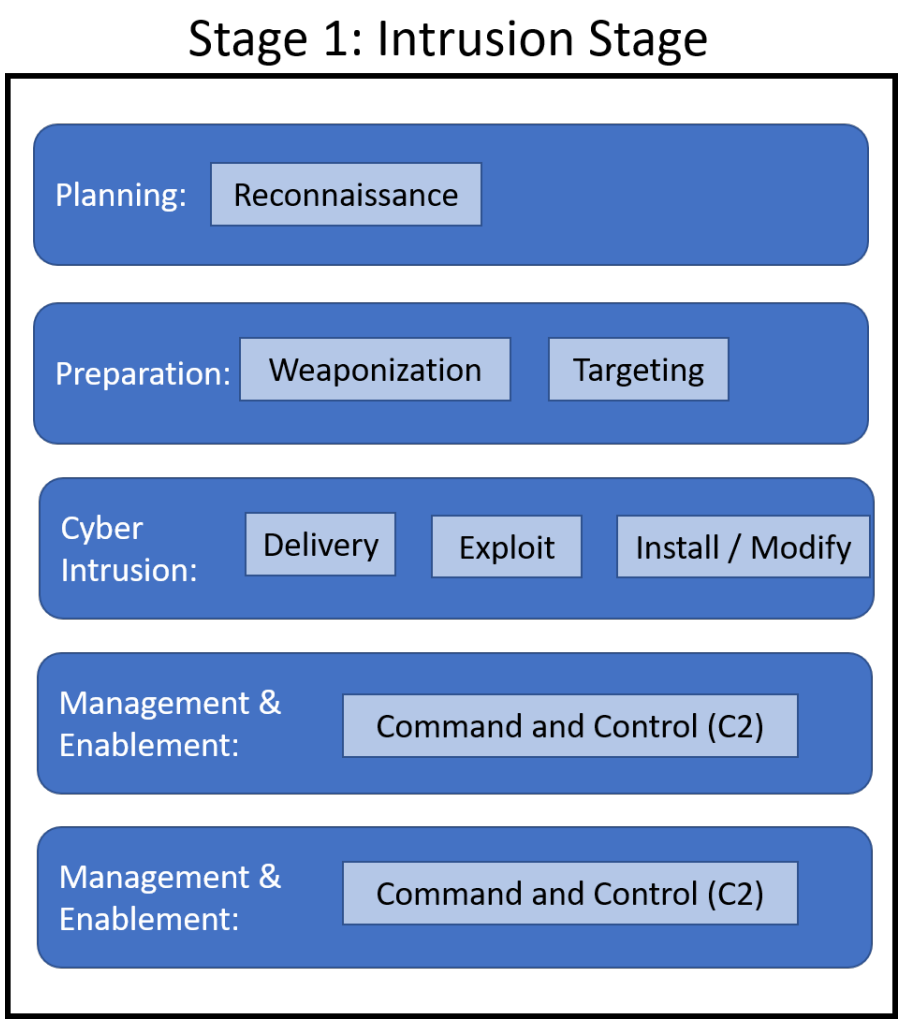

**Figure 2.1:** ICS Cyber Kill Chain Stage 1 [Assante and Lee, 2015]

and control (C2) channel for a remote attacker.

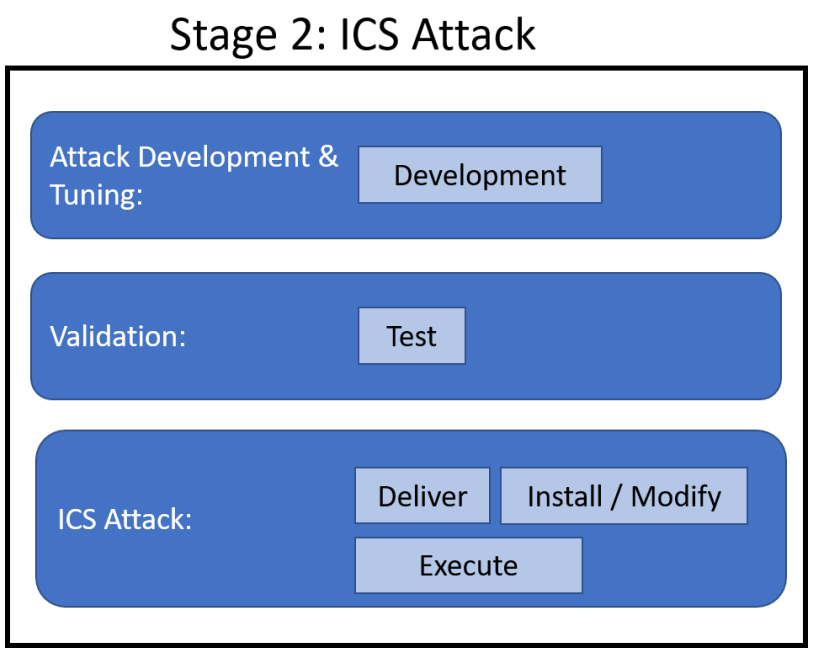

**Figure 2.2:** ICS Cyber Kill Chain Stage 2 [Assante and Lee, 2015]

The ICS attack phase is for an attacker to develop the capability of attacks against the specific ICS. The attack development phase exploits the information collected from the ICS to develop specific attack capabilities in order to cause the desired impact. Once the capability is developed, an attacker may often test it before mounting the actual attack. The last phase of the stage is launching attacks. The validated capability would be delivered and installed in the target

ICS to maliciously change process states, modify device configurations (e.g., setpoints and threshold) as well as spoof state information or measurements.

The MITRE Corporation formulated MITRE ATT&CK (Adversarial Tactics, Techniques, and Common Knowledge)) Matrix [*MITRE ATT&CK* n.d.]. MITRE ATT&CK Matrix was developed based on ICS Cyber Kill Chain, by adding attack tactics at each phase of the cyber attack against ICS, which are derived based on the study of real-world cyber incidents, such as Stuxnet malware, Ukraine power plant attack, and other high-profile ICS cybersecurity incidents occurred. Below we provide a brief overview of the MITRE ATT&CK Matrix for ICS.

As seen in Figure 2.3, MITRE defined stages of cyber attacks against industrial control systems, starting from "Initial Access" to "Impact". Then, for each stage, a number of attack tactics are listed. For instance, initial access can be attempted by exploiting "External Remote Services", "Remote Devices", and so forth. The matrix is mainly designed for modeling attack procedures, but at the same time, it can be utilized to evaluate the coverage of cybersecurity measures deployed. Thus, in the next chapter, we discuss emerging cybersecurity measures that are designed for smart grid systems, and for each type of defense measure, we use this matrix to show which attack tactics can be countered.

| Initial Access | Execution | Persistence | Privilege Escalation | Evasion | Discovery | Lateral Movement | Collection | Command and Control | Inhibit Response Function | Impair Process Control | Impact |
|---|---|---|---|---|---|---|---|---|---|---|---|
| Drive-by compromise | Change operating mode | Modify program | Exploitation for Privilege Escalation | Change operating mode | Network connection enumeration | Default credentials | Automated collection | Commonly used port | Activate firmware update mode | Brute force I/O | Damage to property |
| Exploit public facing application | Command line interface | Module firmware | Hooking | Exploitation for evasion | Network sniffing | Exploitation of remote services | Data from information repositories | Connection proxy | Alarm suppression | Modify parameter | Denial of control |
| Exploitation of remote services | Execution through API | Project file infection | | Indicator removal on host | Remote system discovery | Lateral tool transfer | Detect operating mode | Standard application layer protocol | Block command message | Module firmware | Denial of view |
| External remote services | Graphical user interface | System firmware | | Masquerading | Remote system information discovery | Program download | I/O image | | Block reporting message | Spoof reporting message | Loss of availability |
| Internet accessible device | Hooking | Valid accounts | | Rootkit | Wireless sniffing | Remote services | Man in the middle | | Block serial COM | Unauthorized command message | Loss of control |
| Remote services | Modify controller tasking | | | Spoof reporting message | | Valid accounts | Monitor process state | | Data destruction | | Loss of productivity and revenue |
| Replication through removable media | Native API | | | | | | Point & tag identification | | Denial of service | | Loss of production |
| Rogue master | Scripting | | | | | | Program upload | | Device restart/shutdown | | Loss of safety |
| Spear-phishing attachment | User execution | | | | | | Screen capture | | Manipulate I/O image | | Loss of view |
| Supply chain compromise | | | | | | | Wireless sniffing | | Modify alarm settings | | Manipulation of control |
| Transient cyber asset | | | | | | | | | Rootkit | | Manipulation of view |
| Wireless compromise | | | | | | | | | Service stop | | Theft of operational information |
| | | | | | | | | | System firmware | | |

**Figure 2.3:** Mitigation by Honeypot on MITRE ATT&CK Matrix for ICS.

When discussing cybersecurity solutions, it is crucial to map out which attack tactics can be mitigated by each cybersecurity solution. At a high level, cybersecurity solutions can be categorized into three types: deterrence, prevention, detection, containment, and recovery. The first line of defense, deterrence, is to discourage attackers from attacking the system. Deterrence can be attained by making attackers think that the attack is not feasible. Prevention can be implemented by means of cryptographic protection for messages, deployment of firewall or data diode devices for enforcing network flow control policies. Once these prevention measures are bypassed, we should detect the occurrence of the attack as quickly as possible, so that containment and recovery measures are triggered in a timely manner. In the following section, we discuss representative cybersecurity solutions for smart grid systems of each category and evaluate them based on the MITRE ATT&CK Matrix for ICS.

# 3

# Deterrence of Cyber Attacks against Smart Grid

Deterrence in cybersecurity aims at discouraging attackers from mounting attacks. For instance, an attacker may retreat if he finds something suspicious (an indication of deployment of cyber defense measures that are not familiar to him) in the system. Alternatively, a defender could consider misleading attackers to a dummy system and bring the attacker away from the real system to be protected.

Deception technologies for cybersecurity are, in general, categorized into two types: honeypot and decoy network (also called in-network deception). Both are typically implemented as a virtual system or device that appears and behaves like the real counterpart, but the deployment model and purpose are different.

## 3.1 Honeypot

Honeypot is the most popular example of deception technology, and it is a dummy system that imitates the characteristics and behavior of a real system or device. The honeypot is intentionally exposed to potential attackers to attract them. For instance, it is intentionally connected to the Internet and/or configured with weak or default login credentials. Such a honeypot can be found by the exhaustive network scanning of attackers or via a search engine like Shodan (http://shodan.io). The main purpose of a honeypot is to collect information and intelligence about attackers, including who they are, where they come from, what tactics/tools they use, and/or how the attack progresses. The collected threat intelligence can be utilized to fine-tune the cybersecurity measures deployed in the real system, such as firewalls and intrusion detection systems. Honeypot is also effective in delaying attacks by misleading them into the dummy system, which is isolated from the real system infrastructure to be protected. This way, the defender buys time before the attacker starts attacking the real system.

Honeypot is categorized into two types: low-interaction honeypot and high-interaction honeypot. Low-interaction honeypot allows limited access and interaction for attackers. For instance, a low-interaction honeypot would only imitate opened network ports of real devices. This way, an attacker could attempt to connect to such ports. However, the services running at the port are either empty (i.e., only listening at the port) or minimal (e.g., returns only the static replies) and don't allow the attacker to further interact with the device. On the other hand, a high-interaction honeypot provides more detailed internals of the device or system.

Mitigation provided by honeypot in terms of MITRE ATT&CK Matrix for ICS is summarized in Figure 3.1. In terms of mitigation of attack tactics, the contribution of the honeypot can be found mainly in the "Initial Access" and "Execution" stages because the honeypot can mislead attackers, which attempt to access and compromise the exposed user interfaces from the external network to the dummy system. While a high-interaction honeypot could be used to collect detailed threat intelligence by observing attackers' activities in the honeypot, it is not regarded as a direct contribution to mitigating attacks. Instead, such findings are utilized to configure other cybersecurity measures, such as intrusion detection systems discussed later.

While there are many efforts devoted to developing a honeypot for enterprise IT systems and web systems, a honeypot for smart grid systems is still in the early stage. In this section, we discuss a few examples of each category of honeypot systems designed for smart grid systems.

| Initial Access | Execution | Persistence | Privilege Escalation | Evasion | Discovery | Lateral Movement | Collection | Command and Control | Inhibit Response Function | Impair Process Control | Impact |
|---|---|---|---|---|---|---|---|---|---|---|---|
| Drive-by compromise | Change operating mode | Modify program | Exploitation for Privilege Escalation | Change operating mode | Network connection enumeration | Default credentials | Automated collection | Commonly used port | Activate firmware update mode | Brute force I/O | Damage to property |
| Exploit public facing application | Command line interface | Module firmware | Hooking | Exploitation for evasion | Network sniffing | Exploitation of remote services | Data from information repositories | Connection proxy | Alarm suppression | Modify parameter | Denial of control |
| Exploitation of remote services | Execution through API | Project file infection | | Indicator removal on host | Remote system discovery | Lateral tool transfer | Detect operating mode | Standard application layer protocol | Block command message | Module firmware | Denial of view |
| External remote services | Graphical user interface | System firmware | | Masquerading | Remote system information discovery | Program download | I/O image | | Block reporting message | Spoof reporting message | Loss of availability |
| Internet accessible device | Hooking | Valid accounts | | Rootkit | Wireless sniffing | Remote services | Man in the middle | | Block serial COM | Unauthorized command message | Loss of control |
| Remote services | Modify controller tasking | | | Spoof reporting message | | Valid accounts | Monitor process state | | Data destruction | | Loss of productivity and revenue |
| Replication through removable media | Native API | | | | | | Point & tag identification | | Denial of service | | Loss of production |
| Rogue master | Scripting | | | | | | Program upload | | Device restart/shutdown | | Loss of safety |
| Spear-phishing attachment | User execution | | | | | | Screen capture | | Manipulate I/O image | | Loss of view |
| Supply chain compromise | | | | | | | Wireless sniffing | | Modify alarm settings | | Manipulation of control |
| Transient cyber asset | | | | | | | | | Rootkit | | Manipulation of view |
| Wireless compromise | | | | | | | | | Service stop | | Theft of operational information |
| | | | | | | | | | System firmware | | |

**Figure 3.1:** Mitigation by Honeypot on MITRE ATT&CK Matrix for ICS.

Implementation of low-interaction honeypots that can be used to imitate smart grid systems or devices can be found in the research work from the open-source community as well as academic research. One representative example of the former is Conpot [Jicha *et al.*, 2016] and Honeyd [Provos, 2003]. Conpot is an easy-to-deploy honeypot implementation that supports popular protocols used in industrial control systems, such as Modbus. One limitation is that, because of its popularity, it is relatively easy to get detected or fingerprinted by attackers. For instance, based only on its network characteristics, Conpot could be detected relatively easily [Mirian *et al.*, 2016]. Also, it would be easily flagged by the Honeyscore system implemented by Shodan. Honeyd allows us to deploy a set of virtual devices with different network characteristics (e.g., IP and MAC addresses, port opened, etc.). The traffic received by Honeyd is forwarded to the back-end server according to the configuration, and thus we can imitate smart grid devices by running servers for smart grid protocols. One notable feature of Honeyd is that it has the feature to deceive device/OS fingerprinting by a scanning tool like Nmap (https://nmap.org/). These tools are also utilized as a building block for other smart grid honeypot implementations such as SCADA Honeynet [Pothamsetty and Franz, n.d.], GridPot [Bieker and Pilkington, 2020], and so forth.

In order to demonstrate the value of a low-interaction honeypot, let us discuss one effort from academia. In [Mashima *et al.*, 2019], the authors implemented a honeypot that is listening at the popular smart grid and ICS protocols, such as IEC 61850 Manufacturing Message Specification (MMS), IEC 60870-50104, Modbus, DNP3, and so forth. While most of the ports are simply opened, some services, such as IEC 60870-5-104, run a dummy server program that can provide minimal interaction with clients. Such honeypot instances are deployed on different geographic locations on Amazon cloud and operated over 6 months to collect data. The paper also demonstrated how the data collected by the honeypot can be used to configure cybersecurity tools such as firewalls and intrusion detection systems.

High-interaction honeypot for smart grid is still rare. At a high level, a high-interaction honeypot should imitate not only cyber-side characteristics but also consistent physical system behaviors so that an attacker interacting with dummy devices or systems would see the realistic response of the physical plants. Because of this reason, most of the efforts rely on some sort of power system simulators behind the cyber components, which can be implemented by using the aforementioned low-interaction honeypots.

One open-source effort is called GridPot [Bieker and Pilkington, 2020]. GridPot employs Conpot for the cyber side and utilizes an open-source power system simulator called GridLab-D. Similar efforts are found in academia. For instance, [Mashima *et al.*, 2017a] proposes a honeypot imitating an IEC 61850 compliant substation, which incorporates a number of virtual IEDs provided by an open-source project called SoftGrid [Gunathilaka *et al.*, 2016b] that are connected to a back-end PowerWorld simulator. This implementation further utilizes an open-source network emulator called Mininet [Kaur *et al.*, 2014] to present a realistic network topology on the cyber side. The architecture is shown in Figure 3.2. At the entrance of each honeypot substation, a substation gateway, which supports IEC 60870-5-104 and IEC 61850 protocols are located, and the virtual IEDs (“vHost” in the figure) are organized in a ring topology with multiple virtual network switches (“vSwitch” in the figure)in the substation local area network, which is often used in real systems. For the sake of presenting consistent cyber-physical system view, a power flow simulator is run behind the scene and is connected to the virtual IEDs.

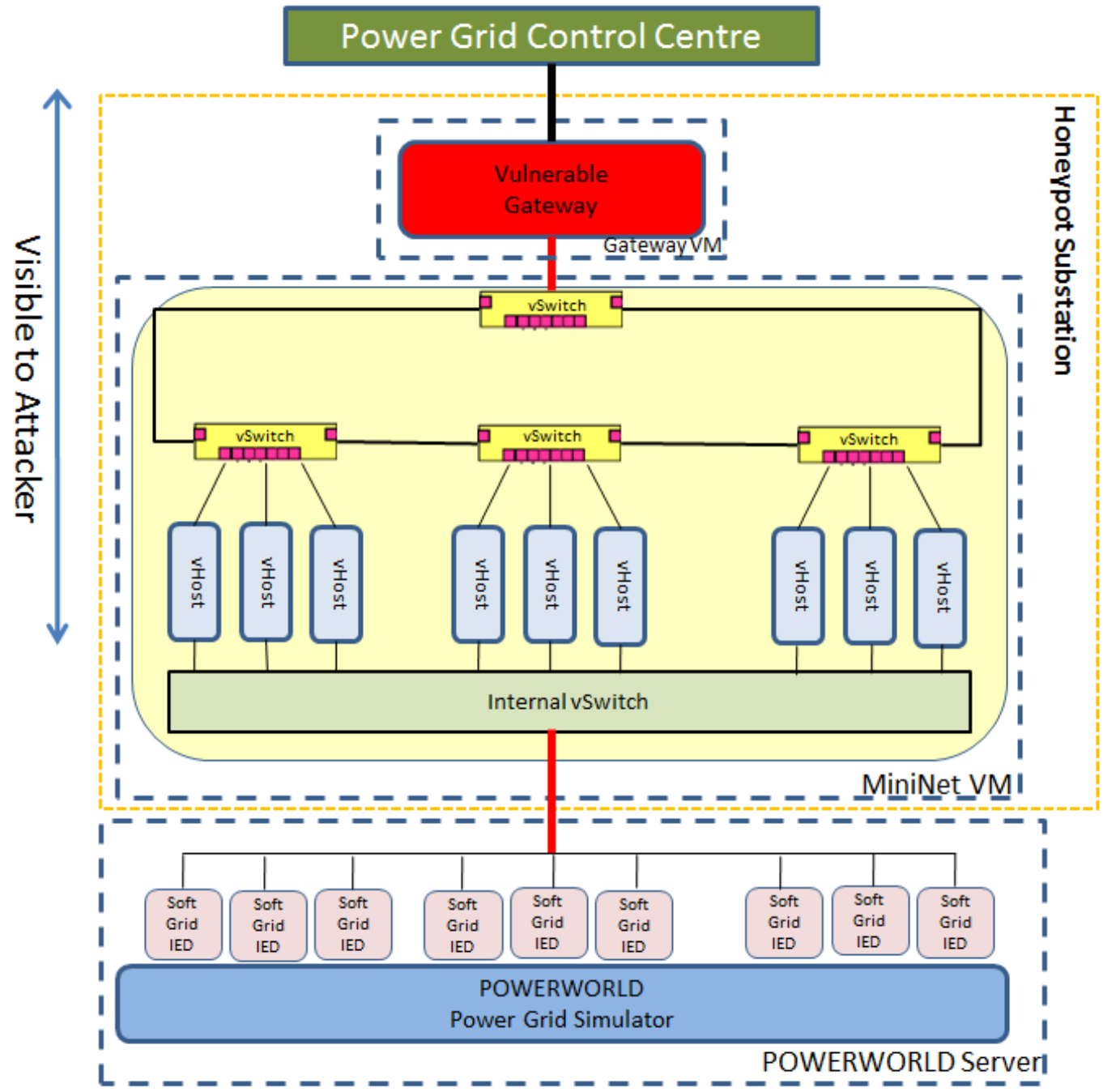

**Figure 3.2:** Architecture of IEC 61850 Compliant Substation Honeypot [Mashima *et al.*, 2017a]

There are some limitations found in these systems. For instance, both essentially implement only IEDs in the cyber side of the system, while in reality, there are other devices or servers, such as SCADA human-machine interface (HMI), historian database, engineering workstations, programmable logic controllers, and so forth. Moreover, device-level imitation is limited, and thus sophisticated attackers could easily identify that they are dummy devices. In order to address these challenges, [Mashima *et al.*, 2020] developed a honeypot that emulates comprehensive smart grid infrastructure, including the control center and field substation connected via a wide-area network. The honeypot consists of a number of Windows virtual machines running real SCADA and historian database software, Linux virtual machines (VMs) running firewalls, substation gateways, and virtual intelligent electronic devices (IEDs) and programmable logic controllers (PLCs). To enhance the realism, virtual IEDs are fronted by Honeyd to counter device/OS fingerprinting mounted by attackers. As the entry point for the attackers, it implements a VPN service, which is intentionally configured with weak password and known vulnerability as well as a jumpbox server in the DMZ (demilitarized zone) of a corporate network, imitating the typical remote access configuration implemented by real-world power grid operators. Another essential feature of a honeypot is secure and hidden logging mechanisms. In this direction, the honeypot implements a transparent proxy box running network scanners, which is not addressable by attackers in the honeypot. The overall architecture is shown in Figure 3.3. The realism of the honeypot was evaluated by means of penetration testing to eliminate any clue that would help attackers.

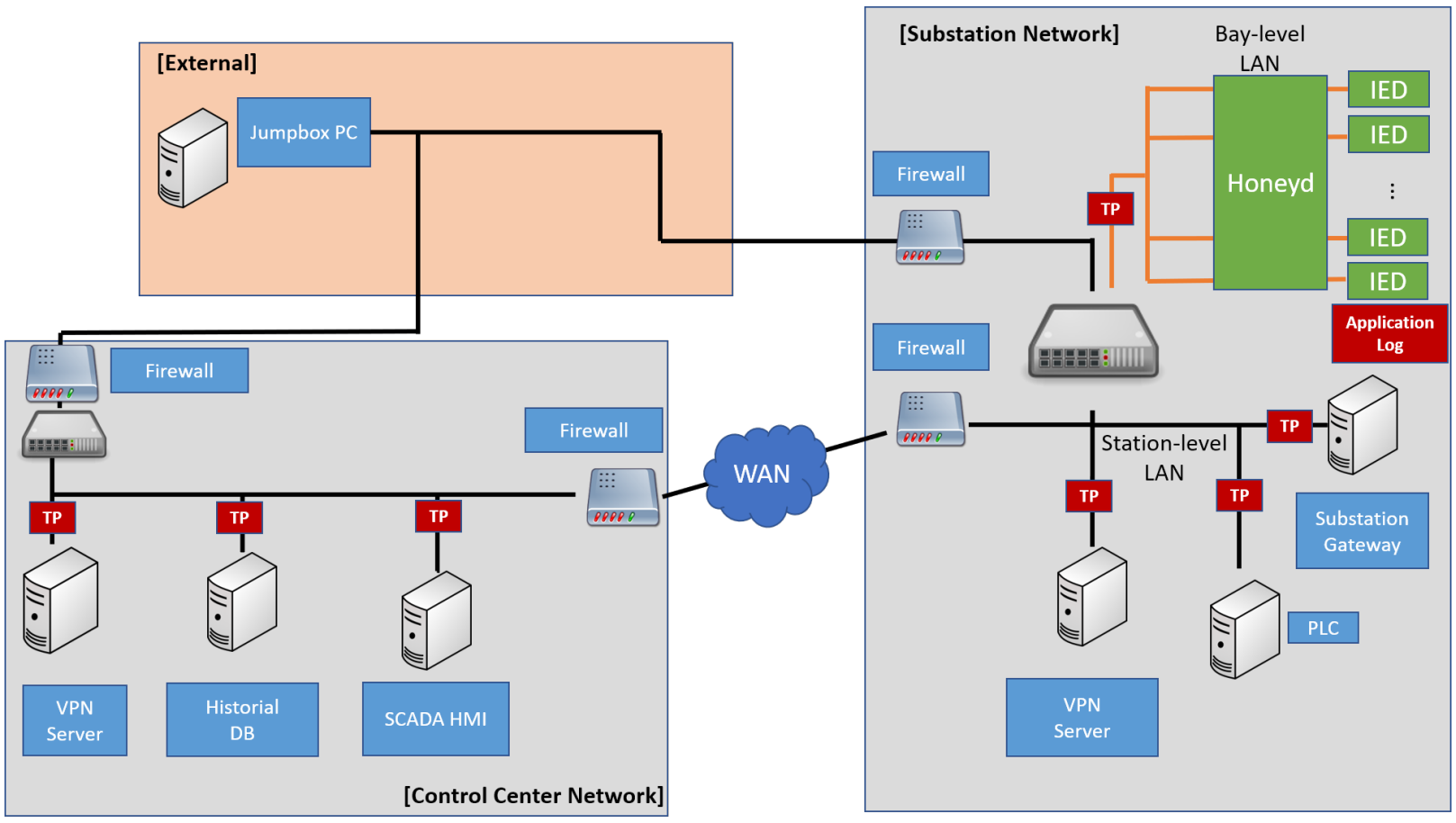

**Figure 3.3:** Architecture of Comprehensive Smart Grid Honeypot [Mashima *et al.*, 2020]

## 3.2 Decoy Network / In-network Deception

Unlike honeypots, decoy networks or in-network deception solutions are deployed in the real system infrastructure. In other words, dummy systems or devices are blended into the real system.

Decoy network typically aims at countering attackers (or malware) that already have a footprint in the system infrastructure. Such persistent attackers would do reconnaissance by passively sniffing the network traffic and/or by actively, but under the radar, sending query/interrogation commands to collect system information to prepare for the attack. By deploying a large number of dummy devices in the infrastructure that looks and behaves in a way indistinguishable from real devices, it would become difficult for attackers to pinpoint the target device. Moreover, by making them send fake, perhaps also intelligently crafted, data or measurements, we can prevent attackers from learning the system topology and configuration. In addition to deterring passive attackers who are only sniffing the network traffic without any actions, a decoy network can contribute to the early detection of attackers' activity. Namely, when an attacker would send innocuous control/interrogation commands for reconnaissance, if such access hits any of the deception devices deployed, the decoy device can raise the alarm to bring the situation to the operators' attention to initiate cybersecurity incident response measures before the serious attack is launched.

In terms of MITRE ATT&CK Matrix for ICS, Decoy network / in-network deception technology can counter the tactics highlighted in Figure 3.4. As seen, the coverage is orthogonal to the honeypot, mainly because of the difference in deployment models. More specifically, decoy network/in-network deception counters tactics in "Discovery", "Lateral Movement", "Collection", "Command and Control", and "Impair Process Control" stages. Security solutions of this category can make it difficult for attackers to pinpoint target devices and acquire knowledge of smart grid system topology. Lateral movements, which are activities of attackers or malware to propagate, and malicious command injection can also be mitigated because once decoys are touched by attackers, an alarm is raised. Likewise, attackers' attempts to upload malicious firmware to IEDs can also hit decoy devices and thus be detected.

One representative implementation of the decoy network designed for the IEC 61850 compliant smart grid system is DecIED [Yang *et al.*, 2020], which stands for deception IED. As the name implies, DecIED runs a number of dummy (virtual) IEDs that imitate device characteristics and behavior of real IED devices deployed in the system. As seen in

| Initial Access | Execution | Persistence | Privilege Escalation | Evasion | Discovery | Lateral Movement | Collection | Command and Control | Inhibit Response Function | Impair Process Control | Impact |
|---|---|---|---|---|---|---|---|---|---|---|---|
| Drive-by compromise | Change operating mode | Modify program | Exploitation for Privilege Escalation | Change operating mode | Network connection enumeration | Default credentials | Automated collection | Commonly used port | Activate firmware update mode | Brute force I/O | Damage to property |
| Exploit public facing application | Command line interface | Module firmware | Hooking | Exploitation for evasion | Network sniffing | Exploitation of remote services | Data from information repositories | Connection proxy | Alarm suppression | Modify parameter | Denial of control |
| Exploitation of remote services | Execution through API | Project file infection | | Indicator removal on host | Remote system discovery | Lateral tool transfer | Detect operating mode | Standard application layer protocol | Block command message | Module firmware | Denial of view |
| External remote services | Graphical user interface | System firmware | | Masquerading | Remote system information discovery | Program download | I/O image | | Block reporting message | Spoof reporting message | Loss of availability |
| Internet accessible device | Hooking | Valid accounts | | Rootkit | Wireless sniffing | Remote services | Man in the middle | | Block serial COM | Unauthorized command message | Loss of control |
| Remote services | Modify controller tasking | | | Spoof reporting message | | Valid accounts | Monitor process state | | Data destruction | | Loss of productivity and revenue |
| Replication through removable media | Native API | | | | | | Point & tag identification | | Denial of service | | Loss of production |
| Rogue master | Scripting | | | | | | Program upload | | Device restart/shutdown | | Loss of safety |
| Spear-phishing attachment | User execution | | | | | | Screen capture | | Manipulate I/O image | | Loss of view |
| Supply chain compromise | | | | | | | Wireless sniffing | | Modify alarm settings | | Manipulation of control |
| Transient cyber asset | | | | | | | | | Rootkit | | Manipulation of view |
| Wireless compromise | | | | | | | | | Service stop | | Theft of operational information |
| | | | | | | | | | System firmware | | |

**Figure 3.4:** Mitigation by Decoy Network / In-network Deception on MITRE ATT&CK Matrix for ICS.

Figure 3.5, DecIED is implemented as a security appliance box that is connected to the station bus, which connects SCADA HMI, engineer workstation, and substation-local control center devices, and process bus, which connects IEDs and merging units (sensors), of the substation local area network. On a commodity industrial PC, around 200 virtual (decoy) IEDs can be run with different IP addresses.

The solution aims at deploying “$k$-anonymous smoke screen”, which presents $k-1$ decoy IEDs that look and behave identically to the real IED. For example, when $k$=10, an attacker who has a footprint in the infrastructure would see 10 identically looking IEDs that are sending the same data at the same timings. In order to imitate device fingerprints, DecIED utilizes an open-source software called Honeyd [Provos, 2003], and network services run on each virtual IED are configured based on the study of real Siemens IEDs. Imitation of behaviors (e.g., timing and content of network packet to be sent out by virtual IEDs) requires synchronization of internal state and physical system view between the real IED and virtual IED instances. DecIED addresses this challenge (without obvious communication for coordination among them) by utilizing link-layer broadcast communication employed by standard protocols, namely IEC 61850 Generic Object Oriented Substation Events (GOOSE) and Sampled Values (SV). These protocols carry status updates and power grid measurements and also can be heard by the real IED and virtual IEDs at the same time. Thus, virtual IEDs can maintain the synchronized system view as the real IED and can further reply to IEC 61850 MMS queries sent by the SCADA HMI just like the real IED. These imitations make it difficult for the attacker to send a malicious control command (e.g., to open/close circuit breakers) to the real device connected to the physical power grid component (e.g., a circuit breaker in this example).

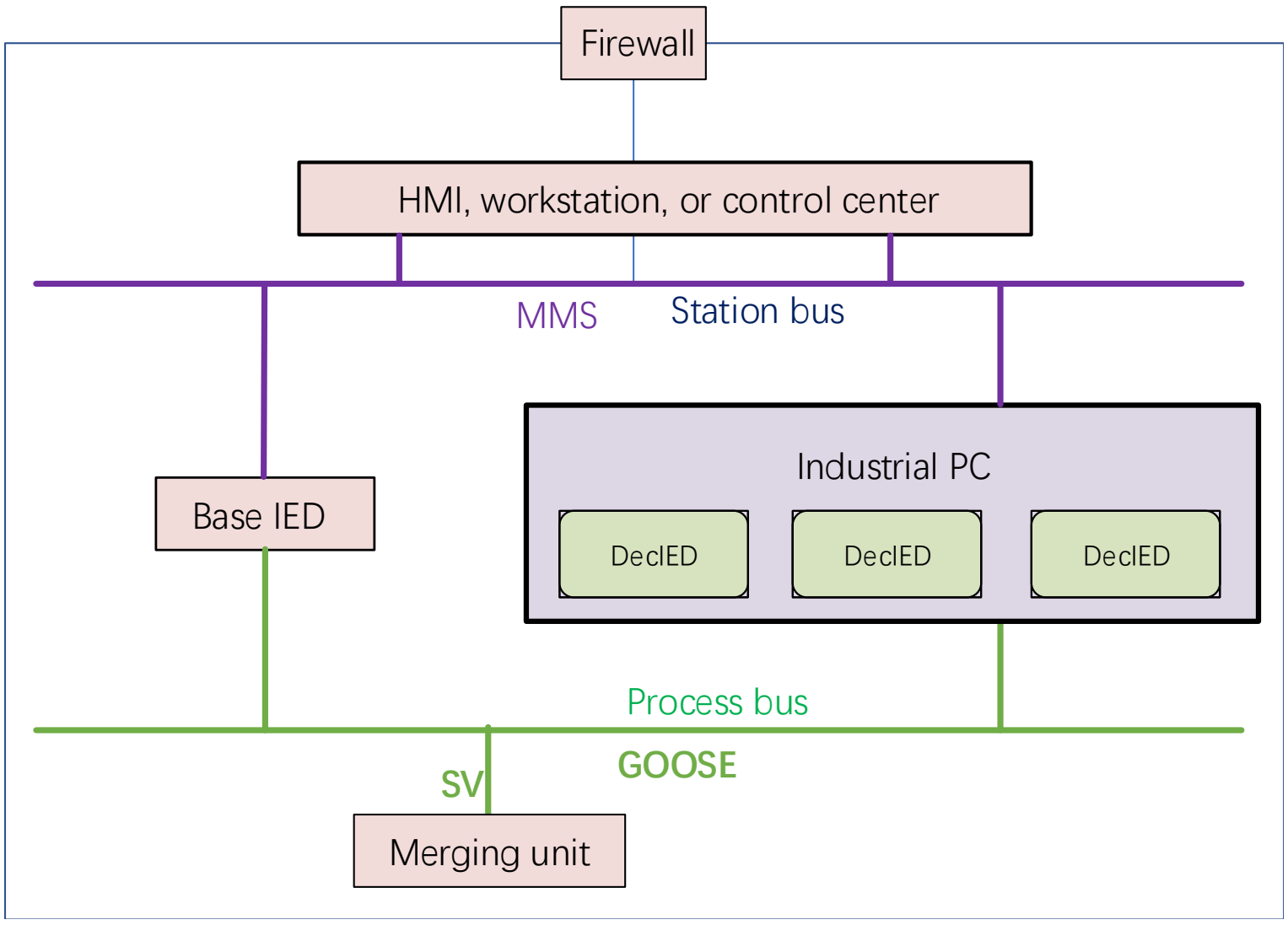

**Figure 3.5:** Deployment of DecIED in IEC 61850 Compliant Substation [Yang *et al.*, 2020]

## 3.3 Future Directions

The deception technologies for smart grid systems are still in the nascent stage. Thus, the enhancement of fidelity is one challenging work to be pursued. For instance, the honeypot or decoy network technologies discussed in this chapter provide only an imitation of generic device functionality and internals, and they don't emulate characteristics or vulnerabilities of a specific device model in the market. A lack of such imitation would hint to attackers to tell whether the system is a dummy or real. Considering such limitations, it is important to have a comprehensive framework to evaluate the realism and fidelity of the implementation qualitatively or quantitatively, which is then used to improve the degree of deception. Such a framework may require the enumeration of a wide range of honeypot/deception fingerprinting techniques that could be mounted by attackers. This is also an important future work, and formulating a taxonomy of honeypot fingerprinting (i.e. anti-honeypot) tactics for smart grid honeypot is the first step towards this direction [Tay *et al.*, 2023].

Design and implementation of comprehensive smart grid honeypot or honeynet, instead of a deception technology imitating a single device, is a challenging task that would require intensive domain knowledge in addition to cybersecurity expertise. It would be interesting to implement a toolkit or library for the essential building blocks for smart grid honeypots and further generate the dummy system according to the user configuration automatically.

Another promising research direction is an integration of artificial intelligence (AI) technologies. For instance, the configuration or topology of the deception system can be changed according to the observed activity of attackers in an adaptive manner by using generative AI technologies. It would not only help provide better deception but also collect threat intelligence more effectively.

# 4

# Prevention of Cyber Attacks against Smart Grid

Prevention aims at preventing cyber attacks from causing an impact on the system. To ensure the availability and integrity of power grid operation under the presence of attackers, assurance of the correctness and authenticity of messages exchanged among devices is crucial. However, many of the communication protocols used in smart grid systems, such as Modbus, DNP3, IEC 60870, and IEC 61850, are not equipped with built-in security features. Thus, smart grid devices are vulnerable to message manipulation on the network, injection of fake messages (false data injection and false command injection), and impersonation.

In order to prevent such attacks, in the general cybersecurity domain, the use of cryptography to protect message authenticity, integrity, and confidentiality is popular. In the smart grid domain, IEC 62351 was published as the supplementary cybersecurity specifications for widely-used smart grid communication protocols. For instance, it recommends the use of Transport Layer Security (TLS) for all communication protocols over the transport layer (e.g., TCP or UDP) of the protocol stack. This recommendation covers protocols such as Modbus TCP, DNP3, IEC 60870-5-104, and IEC 61850 MMS. For the protocols that work on the link layer, for instance, IEC 61850 GOOSE and SV, the use of a digital signature or message authentication code for message authentication is defined. While these technologies are very common in typical IT systems, it is not yet widely adopted in smart grid systems. One major barrier is resource constraints on industrial control systems devices, such as PLCs and IEDs used in smart grid systems as well as stringent latency requirements.

It is unavoidable that cryptographic protection requires additional time for encryption and decryption (or signing and signature verification). In particular, public-key cryptography, such as RSA and Elliptic-curve cryptography, requires very heavy computation, such as modular exponentiation. Although such computation can be done within manageable latency on commodity computers and servers, it is still costly for many industrial control systems in the market. Based on the experiment reported in [Tefek *et al.*, 2022], the total latency for Elliptic Curve Digital Signature Algorithm (ECDSA) signing and verification is over 9ms on an embedded platform. On the other hand, according to IEEE Power and Energy Society guideline [IEEE Power and Energy Society, 2004], the most latency stringent communication cannot tolerate latency over 2ms, which practically prohibits the use of public-key cryptography. We should note that this latency requirement counts not only time for cryptographic protection but also pure transmission time. Thus, even for symmetric-key cryptography, such as message authentication code (MAC), meeting this requirement might be difficult depending on the device specification and or communication models (e.g., multi-cast communication). Recent products may support advanced cryptographic operations, but unfortunately it is very common in industrial control system that outdated, legacy devices are prevalently utilized. This is because once the infrastructure and devices are deployed, typically they continue to be utilized for a decade or longer, Moreover, infrastructure operators are often reluctant to conduct device upgrades or firmware updates of devices for fear that such changes may affect the stability and availability of the system.

With cryptographic protections for ICS message authenticity, integrity, and confidentiality, we can counter a set of attack tactics listed in MITRE ATT&CK Matrix shown in Figure 2.3. By encrypting the communication, the Discovery stage of the attack procedure is made difficult. For instance, network sniffing for attackers' reconnaissance activities is prevented. Likewise, many of the tactics in the Collection stage can also be countered. For instance, Man in the Middle can be prevented by means of message authentication, such as MAC or digital signatures. Program Upload, which aims at uploading malicious code to smart grid devices, can also be countered by appropriately signing the software. Moreover, attack tactics in Command and Control stage, such as Standard Application Layer Protocols, which abuse standard ICS communication protocols, can also be countered by implementing source and message authentication. Other tactics in the same stage can be countered likewise. Cryptographic protection is also considered effective in Impair Process Control stage. For instance, by implementing message authentication, both Spoof Reporting Message and Unauthorized Command Message are prevented. Based on our assessment, the cryptographic protections are effective in mitigating the tactics found in Figure 4.1.

| Initial Access | Execution | Persistence | Privilege Escalation | Evasion | Discovery | Lateral Movement | Collection | Command and Control | Inhibit Response Function | Impair Process Control | Impact |
|---|---|---|---|---|---|---|---|---|---|---|---|
| Drive-by compromise | Change operating mode | Modify program | Exploitation for Privilege Escalation | Change operating mode | Network connection enumeration | Default credentials | Automated collection | Commonly used port | Activate firmware update mode | Brute force I/O | Damage to property |
| Exploit public facing application | Command line interface | Module firmware | Hooking | Exploitation for evasion | Network sniffing | Exploitation of remote services | Data from information repositories | Connection proxy | Alarm suppression | Modify parameter | Denial of control |
| Exploitation of remote services | Execution through API | Project file infection | | Indicator removal on host | Remote system discovery | Lateral tool transfer | Detect operating mode | Standard application layer protocol | Block command message | Module firmware | Denial of view |
| External remote services | Graphical user interface | System firmware | | Masquerading | Remote system information discovery | Program download | I/O image | | Block reporting message | Spoof reporting message | Loss of availability |
| Internet accessible device | Hooking | Valid accounts | | Rootkit | Wireless sniffing | Remote services | Man in the middle | | Block serial COM | Unauthorized command message | Loss of control |
| Remote services | Modify controller tasking | | | Spoof reporting message | | Valid accounts | Monitor process state | | Data destruction | | Loss of productivity and revenue |
| Replication through removable media | Native API | | | | | | Point & tag identification | | Denial of service | | Loss of production |
| Rogue master | Scripting | | | | | | Program upload | | Device restart/shutdown | | Loss of safety |
| Spear-phishing attachment | User execution | | | | | | Screen capture | | Manipulate I/O image | | Loss of view |
| Supply chain compromise | | | | | | | Wireless sniffing | | Modify alarm settings | | Manipulation of control |
| Transient cyber asset | | | | | | | | | Rootkit | | Manipulation of view |
| Wireless compromise | | | | | | | | | Service stop | | Theft of operational information |
| | | | | | | | | | System firmware | | |

**Figure 4.1:** Mitigation by Cryptographic Protection on MITRE ATT&CK Matrix for ICS.

There are two directions to address the challenges that prevent cryptographic protections from being widely deployed. In order to address the difficulty of system and device upgrades, we can consider introducing bump-in-the-wire security appliances, which are add-on devices that transparently provide security, including cryptographic protection, without requiring changes on smart grid devices that are already deployed. Concerning the challenge caused by computational complexity and latency requirements, very low-latency, light-weight message authentication mechanisms that offer security equivalent to the traditional symmetric-key and public-key encryption schemes are proposed. For the rest of this section, we discuss the technical details of solutions in these categories.

## 4.1 Bump-in-the-wire Security Appliance

The universal challenges in industrial control systems (ICS) are resource constraints and the difficulty of modification and upgrade of the system and devices. It is common that, once deployed, ICS devices are used for over a decade. This implies that devices developed 10 years ago might still be in use, which poses a challenge to support advanced cryptographic technologies. Furthermore, infrastructure operators are usually reluctant to replace ICS devices to support security measures, even if new models would be capable of conducting heavier cryptographic computations with a short latency.

The promising solution to address these is to insert additional security appliances in a "transparent" manner. Such devices work as part of the network medium from the perspective of end ICS devices and intercept ICS messages from/to ICS devices to provide cryptographic protection. For instance, a security appliance device located near the sender would intercept the message and calculate and append a digital signature for it, which is then verified and stripped by another security appliance located in front of the receiver. This way, no device upgrade or configuration change is needed. Another advantage of this approach is that the security appliance can be equipped with the latest hardware with better computation capabilities to support computationally-intensive operations, such as public-key cryptography.

One example of the bump-in-the-wire (BITW) security appliance to implement light-weight message authentication for legacy smart grid communication protocol, called F-pro, was discussed by Esiner et al. [Esiner *et al.*, 2019]. F-pro defines a security protocol using a cryptographic hash function using pre-shared symmetric encryption keys.

F-pro can be implemented on low-cost embedded devices (Figure 4.2) and are deployed in front of each ICS device in the infrastructure as shown in Figure 4.3. This way, BITW F-pro devices can intercept incoming and outgoing messages from the ICS devices.

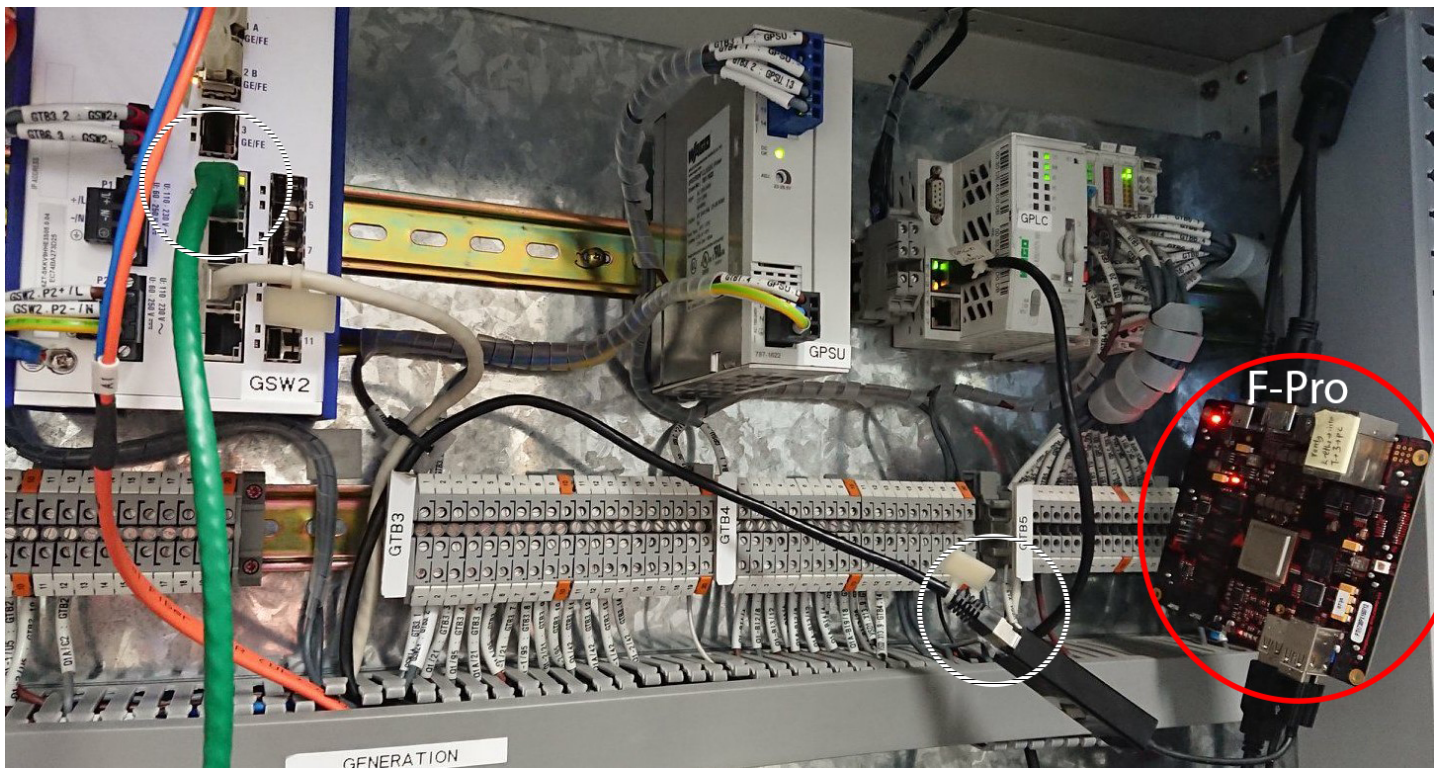

**Figure 4.2:** F-pro Device Deployed in Front of a PLC in the Smart Grid Testbed [Esiner *et al.*, 2019]

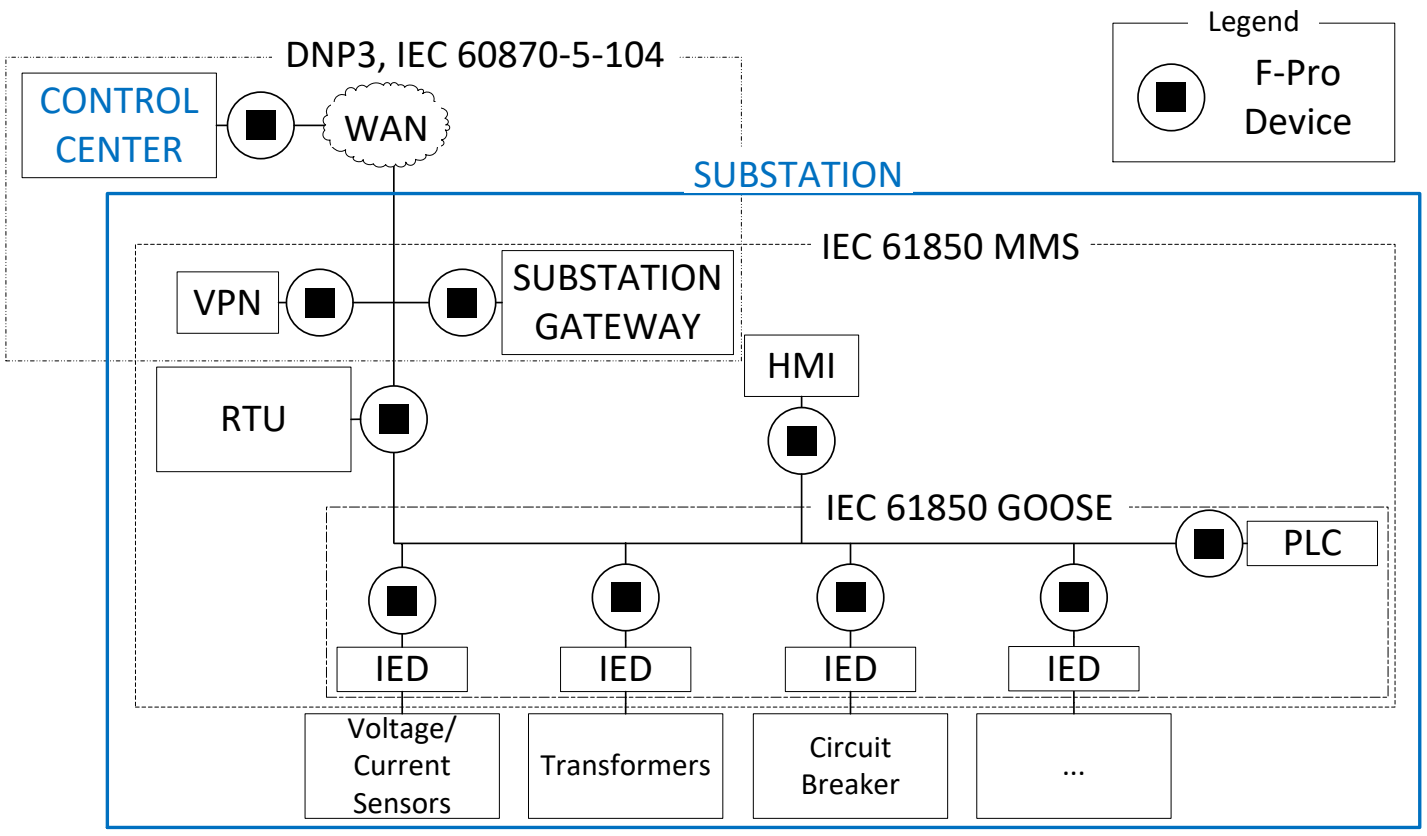

**Figure 4.3:** Deployment of F-pro Devices in a Modernized Substation [Esiner *et al.*, 2019]

In addition to the source authentication and integrity of messages, F-pro further provides cryptographically verifiable provenance of messages, including a path through which the message has traveled before reaching the destination. More specifically, an F-pro device located in front of the sender device (e.g., the SCADA HMI in the control center) initiates a chain of the cryptographic evidence, and then, by the F-pro devices deployed at each intermediate device (e.g., the substation gateway and remote terminal unit) or security appliance (e.g., firewall), the cryptographic evidence is updated to indicate which devices witnessed the message. Such a chain can be cryptographically verified at the F-pro device in front of the destination ICS device (e.g., an IED) and the derived provenance information can be checked against the security policy implemented on the F-pro device.

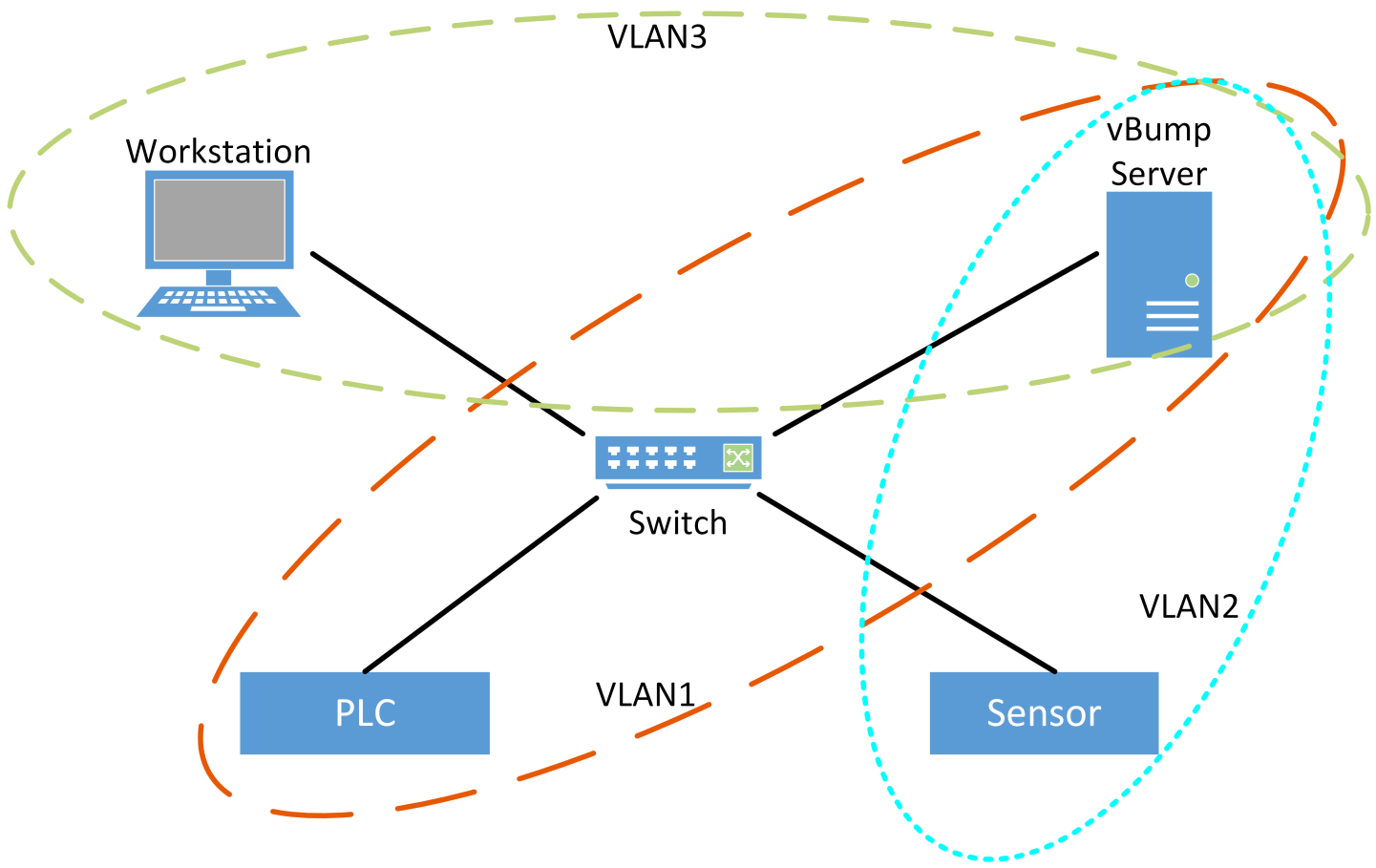

**Figure 4.4:** VLAN Configurations for Virtual Bump-in-the-wire Security Appliance [Tippenhauer *et al.*, 2021]

Deploying additional devices into the infrastructure is often costly. In such a case, infrastructure operators can consider a virtual BITW device. It is imperative that the BITW security module cannot be bypassed and BITW devices can reliably mediate all incoming and outgoing traffic to enforce the security checking and policy enforcement. In order to implement such a reliable mediation in a virtualized way, there is a technology called service insertion [Breslin *et al.*, 2014; Naiksatam *et al.*, 2017; Tippenhauer *et al.*, 2021], which takes advantage of the virtual LAN (VLAN) technology that is widely implemented in commercial switching hub for industrial control systems. We here use a technology called vBump [Tippenhauer *et al.*, 2021] for illustrating the concept.

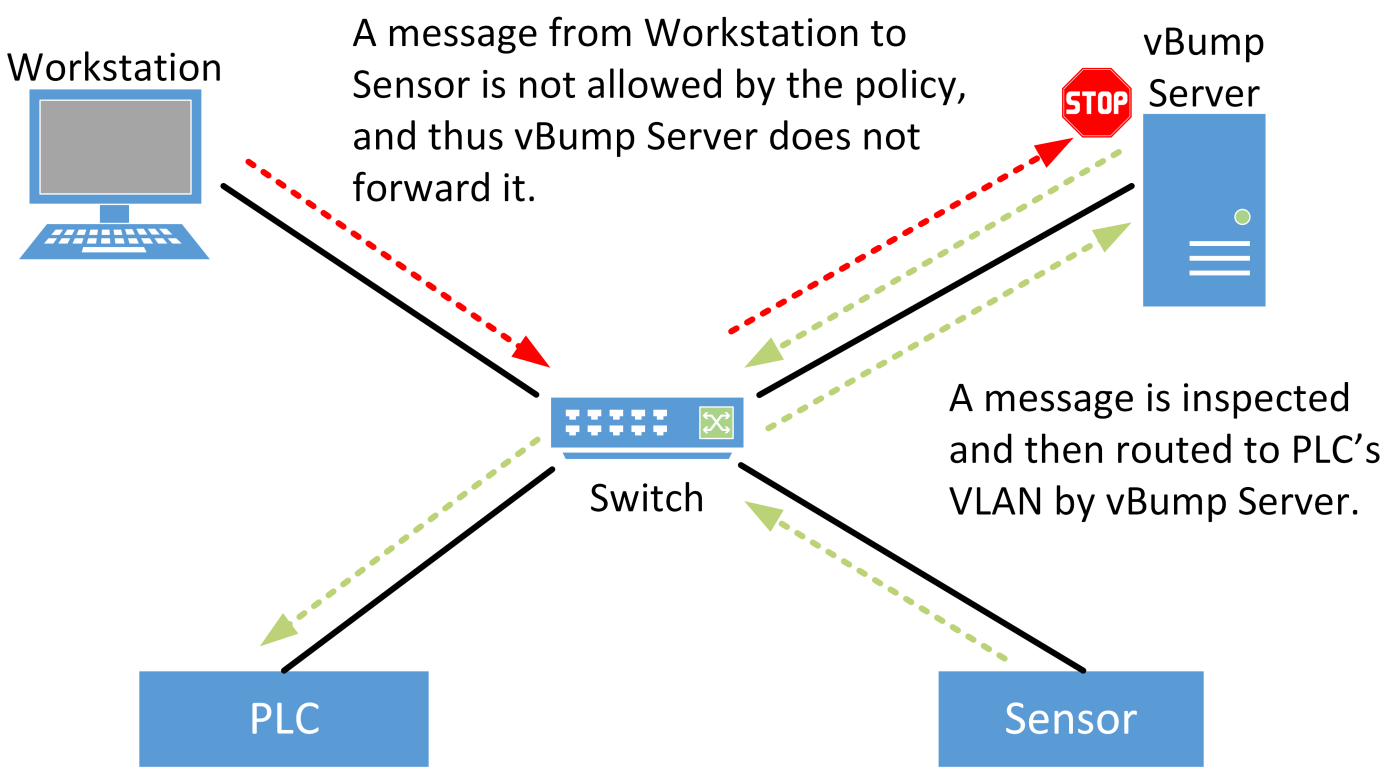

**Figure 4.5:** Network Traffic Policing by Virtual Bump-in-the-wire Security Appliance [Tippenhauer *et al.*, 2021]

The main idea of vBump is to automatically configure VLANs on all network infrastructure to redirect all traffic through a central server called vBump Server. At the high level, VLANs are defined in a way that each VLAN contains only one ICS device connected to the switch and the vBump Server (see Figure 4.4). In other words, all ICS devices belong to different VLANs, while the vBumnp server belongs to all the VLANs. Figure 4.5 demonstrates the idea of network traffic policing by vBump. In the figure, network traffic from a sensor to a PLC needs to be forwarded from one VLAN to another, which can be done only by vBump Server that belongs to both VLANs. Therefore, all the network traffic among the ICS devices can be reliably mediated by vBump Server. If an attacker on the workstation attempts to inject a malicious message to a sensor, the message must go through the security checking done by vBump Server, and thus it can be blocked before reaching the target device. Because traffic passing through the switch is aggregated on vBump Server, a single vBump server can provide security for multiple ICS devices in the network. Besides the traffic policing based on the security policy, vBump Server can further provide intrusion detection using the aggregated traffic information as well as cryptographic protection, such as encryption and message authentication.

## 4.2 Low-latency Message Authentication

While message authentication using cryptography is promising to counter multiple attack tactics, we need to address challenges in resource constraints and latency requirements. While approaches using symmetric key cryptography and keyed hash function would have the advantage of meeting such constraints, public verifiability, which is usually provided by digital signature schemes using asymmetric key cryptography, is often demanded (e.g., for multicast communication). For instance, IEC 61850 GOOSE and SV protocols utilize a publisher-subscriber model using multicast communication. If a symmetric key encryption scheme is utilized for message authentication, a sender (publisher) needs to calculate a message authentication code (MAC) for each recipient to ensure security. For instance, if there are 20 subscribers, the time needed by the publisher to process a single message would be 20 times longer than the single MAC calculation time, and the latency would grow linearly to the number of subscribers. On the other hand, if we use a digital signature scheme, a single signature for the message can be verified by all subscribers. However, the time needed to process (sign d and verify) a digital signature takes over 9ms for Elliptic Curve Digital Signature Algorithm (ECDSA) and over 11ms for the RSA signature scheme on an embedded platform [Tefek *et al.*, 2022], while latency requirement for the most urgent type of communication in a smart grid, according to the IEEE's guideline [IEEE Power and Energy Society, 2004], is 2ms. This was the main reason why the IEC 62351 standard recently relaxed the requirement mandating the use of digital signature schemes for protecting IEC 61850 GOOSE and SV communication.

A promising solution to address these while benefiting security guarantees provided by asymmetric key cryptography is to do precomputation. In other words, all or majority of the heavy computation for generating a digital signature or message authentication code "before" a message to be authenticated becomes ready. Digital signature scheme following this design principle is called online/offline signatures [Shamir and Tauman, 2001]. By designing this way, the majority of the time-consuming operations are done in advance, and thus latency for real-time communication can be minimized. This may sound infeasible, but by taking advantage of the sufficiently large bandwidth of the network medium (LAN with 1Gbps or above) in substations, such a solution can be designed.

A state-of-the-art scheme that implements the pre-computation concept for smart grid systems is called LoMoS (Less Online More Offline Signatures) [Esiner *et al.*, 2022]. LoMoS is an advanced version of an established concept called online/offline signatures. LoMoS concept is illustrated in Figure 4.6. Here, $KeyGen$ (key generation and setup) is a one-time task upon the system configuration. $Setup$ and $VerifySetup$ are the tasks to be done in an offline phase, while $Prove$ and $Verify$ are the tasks done in an online phase.

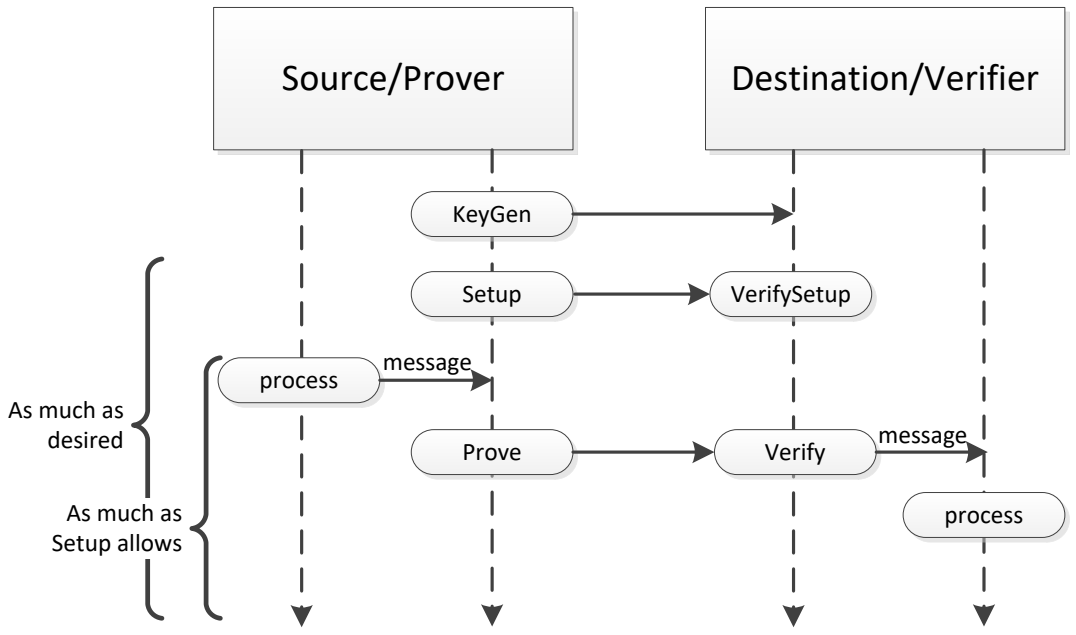

**Figure 4.6:** Concept of Less Online More Offline Digital Signature Scheme (LoMoS) [Esiner *et al.*, 2022]

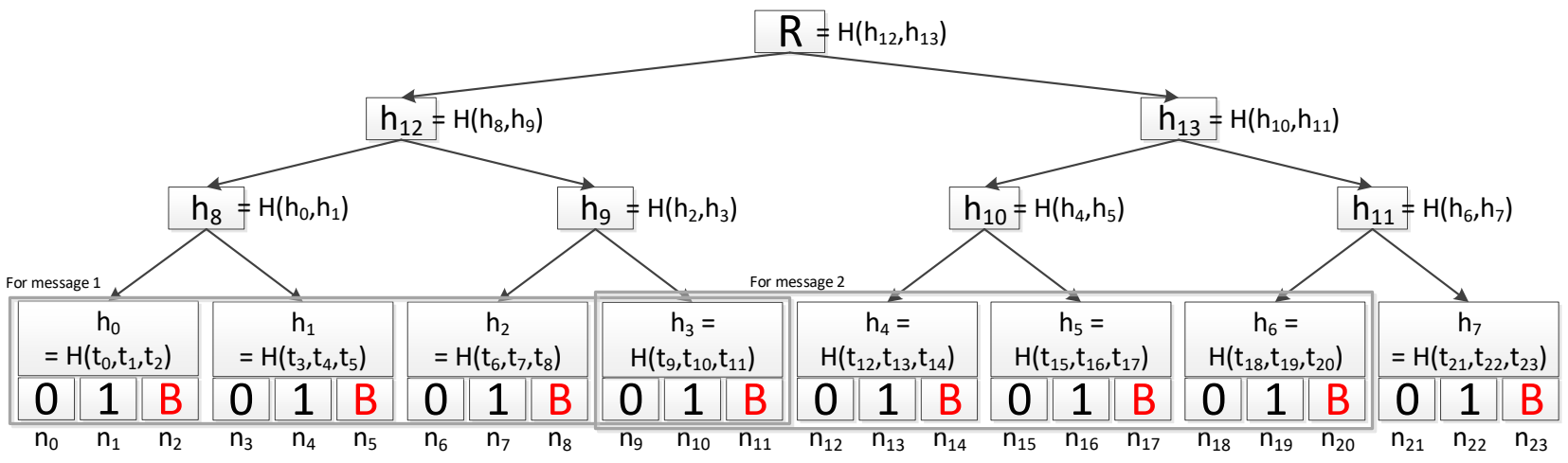

**Figure 4.7:** A Tri-leaf Tree example, where $t_i = H(n_i)$ [Esiner *et al.*, 2022]

$Setup$, which can be done before a message to be authenticated becomes available, generates an authenticated data structure based on Merkle Hash Tree, called Tri-leaf Tree (Figure 4.7), signs on its root hash value $R$ using any digital signature schemes (e.g., ECDSA), and then distributes it to all the subscribers. A tri-leaf tree is a hash tree calculated by using a public, secure hash function $H$ and each leaf node contains 3 random numbers corresponding to 0-bit, 1-bit, and "break" (shown as "B" in the figure), which is used to indicate the ending of each message. All the random numbers ($n_0, \ldots, n_{23}$ in the figure) are kept secret in the publisher until they are used in an online phase. $VerifySetup$ is done by each subscriber, and it includes verification of the digital signature. While these offline tasks rely on computationally heavy digital signature schemes, they can be done in advance, and one data structure generated and verified at $Setup$ and $VerifySetup$ can be used to authenticate multiple future messages. Moreover, the offline tasks can be executed even in parallel to the online message authentication.

Once the message to be authenticated is made available, $Prove$ is executed by the publisher. At the high level, $Prove$ picks random numbers corresponding to the bits in the message to be authenticated. For instance, in Figure 4.7, if a message to be sent is 010 in a binary representation, $n_0, n_4, n_6, n_{11}$ are picked. Note that $n_{11}$ is needed to indicate the ending of the message. A proof to be sent to the subscriber also contains hash values corresponding to the remaining random numbers of each node. In addition, hash values corresponding to non-leaf nodes that are required by a subscriber to calculate the root hash value are also included. In the earlier example, $h_{13}$ is included in the proof. Note that this step only requires memory look-up and calculation of hash values, whose latency is almost negligible. This proof is sent along with the message to be authenticated. In $Verify$, a subscriber calculates the root hash value by placing the disclosed random numbers to the position corresponding to the message bits along with the non-leaf hash value(s) provided. If the calculated root hash value is the same as the one verified at the $VerifySetup$, the message is authenticated. Otherwise, the message should be rejected. Again, $Verify$ also involves light-weight hash calculations, and thus the latency is negligible. We also note that the remaining part of the trees (in this example, the right half of the tree) can be utilized for the next message authentication without redoing the offline tasks. Based on the performance experiments conducted by the authors, the proposed message authentication scheme can handle 4,000 messages per second for a practical data size carried in messages, which is equivalent to the number of IEC 61850 SV messages that are expected in 50Hz power grid system [Esiner *et al.*, 2022].

## 4.3 Future Directions

In order to prevent attacks against the smart grid, it is essential to define comprehensive security policies to be enforced by firewalls, intrusion detection systems, and/or bump-in-the-wire security appliances. However, the challenge is to define a comprehensive but optimal set of policies (i.e., the minimal number of policies for the desired security guarantee). Furthermore, since such policies are to be enforced at various places in the infrastructure (e.g., bump-in-the-wire security appliances deployed throughout the infrastructure, substation gateway/firewall, and the control center), it is essential to ensure that policies are not contradicting each other or conflicting with normal control and protection functions. On the other hand, the behavior of smart grid systems is more consistent and predictable than general IT systems. Taking advantage of this fact, it is valuable to have a mechanism to automatically derive security policies based on the normal (or expected) system behaviors and address dependency and priority among them. For instance, network traces collected in the normal system state could be utilized for deriving communication models (protocols used in the system, nodes that are supposed to talk with each other, etc.). We can also take advantage of system configuration models, such as PLC logic codes and IEC 61850 SCL (System Configuration Description Language) files.

Moreover, it is expected that emerging technologies are to be incorporated into the smart grid infrastructure, such as cloud, 5G, and satellite communication, and more players to be involved in the operation, such as distributed energy resources (DERs) owners and aggregators, and virtual power plant (VPP) operators. Therefore, it is essential to design end-to-end security protocols and policy enforcement for cross-system, multi-hop communication across multiple management domains. It is also important to have a trustworthy, secure log or data repository that heterogeneous entities could use for maintaining and sharing data for policy enforcement. For this purpose, technologies like blockchain would be promising.

# 5

# Detection of Cyber Attacks against Smart Grid

Once the attack is launched, we should be able to detect the indication or occurrence of the attack as quickly as possible, ideally before it causes a significant impact. This chapter discusses intrusion detection systems (IDSs) as well as moving target defense that could strengthen the IDS by misleading attackers to result in less stealthy attacks. We then discuss containment and recovery measures to be taken once an attack is detected.

## 5.1 Intrusion detection

Intrusion detection is a cybersecurity solution that monitors system and network status to detect any indication of cyber attacks as well as system anomalies. IDSs have a long history in enterprise IT systems. An IDS can be categorized into two types, based on what it monitors and where it runs, namely host-based IDS and network-based IDS. Host-based IDS runs on a device (e.g., a workstation and server) to monitor the system's behavior. On the other hand, network-based IDS runs on the network to monitor network traffic. In the industrial control systems (ICS) domain, the latter is more popular since the former requires the installation of an agent module to collect and analyze data, which is often not suitable for ICS devices in smart grid systems, namely PLCs and IEDs. ICS devices are mostly resource-constrained embedded devices, and thus running additional processes on them would affect not only performance but also stability. On the other hand, network-based IDS is not intrusive to existing devices and infrastructure because it passively sniffs network traffic to detect indications of cyber attacks.

| Initial Access | Execution | Persistence | Privilege Escalation | Evasion | Discovery | Lateral Movement | Collection | Command and Control | Inhibit Response Function | Impair Process Control | Impact |
|---|---|---|---|---|---|---|---|---|---|---|---|
| Drive-by compromise | Change operating mode | Modify program | Exploitation for Privilege Escalation | Change operating mode | Network connection enumeration | Default credentials | Automated collection | Commonly used port | Activate firmware update mode | Brute force I/O | Damage to property |
| Exploit public facing application | Command line interface | Module firmware | Hooking | Exploitation for evasion | Network sniffing | Exploitation of remote services | Data from information repositories | Connection proxy | Alarm suppression | Modify parameter | Denial of control |
| Exploitation of remote services | Execution through API | Project file infection | | Indicator removal on host | Remote system discovery | Lateral tool transfer | Detect operating mode | Standard application layer protocol | Block command message | Module firmware | Denial of view |
| External remote services | Graphical user interface | System firmware | | Masquerading | Remote system information discovery | Program download | I/O image | | Block reporting message | Spoof reporting message | Loss of availability |
| Internet accessible device | Hooking | Valid accounts | | Rootkit | Wireless sniffing | Remote services | Man in the middle | | Block serial COM | Unauthorized command message | Loss of control |
| Remote services | Modify controller tasking | | | Spoof reporting message | | Valid accounts | Monitor process state | | Data destruction | | Loss of productivity and revenue |
| Replication through removable media | Native API | | | | | | Point & tag identification | | Denial of service | | Loss of production |
| Rogue master | Scripting | | | | | | Program upload | | Device restart/shutdown | | Loss of safety |
| Spear-phishing attachment | User execution | | | | | | Screen capture | | Manipulate I/O image | | Loss of view |
| Supply chain compromise | | | | | | | Wireless sniffing | | Modify alarm settings | | Manipulation of control |
| Transient cyber asset | | | | | | | | | Rootkit | | Manipulation of view |
| Wireless compromise | | | | | | | | | Service stop | | Theft of operational information |
| | | | | | | | | | System firmware | | |

**Figure 5.1:** Mitigation by (Network-based) Intrusion Detection Systems on MITRE ATT&CK Matrix for ICS.

In general, the network based intrusion detection systems can detect attacks based on the various characteristics of the network traffic (e.g., sender and receiver addresses, communication protocol used, timing and frequency of communication, size of packets, message payload, files, and so forth). Therefore, the coverage is broad. Although the coverage and effectiveness of IDSs are different for each IDS implementation, we here provide general discussion without focusing on a specific IDS. At Initial Access stage, the majority of the attack tactics involve communication over the network, thus being able to be detected by monitoring communication patterns (e.g., who is talking to whom), packet size, and so on. In Execution stage, interaction with ICS devices (Command Line Interface, Graphical User Interface) can be detected, if the interaction is initiated by unauthorized devices. Change Operating Mode often results in changing communication model, which may exhibit deviation from normal patterns. Regarding Persistence stage, transferring software or firmware to the ICS devices can also monitored, and thus can be flagged. In Evasion stage, by monitoring the payload of the messages and checking it against physical system status, Spoof Reporting Message can be countered by IDSs. Tactics under Discovery stage would involve active scanning of the system, and thus can be detected. Lateral Movement stage involves tactics for attackers or malware to propagate to other nodes. Such activities involves network communication, and thus they can be observed. Tactics under Collection stage often involves active probing, which would be flagged by IDSs. Man-in-the-Middle attack typically involves ARP spoofing to mislead and intercept the network traffic, and it can be detected relatively easily. Command and Control stage injects malicious interrogation/control commands to the target device. Again by monitoring the communication pattern and/or evaluating legitimacy of a control command based on the power grid status, we can counter the attacks. Regarding the tactics under Inhibit Response Function stage, tactics like Alarm Suppression, Block Command/Reporting Message could be mounted by means of Man-in-the-Middle attacks, and thus can be detected. Denial of Service (DoS) could be mounted by flooding large amount of traffic or exploiting the protocol specification. For instance, as discussed in [Biswas *et al.*, 2019], an attacker could mount DoS attack by injecting a fake GOOSE message with a manipulated status number or sequence number. Former can be detected by checking the amount of traffic against a certain threshold, and the latter can be detected by inspecting the network traffic in a stateful manner. Lastly, Impair Process Control stage, Brute Force I/O involves repetitive communication, which can be flagged by an IDS. Unauthorized Command Message can be detected by monitoring the communication peers and/or type of commands sent against a security policy. For the rest, the similar argument done in Persistence stage and Evasion stage holds.

The mapping to MITRE ATT&CK Matrix is summarized in Figure 5.1. For the rest of this section, we introduce representative network-based IDSs of different approaches.

### 5.1.1 Intrusion Detection Systems Using Cyber Side Information

In the enterprise IT domain, network-based IDS can be categorized into rule-based, signature-based, and machine-learning-based approaches. These are based on communication models and patterns, and many of them are applicable to smart grid systems.

**Rule-based IDS** utilizes user-defined attack detection rules. Rules can be defined by the system operator based on the normal system behavior and traffic pattern. For instance, if the number of packets in a unit of time exceeds a certain threshold, which is set based on the amount of traffic in a normal state, the IDS could raise an alarm. However, rules that can be defined on IDS are usually simple ones. In addition, the definition of rules requires extensive domain knowledge about the system, and thus attaining sufficient coverage is a major challenge.

**Signature-based IDS** utilizes a database of patterns in network traffic that are associated with known attacks. One representative example of the signature-based IDS is Snort, an open-source IDS software. While this approach can detect known attacks with good accuracy, the detection is limited only to attack vectors that are known already. In other words, Signature-based IDS is not effective against a zero-day attack that exploits unknown or very fresh vulnerabilities. Moreover, compared to the enterprise IT domain, available information in the industrial control systems domain is limited, and the communication protocols utilized are often domain-specific and heterogeneous.

**Specification-based IDS** utilizes some domain knowledge about the communication protocols utilized and the design of the system. The normal system behavior is modeled as a state machine (specification), and then the observed system behavior (e.g., network traffic among devices) is checked against the specification. If any deviation is detected, the system raises an alarm. Unlike enterprise IT systems, in an industrial control system, the behavior of the system is more static and regular, and thus such a specification-based approach is considered effective [Berthier and Sanders, 2011]. One of the specification-based approaches proposed for IEC 61850 GOOSE protocol is discussed in [Bohara *et al.*, 2020]. Another example for DNP3 protocol is found in [Lin *et al.*, 2013].

**Machine learning based IDS**, in general, utilizes a model of normal system/device behaviors, which are trained with historical data, to detect deviation from the normal pattern. This way, anomaly detection has the potential to detect even unknown, zero-day attack vectors. However, the machine-learning-based approach usually suffers from relatively high false positive and negative rates. Besides, the universal challenge is the lack of well-curated data for training the machine learning models. In particular, real attack examples are usually very scarce.

### 5.1.2 Physics-based Intrusion Detection Systems

While IDSs for general IT systems rely on network traces, communication patterns, and analysis of packet payloads, IDSs for cyber-physical systems, including smart grid, can also utilize information derived from physical plants. We next discuss mechanisms to counter cyber attacks using power system physics.

The first example is the bad data detection (BDD) algorithm based on power system state estimation [Handschin *et al.*, 1975]. This is not originally developed as a cybersecurity measure, but it has the capability to detect anomalies in power grid measurements, which could be caused by false data injection attacks. State estimation, as the name implies, is a methodology to estimate the state of the power grid, including the voltage and voltage angles of system buses based on the collected power grid measurements. In reality, even under the normal situation, power grid measurements are often incomplete or erroneous owing to the issue in the communication network or sensor failure. The state estimation accounts for the measurement errors by performing a regression analysis with all the available measurement data (including a certain level of redundancy).

The relationship between state variables and measurements is given by [Schweppe and Wildes, 1970; Liu *et al.*, 2011]:

$$z = h(x) + e \tag{5.1}$$

where $z$ is the measurement vector of $m$ measurements (including power flow, nodal power injection, voltage, etc.), $x$ is the state vector of $n$ states, $e$ is the measurement error vector, and $h(.)$ is a set of non-linear functions of the state vector. The direct current (DC) approximation is widely used in power grid security literature for analysis [Liu *et al.*, 2011]. Under this model, the non-linear mapping $h(x)$ can be approximated using a linear model as

$$z = Hx + e, \tag{5.2}$$

where $H$ is known as the measurement matrix (MM) that depends on the bus connectivity and the transmission line reactance. The state vector $x$ is estimated (as $\hat{x}$) using the weighted least square (WLS) method by solving the following equation:

$$\hat{x} = arg \min_{x} [z - Hx]^T W^{-1} [z - Hx] \tag{5.3}$$

where $W$ is the matrix of reciprocals of the measurement error variances. The solution to Equation (5.3) is determined through an iterative approach. To find bad data in measurements, traditionally, a chi-squared test is carried out on the performance index, which quantifies the accuracy of the estimate, defined as:

$$J(\hat{x}) = \sum_{i=1}^{m} \frac{[z_i - H_i\hat{x}]^2}{\sigma_i^2} \tag{5.4}$$

where $\sigma_i^2$ is the variance of $i$-th measurement $(1 \leq i \leq m)$. With the assumptions that state variables are mutually independent and the measurement errors follow a normal distribution, it can be proved that $J(\hat{x})$ follows $\chi^2_{(m-n)}$, chi-squared distribution with $(m - n)$ degrees of freedom. The threshold of identifying bad data is $\chi^2_{(m-n),p}$, which is the value of the chi-squared distribution for the probability $p$ and $(m - n)$ degree of freedom. If $J(\hat{x})$ exceeds this threshold, bad data is suspected in measurements.

While the bad data detection mechanism can detect erroneous measurements, it is not fully effective if the data is intelligently manipulated. For instance, Liu et al. [Liu *et al.*, 2011] discussed stealthy false data injection attacks that can bypass bad data detection. In particular, attacks of the form $a = Hc$, where $a$ is the false data injection (FDI) attack vector, and $c$ is an arbitrary vector of dimension $n$, then the attacks remain undetected by the BDD. In essence, any attack vector that lies in the column space of the measurement matrix $H$ will leave the value of the residual $J(\hat{x})$ unchanged (as compared to its value with the unattacked measurements), and hence cannot be detected by the BDD. Such undetectable attacks can also be crafted considering the AC power flow model [Hug and Giampapa, 2012]. Another issue in BDD is the latency for detection. Sourav et al. [Sourav *et al.*, 2023] argued that the BDD is not suitable for high-rate communication in an IEC 61850-based substation, and thus the authors proposed use of machine learning technologies to reduce the overall latency.

To overcome the vulnerability of the BDD, several recent works have proposed the use of machine learning (ML) techniques to detect the aforementioned attacks. The basic idea is to train an ML model to differentiate between normal and attacked measurement samples. Different ML approaches have been proposed in this context, such as support vector machines (SVM) [Ozay *et al.*, 2016], deep belief networks (DBNs) [He *et al.*, 2017], and deep neural networks [Yu *et al.*, 2018]. However, the aforementioned works propose a supervised learning approach that has drawbacks, such as the lack of labeled training data, unbalanced datasets, etc. To overcome this issue, several alternative approaches are proposed, such as the use of autoencoders [Wang *et al.*, 2020; Zeng *et al.*, 2022] and semi-supervised learning approaches [Zhang *et al.*, 2021]. In the context of *load-altering attacks* that target the end-user Internet-of-Things (IoT) enabled loads in the system, [Lakshminarayana *et al.*, 2022] proposed physics-informed machine learning approaches that explicitly use the knowledge of the underlying power system model to detect attacks, without having to rely on offline training.

The power system simulators, which calculate the power flow status given the power grid topology and load profile, can also be utilized to detect malicious control commands. In [Mashima *et al.*, 2017b; Mashima *et al.*, 2018], the authors proposed the use of on-the-fly power system simulation to evaluate the correctness and legitimacy of remote control commands. At the high level, the idea is to run the power system simulation with up-to-date power grid measurements and topology information to predict the consequence of command execution. If any violation of stability conditions is observed in the simulation result, the corresponding command is flagged anomalous, and then it is either blocked or reported to the operator.

According to [Mashima *et al.*, 2017b], there is a substation gateway device that is responsible for protocol translation (between IEC 61850, which is used in the substation, and IEC 60870-5-104, which is used between the control center and substation), and thus it can offer reliable mediation of all incoming remote control commands. A command validation (or also called command authentication) system can be deployed on such a substation gateway. A gateway can also observe measurement reports sent from field systems, which can be used for updating the simulation model. The power system simulator for the validation can be deployed either on the substation gateway device or on a remote server. For instance, light-weight, steady-state simulation for a single substation, can be deployed on a commodity industrial PC or even on embedded device. On the other hand, when computationally intensive, power system dynamics simulation is used, a simulator can be outsourced to other resourceful node, which is consulted over network by the validation module deployed on on the substation gateway, as done in [Mashima *et al.*, 2018]. One concrete design and implementation of such a simulation-based command validation system integrated into a substation gateway is proposed as the active command mediation defense system (A*CMD) [Mashima *et al.*, 2017b] shown in Figure 5.2. Although promising attack detection accuracy is demonstrated, command validation solutions using on-the-fly power flow simulation must incur latency for the evaluation of incoming remote control commands. Therefore, command messages are to be held by the command validation module on the substation gateway till the simulation and validation complete. [Mashima *et al.*, 2017b] also discusses mechanisms to decide the amount of the artificial delay to wait for the simulation results without affecting legitimate operations.

Command validation systems can work as an additional life of defense against the situation seen in the Ukraine power plant attack in 2015, where a legitimate control center system was manipulated to send out a large number of malicious control commands. In such a situation, it is not feasible to detect the attack by using the firewall or network-based intrusion detection systems based on cyber-side information because the source of the control command is the legitimate control center server or SCADA HMI and communication model would appear completely normal. Command validation of control commands based on up-to-date power grid status effectively complements these measures.

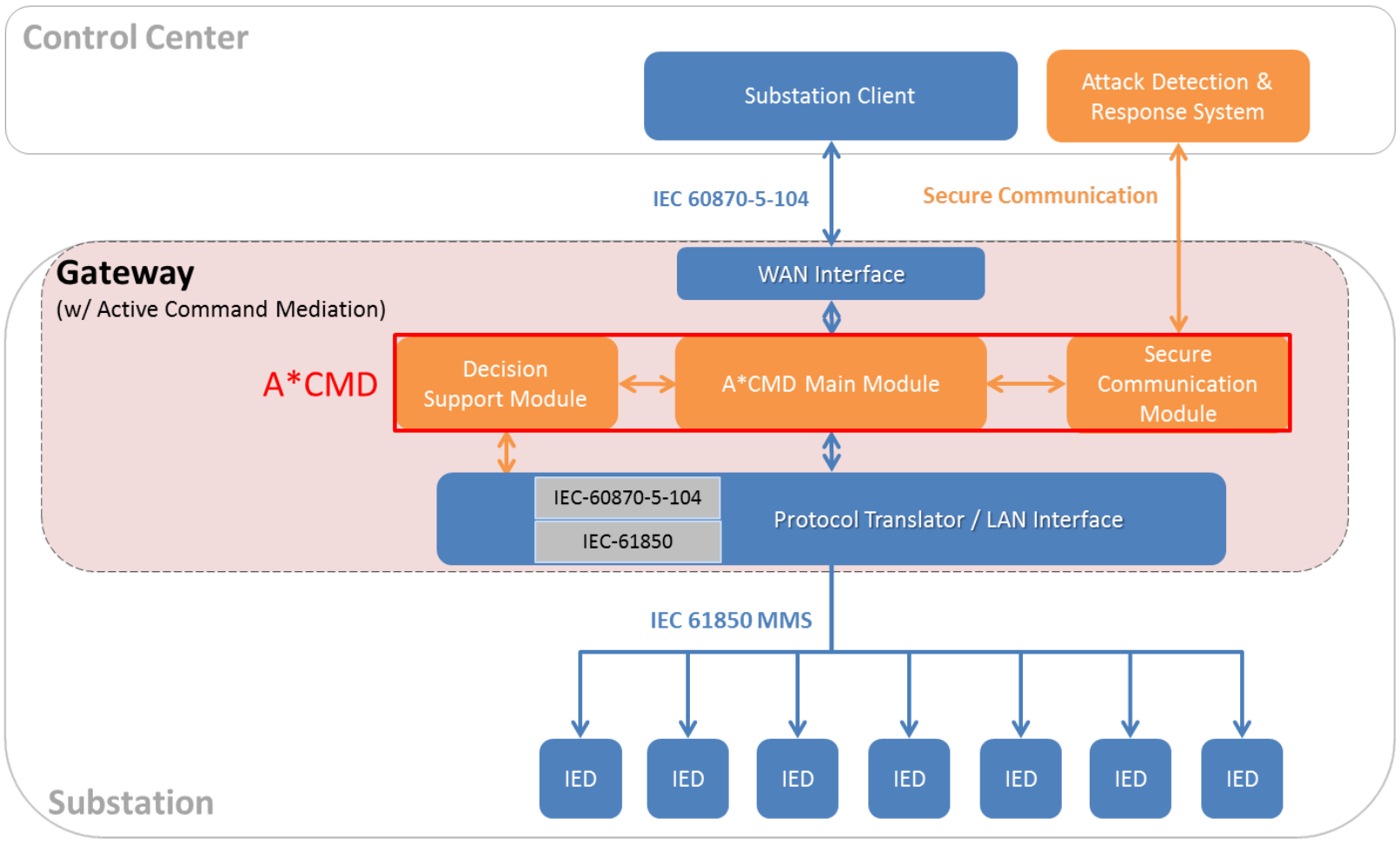

**Figure 5.2:** Deployment Model of Command Authentication System [Mashima *et al.*, 2017b]

### 5.1.3 Ensemble, Multi-layer Intrusion Detection Systems

In [Kang and McLaughlin, 2014; Ren *et al.*, 2018], multi-layer intrusion detection systems are proposed. Their systems utilize an ensemble of different detection logic and attributes to detect cyber attacks against the smart grid.

For instance, the intrusion detection system proposed by the SPARKS project in Europe [Kang and McLaughlin, 2014] combines three different intrusion detection mechanisms: whitelisting, stateful analysis, and anomaly detection using machine learning.

**Whitelisting**: Whitelisting is positioned as the first line of defense and aims at enforcing network traffic control policies. The whitelist is defined based on network-level characteristics, such as MAC address, IP address, protocol used, and port number. The whitelist can be generated either manually or systematically based on the configuration of devices in substations. For example, in IEC 61850-compliant substations, information about each device as well as their connectivity, is defined in standardized substation configuration language files, and therefore, by parsing those files, the list of authorized nodes and communication patterns can be automatically generated.

**Stateful Analysis**: The second layer of detection is based on protocol-specific information. For instance, IEC 61850 MMS has the request-response communication model, and each pair of requests and responses is coupled by "InvokeID." If such a pattern or rule is violated, an alarm can be raised. This layer also involves behavior inspection, which investigates the payload of each packet (e.g., power grid measurements such as voltage) and enforces rules in terms of such measurements to detect misbehavior of monitored power grid devices.

**Anomaly Detection**: The third layer of the proposed scheme employs anomaly/outlier detection based on the aforementioned network-level and application-level (i.e., protocol-specific) characteristics using machine-learning technologies, such as one-class SVM. The advantage of this approach is that the system can detect novel intrusions or anomalies that are not yet coded into rules used in the previous layers.

Another example, EDMAND [Ren *et al.*, 2018], implements the 3-level anomaly detector, consisting of transport level, operation level, and content level. The output from all anomaly detection modules is aggregated into a meta-alert. More details of the detection at these three levels are provided below.

**Transport Level**: The transport-level anomaly detection relies on statistics of features, such as packet size, inter-arrival time, and packet count, calculated based on network traces of each time window. The anomaly score is calculated based on the deviation of the statistics from the normal profile.

**Operation Level**: The operation-level anomaly detection utilizes the payload of the ICS protocol payloads to detect anomalous operations. The detection module utilizes features including the origin and destination, the protocol used, and the function code, which indicates the type of operation. For instance, invalid function code, wrong direction of the operation, unseen operation observed, and deviation of periodicity in the normal state.

**Content Level**: The content-level detection module utilizes the power grid measurement values (e.g., binary status, voltage, frequency) carried in the packets. For instance, for the binary status, the anomaly score is computed based on the entropy of observed samples. For analog measurements, the anomaly score is calculated based on the mean and standard deviation of historical data.

## 5.2 Moving Target Defense

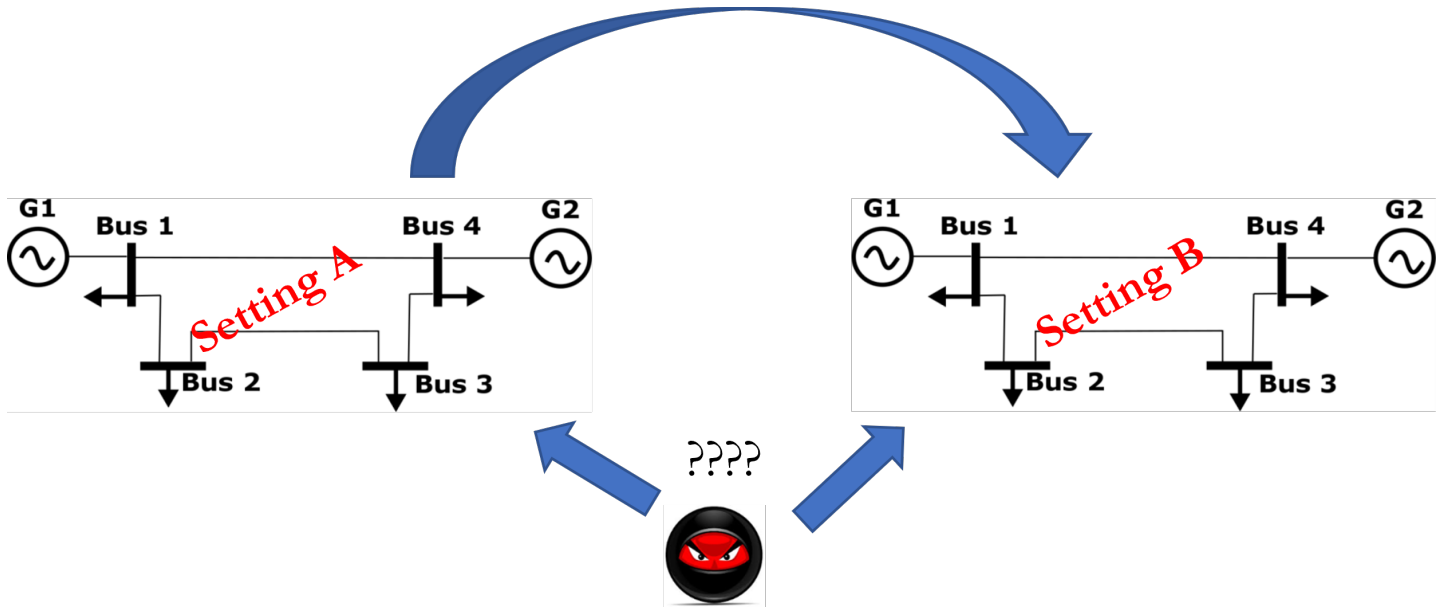

**Figure 5.3:** Pictorial Depiction of MTD in Power Grids. The Settings "A" and "B" represent different configurations of the power grid. The MTD "moves" the system between several such configurations to invalidate the attacker's knowledge.

Moving target defense (MTD) is an emerging defense technique in power grid security. It aims to solve the long-standing problem of securing the power grid against stealthy attacks, such as the undetectable FDI attacks discussed in Section 5.1. A major drawback of static attack detection approaches is that an attacker can learn their operational details by continuous reconnaissance. For instance, in the context of FDI attacks, recent research has shown that an attacker can learn the information required to bypass the BDD by monitoring power grid measurements over a period of time [Lakshminarayana *et al.*, 2021b]. MTD overcomes this drawback by introducing periodic/event-triggered controlled changes to the power grid's SCADA network/physical plant, thereby invalidating the knowledge attackers use for crafting stealthy attacks. A pictorial depiction of MTD in power grids is shown in Fig. 5.3. In many cases, the MTD reconfiguration can be performed by existing devices that are already deployed in the grid (see Section 5.2.1 for more details) and does not require major device upgrades (such as encryption-enabled PLCs or remote hardware/software attestation by a trusted verifier).

| Initial Access | Execution | Persistence | Privilege Escalation | Evasion | Discovery | Lateral Movement | Collection | Command and Control | Inhibit Response Function | Impair Process Control | Impact |
|---|---|---|---|---|---|---|---|---|---|---|---|
| Drive-by compromise | Change operating mode | Modify program | Exploitation for Privilege Escalation | Change operating mode | Network connection enumeration | Default credentials | Automated collection | Commonly used port | Activate firmware update mode | Brute force I/O | Damage to property |
| Exploit public facing application | Command line interface | Module firmware | Hooking | Exploitation for evasion | Network sniffing | Exploitation of remote services | Data from information repositories | Connection proxy | Alarm suppression | Modify parameter | Denial of control |
| Exploitation of remote services | Execution through API | Project file infection | | Indicator removal on host | Remote system discovery | Lateral tool transfer | Detect operating mode | Standard application layer protocol | Block command message | Module firmware | Denial of view |
| External remote services | Graphical user interface | System firmware | | Masquerading | Remote system information discovery | Program download | I/O image | | Block reporting message | Spoof reporting message | Loss of availability |
| Internet accessible device | Hooking | Valid accounts | | Rootkit | Wireless sniffing | Remote services | Man in the middle | | Block serial COM | Unauthorized command message | Loss of control |
| Remote services | Modify controller tasking | | | Spoof reporting message | | Valid accounts | Monitor process state | | Data destruction | | Loss of productivity and revenue |
| Replication through removable media | Native API | | | | | | Point & tag identification | | Denial of service | | Loss of production |
| Rogue master | Scripting | | | | | | Program upload | | Device restart/shutdown | | Loss of safety |
| Spear-phishing attachment | User execution | | | | | | Screen capture | | Manipulate I/O image | | Loss of view |
| Supply chain compromise | | | | | | | Wireless sniffing | | Modify alarm settings | | Manipulation of control |
| Transient cyber asset | | | | | | | | | Rootkit | | Manipulation of view |
| Wireless compromise | | | | | | | | | Service stop | | Theft of operational information |
| | | | | | | | | | System firmware | | |

**Figure 5.4:** Mitigation by Moving Target Defense on MITRE ATT&CK Matrix for ICS.

MTD can be viewed as a mechanism to strengthen both network-based and physics-based IDS by dynamically configuring the power grid's *cyber* and *physical* system components, respectively. Figure 5.4 shows the mitigation provided by MTD in terms of MITRE ATT&CK Matrix for ICS. MTD counters (i) Network Connection Enumeration, Remote System/Information in the "Discovery" phase, (ii) Change Operating Mode, Indicator Removal on Host during the "Evasion" phase, (iii) Monitor Process State during the "Collection" phase, (iv) Modify Parameter, Unauthorized Command Message during the "Impair Process Control" phase. Specifically, MTD invalidates the acquired information of the system by reconnaissance operations due to dynamic system reconfiguration, thereby impeding the collection and discovery phase. Moreover, IDS reinforced with MTD can also detect unauthorized control commands or modifications to control parameters as they require accurate knowledge of the system to evade detection.

The majority of works in power grid security focus on MTD to strengthen physics-based IDS. MTD for strengthening network-based IDS has primarily focused on techniques such as IP-hopping or dynamically changing the communication path of the SCADA traffic in a software-defined SCADA network [Pappa *et al.*, 2017; Abdelkhalek *et al.*, 2022]. However, the literature only considers simple attacks such as single-point denial-of-service (DoS) attacks and lacks an in-depth analysis. In contrast, MTD to strengthen physics-based IDS has been studied extensively. In particular, three fundamental questions are relevant in this context – (a) what to move? (b) when to move? (c) how to move? We elaborate on them in the following subsections.

### 5.2.1 What to Move?

MTD based on reconfiguring the *physical* components of a power system has been primarily used to detect stealthy FDI attacks that aim to disrupt power grid state estimation. As noted in Section 5.1.2, an undetectable attack must lie in the column space of the power grid's measurement matrix, which in turn depends on the power grid's topology and the reactance settings of its transmission lines. Thus, effective MTD must invalidate this knowledge.

In particular, distributed flexible alternating current (AC) transmission system (D-FACTS) devices, which can dynamically vary the power grid transmission line reactance, have been proposed as an effective method to reconfigure the system and invalidate the attacker's knowledge [Divan and Johal, 2005]. D-FACTS devices are easy to deploy, as they can be attached directly to transmission lines and can be used to dynamically control effective line impedance. Moreover, they are equipped with communication capabilities and encrypted commands to change the line reactance from the system operator's control room can be securely transmitted using DNP3, IEC-61850, and 60870-5-104 protocols. It is noteworthy that D-FACTS devices are pre-existing devices in a power grid whose primary purpose is to control power flows in the network to relieve congestion in power grid transmission lines [Rogers and Overbye, 2008]. Thus, the defense methodology does not incur significant deployment costs.

### 5.2.2 How to Move?

The MTD design involves selecting the subset of the grid's transmission lines for reactance perturbation and the corresponding perturbation mag-nitude. Naturally, the MTD perturbation must be selected to effectively invalidate the knowledge that attackers need to craft their undetectable attacks. However, grid operators face a fundamental dilemma since MTD perturbations also lead to a change in power grid operational settings (e.g., alter power flows in the system). Thus, MTD must be carefully designed to balance these two competing criteria. We elaborate on both these aspects in the following.

### Maximizing the Effectiveness of Attack Detection

The initial work designed MTD by introducing *randomly selected* perturbations to the system (that are unknown to the attacker) to invalidate the attacker's knowledge while keeping the perturbation magnitude *low enough* [Morrow *et al.*, 2012; Davis *et al.*, 2012]. However, it was shown that the randomly-selected perturbations cannot offer detection guarantees [Lakshminarayana and Yau, 2018].

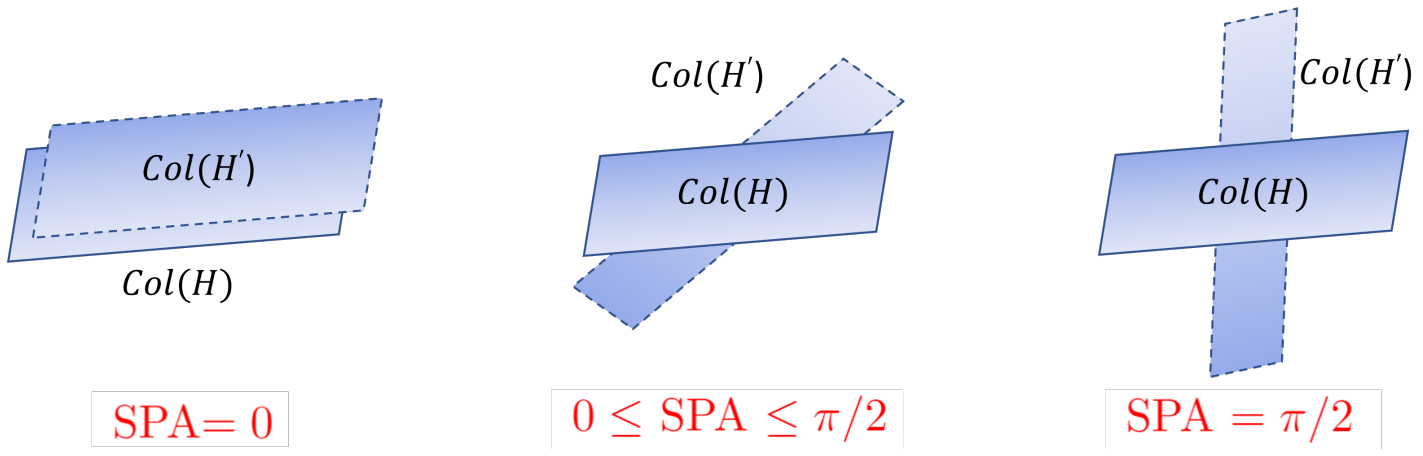

**Figure 5.5:** Oritentation of the Column Spaces of the Pre-perturbation ($H$) and Post-perturbation ($H'$) MMs [Lakshminarayana and Yau, 2018]. SPA of $\pi/2$ Represents the Orthogonality Condition between the Two Column Spaces.

Reference [Lakshminarayana and Yau, 2018] showed that MTD's effectiveness is related to the separation between the column spaces of the measurement matrices before and after the MTD. In particular, if the two-column spaces are orthogonal, then all prior stealthy FDI attacks (i.e., those in the column space of the pre-perturbation MM) will become detectable by the BDD after the perturbation. A similar condition was characterized in [Liu *et al.*, 2018b] based on the rank of the composite matrix (obtained by concatenating the columns of the pre and post-perturbation measurement matrices). Such an MTD strategy is referred to as *complete MTD*. However, [Liu *et al.*, 2018b] also showed that in practice, the sparse topology of power grids (i.e., few transmission lines connecting the buses) can be a major limitation in achieving the rank/orthogonality condition. Thus, in practice, complete MTD is very hard to achieve. Nevertheless, these works also proposed two differing design criteria to maximize MTD's effectiveness – choose reactance perturbation vectors that maximize (i) the smallest principal angle (SPA) between the column spaces of the pre- and post-perturbation MMs and (ii) maximize the rank of the aforementioned composite matrix [Liu *et al.*, 2018b] A pictorial depiction of the MTD design using the SPA criteria is shown in Fig. 5.5. A recent work [Xu *et al.*, 2023] analyzed the two metrics and showed that while MTD designed based on the rank criteria is optimal in the noiseless scenario, they cannot provide performance guarantees in a practical setting with sensor measurement noises. In contrast, MTD designed based on the SPA metric can provide robust performance in noisy scenarios.

Other considerations in MTD design include (i) placement of D-FACTS devices to maximize the effectiveness [Liu and Wu, 2020], (ii) designing MTD such that its activation is hidden from the attacker [Tian *et al.*, 2019], and (iii) design of MTD in distribution networks [Liu *et al.*, 2018a; Jhala *et al.*, 2021]. We omit their details here for brevity.

### Minimizing MTD Implementation Cost

As noted before, MTD perturbations incur a non-zero implementation cost. This can be intuitively understood as follows. Without MTD, the line reactances are adjusted to minimize the system's power losses and/or optimal power flow (OPF) cost [Rogers and Overbye, 2008]. Then, any non-zero perturbation (for MTD) will move the system away from optimality, thereby increasing the OPF cost. In particular, [Lakshminarayana and Yau, 2021] showed that the OPF cost would increase as the SPA between the pre- and post-perturbation MMs increases. Thus, increasing the SPA between the column spaces will result in more effective attack detection capabilities while incurring a higher cost, resulting in a trade-off existing between the two quantities. An illustration of the trade-off implemented using the IEEE-14 bus system is shown in Fig. 5.6. Such cost-benefit curves can guide a power

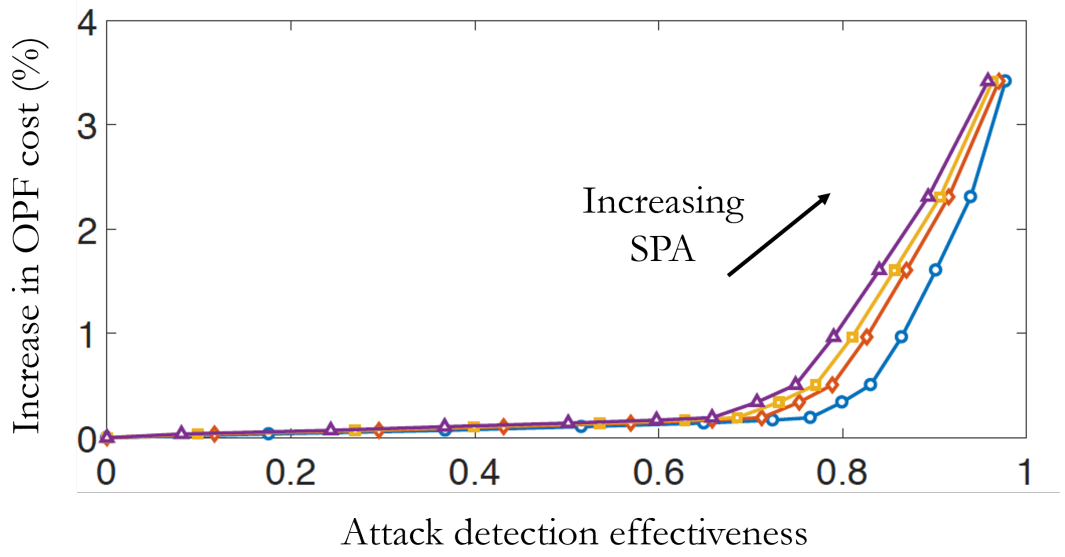

**Figure 5.6:** Trade-off between MTD's Effectiveness and the Implementation Cost [Lakshminarayana and Yau, 2021].

system operator in choosing an effective MTD based on the perceived risk and the security budget.

A pertinent question is how to reduce MTD's operational cost. [Lakshminarayana *et al.*, 2021a] proposed the design of strategic MTD that only protects the critical assets of the system (e.g., sensors and/or transmission lines) that are likely to be targeted by an attacker. In this context, game theory is used as an effective tool to anticipate the attacker's strategy. Then, designing MTD that only partially protects these important assets is shown to reduce the MTD's implementation cost significantly.

### 5.2.3 When to Move?

MTD can either be implemented in a proactive mode or a reactive mode. In proactive mode, MTD is implemented periodically, irrespective of whether an attack occurs or not. In reactive mode, MTD is activated only when suspicious activity is detected by other detectors deployed in the system. While the periodic mode may incur a high cost due to frequent MTD operation, the reactive mode incurs significantly more overheads (implementation of attack detectors, dealing with false positives, etc) and additional delays in detecting the attack. We provide the details below.

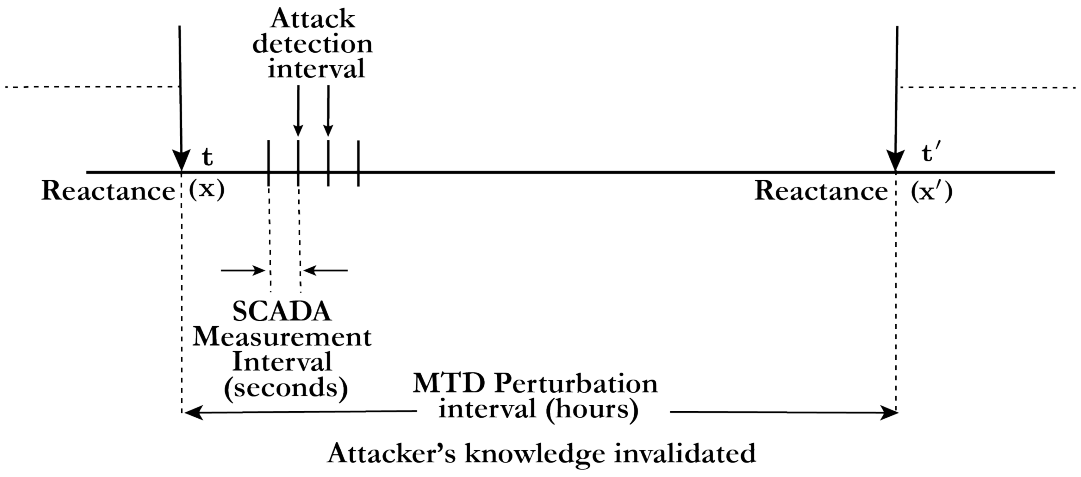

**Figure 5.7:** A Timeline of the Proposed MTD Scheme [Lakshminarayana *et al.*, 2021a].

## Periodic MTD

The main design question in the design of periodic MTD is the perturbation frequency. The question is inherently related to the attacker's learning capabilities. If the system settings are changed before the attacker gathers sufficient information to conduct a successful attack, then MTD can be effective.

Note that to craft FDI attacks that are undetectable to the BDD, the attacker must learn the system's operational details by monitoring the system. In the context of FDI attacks, the attacker can craft attack vectors that lie in the column space of the MM by monitoring the system's measurements (power flow, nodal power injections, etc.) over a period of time [Lakshminarayana *et al.*, 2021b]. The experimental evidence conducted using benchmark IEEE bus systems shows that to construct attacks that bypass the BDD with a high probability, the attacker must monitor the power grid measurements for several hours, which suggests that hourly perturbations are sufficient for practical systems.

Figure 5.7 shows a timeline of the practical implementation of the MTD (hourly perturbations) along with the SCADA measurement frequency (4-6 seconds). Based on the discussion above, the attacker has an outdated knowledge of the system throughout the MTD perturbation interval (since the MTD frequency is faster than the attacker's learning rate). Irrespective of when the attack occurs within the perturbation interval (with outdated knowledge), the attack will be detected once the next set of field measurements reaches the control center (as dictated by the SCADA measurement frequency). Thus, the proposed MTD can detect attacks quickly.

### Event-Triggered MTD

Event-triggered mode implements MTD only after a suspicious activity is detected. [Xu *et al.*, 2022] proposed an event-triggered MTD in which the system operator first detects the attack using a data-driven detector, e.g., a long short-term memory auto-encoder (LSTM-AE). If a positive alarm is raised, MTD will be triggered to verify this result. The event-triggered mechanism is shown to eliminate the false positives raised by data-driven detectors effectively.

## 5.3 Containment and Recovery Measures

Once any attack is detected, from the perspective of cybersecurity, the typical strategy for the containment is to isolate the compromised device or to segment the affected network from the rest to prevent the attack from impacting the system broadly. While such a strategy could be done in some cases (e.g., disconnecting an engineering workstation from the control center or substation network), it is not always the case. For instance, when IEDs and PLCs are compromised, simply disconnecting them would result in the inability to further monitor or control the physical power grid systems.

Thus, in the case of smart grid system, it is necessary to physically send technicians to the site and then to reconfigure/replace the compromised device as quickly as possible. While the system sustains, it is crucial to neutralize the attack and recover the system to its normal state. In order to conduct timely, and effective recovery, it is imperative to conduct contingency analysis and recovery planning on regular basis. It is also necessary to conduct the training of technicians. One of such efforts that practice black start recovery after cyber incidents, using the real power grid equipment, was conducted in Rapid Attack Detection, Isolation and Characterization Systems (RADICS) program led by DARPA (Defense Advanced Research Projects Agency) [Weiss, n.d.]. Use of digital twins or simulation environments would be also effective. While it is outside of the scope of this book, resilience measures and evaluation framework for the critical infrastructures are also formulated in [Cassottana *et al.*, 2023].

Recovery and restoration measures are not fully effective or even possible when the communication channels and integrity of smart grid devices are compromised. For instance, in Ukrainian power plant attack in 2015 [Defense Use Case, 2016], KillDisk malware was utilized at the end of the cyber attack to wipe out or overwrite files of devices in the control system to disable the recovery operations. Such attack tactics are listed in Inhibit Response Function stage and Impair Process Control stages of MITRE ATT&CK Matrix for ICS. Thus it is imperative to deploy cybersecurity measures for these. Intrusion detection systems are deemed effective here but not fully. It is necessary to implement additional defense mechanism at a device level. Since the recent industrial control devices employ ARM-based processors, which offers security feature called ARM TrustZone Trusted Execution Environment (TEE) [Pinto and Santos, 2019]. TEE, at the high-level, offers secure environment for execution of critical computations and storage of sensitive data, which is expected to counter malicious modification of software or applications running on devices and also counter rootkit. Use of ARM TrustZone for securing industrial control system devices without Sacrificing real-time performance has been recently explored [Wang *et al.*, 2022].

In the rest of this subsection, we introduce other approaches to contain the impact of the attack. Namely, even when the attack is successful, the system should sustain itself without causing severe damage. There are multiple directions to consider.

### Command Reversing

Lin et al. [Lin *et al.*, 2016] considers attacks exploiting remote control interface of substations. Their scheme relies on centralized semantic analysis of control commands, based on power flow simulation, for detecting attack attempts injecting malicious control commands. To mitigate impact of attacks, they use command-reversing, which sends reversing control commands shortly after the execution of malicious commands. However, it may not be always ideal because some of the grid control would take non-negligible amount of time to reverse.

### Preemptive Command Authentication

Related to command reversing approaches, we can implement an intelligent module to detect suspicious control commands at or near the IEDs and PLCs [Meliopoulos *et al.*, 2016; Mashima *et al.*, 2018]. Such command authentication solutions can work preemptively (i.e., before the attack impacts the system), unlike command reversing approaches that works in a reactive manner, if they are deployed in front of the target smart grid devices in a bump-in-the-wire manner. While effective, such a solution introduces delays for the smart grid communication to conduct an evaluation of the legitimacy of commands by means of power system simulation, etc. Thus the design of which requires careful consideration to evaluate the impact on legitimate operations [Mashima *et al.*, 2017b].

### Cyber-Resilient Economic Dispatch Framework

An approach to contain the attack impact and increase the system's uptime was introduced in [Chu *et al.*, 2023]. The authors considered the so-called *load-altering attacks* (LAAs) in which an attacker who has compromised a large number of demand-side appliances causes a large-scale surge/drop in the system load. LAAs can disrupt the balance between supply and demand, potentially resulting in unsafe frequency excursions.

Such instances of LAAs can be mitigated if sufficient redundancy is built into the system. For instance, additional generators can be scheduled to cover such events and dispatch them if a cyber attack occurs (i.e., augment unit commitment to cover these cases). However, such an approach could be extremely expensive, since cyber attacks are rare events (note that under this approach, the redundancy must be built in irrespective of whether a cyber attack occurs or not).

To overcome these drawbacks, [Chu *et al.*, 2023] proposed the Cyber-Resilient Economic Dispatch Framework (CRED) framework as shown in Fig. 5.8, in which mitigation actions are triggered only when an attack is detected, e.g., using the anomaly detection approaches discussed in this chapter. The mitigation action is in the form of additional generation dispatch from inverter-based resources (IBRs), such as offshore wind. Note that due to the fast dispatch capabilities, the IBRs need not be scheduled apriori, but can be brought online at short notice. A key challenge under the CRED framework lies in dealing with the inevitable uncertainty associated with attack detection/localization results, namely the false positives, and misdetections (recall that attack detection approaches such as machine learning framework always yield probabilistic results). The CRED framework deals with this challenge by adopting a distributionally robust optimization (DRO) approach. Under this approach, the key idea is to model the estimates of the attack parameters (obtained from the attack detection/localization results) as random variables with unknown distribution but whose moments are known partially. Note that due to the complexity of the attack detection algorithms, such as machine learning-based approaches, estimating the distribution of the attack detection results becomes intractable, but however, their moments can be reasonably accurately (e.g., numerically). Then, the DRO approach minimizes the dispatch cost under the worst-case distribution of the attack detection parameters within the ambiguity set (i.e., potential distributions that follow the estimated moments). Simulation results show that such an approach can quickly correct the imbalance and stabilize the system while incurring minimal cost.

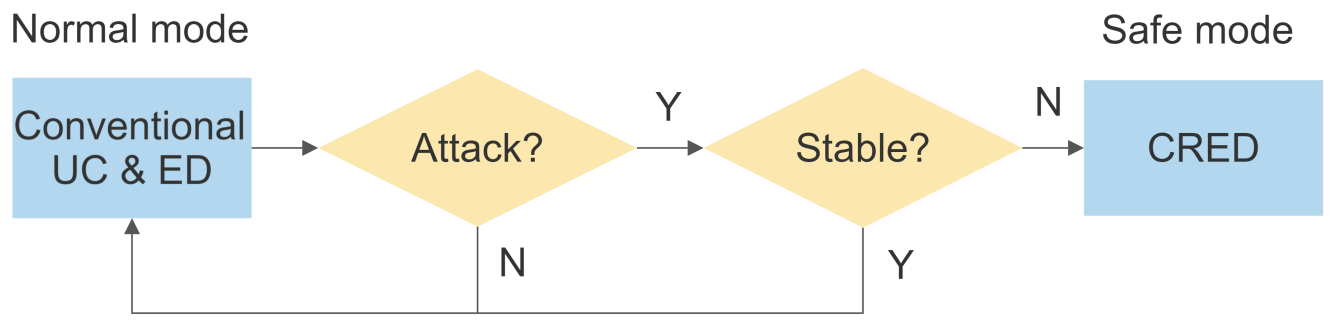

**Figure 5.8:** Cyber-Resilient Economic Dispatch (CRED) Framework Proposed in [Chu *et al.*, 2023]. Abbreviations: UC - Unit commitment, ED - Economic dispatch.

## 5.4 Future Directions

We conclude this chapter by highlighting some open issues that remain in MTD-based IDS and then broaden the discussion to cover future work for power grid IDSs in general. (i) While several works in MTD literature have focused on detecting cyber-physical attacks, localizing attacks using the MTD approach is an open issue. [Chen *et al.*, 2022] presents some initial ideas on this topic by combining MTD with deep learning, yet several open issues remain which we elaborate in the following. (ii) All of the works discussed above focus on MTD for strengthening the BDD, which is a model-based detector. As discussed in Section 5.1.2, several ML-based detectors are being proposed in the literature that have more advanced detection capabilities. Yet, it has been shown recently that such ML-based detectors are also vulnerable to adversarial examples [Sayghe *et al.*, 2020]. How to design MTD to strengthen ML-based detectors is another open issue. (iii) Finally, game-theoretic-based MTD design only considers rational adversaries (i.e., those that follow the policies prescribed by the Nash equilibrium solution), which is a strong assumption. Moreover, real-world attackers operate under several constraints, such as imperfect knowledge of the target system, etc. Thus, designing MTD against such non-rational real-world attackers will be important.

We conclude this chapter by discussing a few open issues in power grid IDSs. (i) Modern-day power grids are witnessing a growing penetration of renewable energy resources (RES) and inverter-based resources (IBRs), all of which have led to significant increases in system oscillations. In this context, differentiating cyber attacks from natural power system fluctuations is a challenging problem. Addressing this problem would require a joint investigation of the cyber and physical signals, which remains an open problem. (ii) Second, while there are several works focussing on *detecting* cyber or physical attacks (i.e., the presence or absence of attack), *localizing* attacks (i.e., identifying the compromised assets) has received limited attention. The problem is challenging since it often involves a combinatorial search across different components of the system that may be compromised by the attacker. (iii) Finally, the growing integration of IoT-enabled devices (e.g., smart electric vehicle charging stations, WiFi-enabled heat pumps) has created a new attack surface for adversaries to target the power grid from end-user customer sites. These devices typically do not have the sophisticated security measures that can be embedded in a SCADA system. In this context, designing IDSs to detect and localize attacks becomes more challenging, since operators typically have poor visibility of the distribution network/end-user sites (unlike transmission networks), due to limited sensing devices.

# 6

# Environment for Cybersecurity Evaluation

## 6.1 Smart Grid Testbeds for Security and Resilience Evaluation

When designing and developing cybersecurity solutions for smart grid systems, universal challenges are: assessment of cyberattack impacts; quantitative evaluation/comparison of cybersecurity solutions; collection of normal and attack data. While it would be ideal to use the real smart grid infrastructure for conducting empirical evaluation, it would never be an option for fear that such experiments would cause a negative and potentially severe impact on the stability and availability of power grid operations. Thus, the research community has been devoting efforts to creating a dedicated testbed, which are isolated from the production environment. Such testbeds are categorized into 3 types: hardware-based testbeds, virtual, software-based testbeds (also called digital twin or cyber range), and hybrid testbeds, which is positioned in between. In this section, we pick some representative implementations from all categories and discuss the pros and cons.

## 6.2 Hardware-based testbed

As the venue for evaluation, the best possible option is to create an isolated, sandboxed environment that uses exactly the same hardware on both the cyber and physical side of the smart grid system. A notable benefit of this approach is fidelity. Using the same PLCs and IEDs (and possibly other networking devices like industrial switches) we can assess and experiment with any possible vulnerability that could be exploited by real-world attackers (e.g., known vulnerabilities of the device models utilized in the system).

### 6.2.1 Electric Power and Intelligent Control (EPIC)

Electric Power and Intelligent Control (EPIC) testbed established and operated by iTrust at Singapore University of Technology and Design is a full-fledged smart grid testbed that is intended for cybersecurity experiments and well depicts a modern-day smart grid. The physical layout of the testbed includes four sectors: generation, transmission, microgrid, and smart homes. Each sector is monitored and controlled by multiple Siemens IEDs and WAGO PLCs that support IEC 61850 standard protocols (namely MMS and GOOSE) as well as Modbus communication protocol.

The generation sector contains Variable Speed Drives (VSDs) and real generators whose rotation speed is governed by VSDs. There are two generators, and their parallel operation (e.g., synchronization, load balancing, and reverse power prevention) is managed by a PLC. The transmission segment is implemented with a transformer. The microgrid sector comprises of photovoltaic (PV) panels and batteries that are controlled by SMA Solar Cluster Controller. The smart home sector includes programmable loads so that various load profiles can be simulated. The loads are monitored by advanced metering infrastructure (AMI), which reports energy consumption to the SCADA using Modbus protocol. Finally, SCADA HMI (human machine interface) and historian database are deployed in the control segment. SCADA HMI collects data via the PLC. The cyber-side network is implemented with a ring topology using fiber-optic cables for redundancy. There are also Wifi access points, and communication among the sectors can be performed over wireless network. The single line diagram of the EPIC testbed is depicted in Figure 6.1 and the network diagram of the EPIC is exemplified in Figure 6.2. The testbed can be utilized as an education and training platform for researchers and also for organizations to train operational technology (OT) personnel.

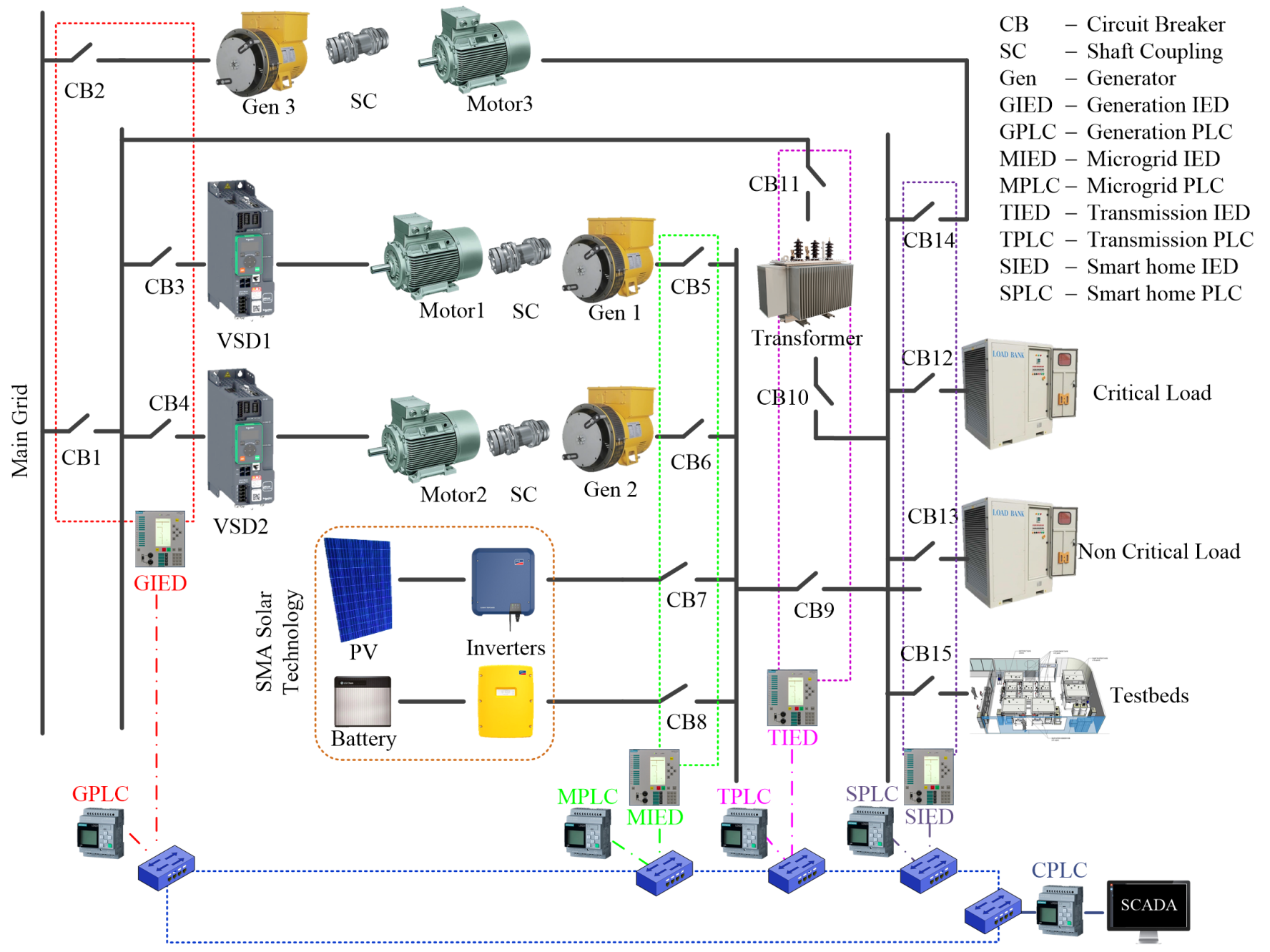

**Figure 6.1:** Layout of EPIC Testbed [Roomi *et al.*, 2023a].

## 6.3 Digital Twin and Cyber Range

### 6.3.1 Simulator

Perhaps the first option people would consider is to use off-the-shelf power system simulators. A vast majority of the research efforts to date rely on power flow simulators. There are wide range of simulators available, from open-source ones such as Pandapower [*Pandapower*

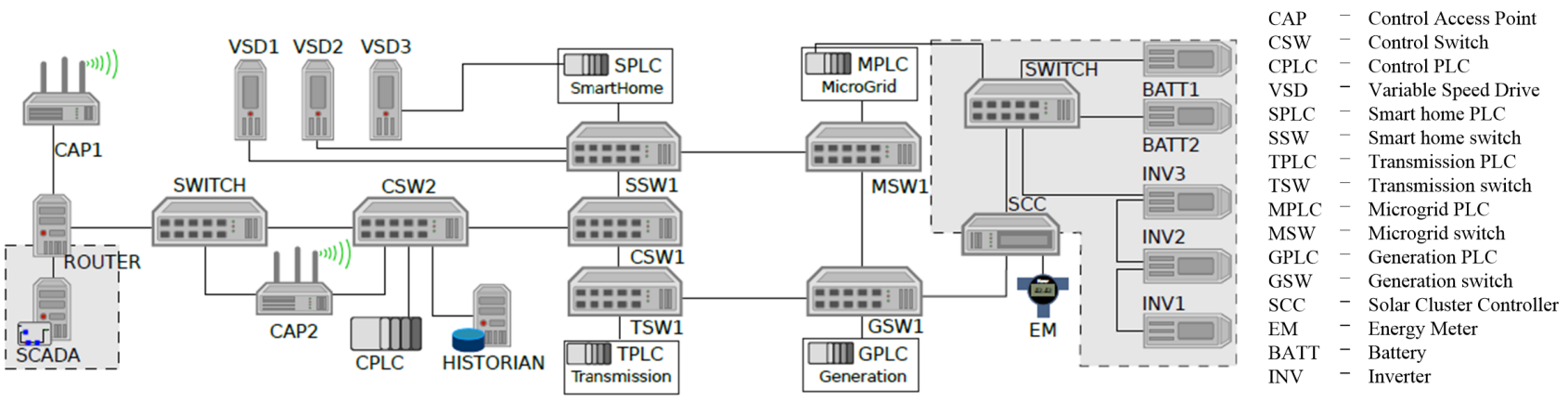

**Figure 6.2:** Overview of EPIC Network [Siddiqi *et al.*, 2018].

n.d.] to commercial products such as PowerWorld [*PowerWorld* n.d.], Matlab [*MathWorks* n.d.], and high-end, real-time simulators including RTDS [*RTDS Technologies* n.d.] and OPAL-RT [*OPAL-RT Technologies* n.d.]. The functionality and fidelity provided by these simulators are highly diverse. For instance, Pandapower and MATPOWER in Matlab only offer steady-state power flow simulation, which is a one-time solver that provides a snapshot of the power grid status. On the other hand, PowerWorld and Matlab Simulink offer a simulation of power system dynamics too. RTDS and OPAL-RT can run dynamics simulations in a continuous and real-time manner.

In order to conduct cyber attack experiments using these simulators, we need to "script" attack scenarios in advance (e.g., a certain circuit breaker is maliciously opened at a certain time). This way, we can evaluate the consequence of cyber attacks. However, they do not allow interactive experiments connected with the cyber side of the smart grid systems. Some commercial simulators offer interfaces for external processes or devices to interact with simulators. However, they are not immediately ready for cyber attack experiments because they are just communication endpoints for the limited protocols (e.g., proprietary TCP or UDP-based protocols or Modbus TCP), and no detailed smart grid device functionality (e.g., protection logic) can be configured. To overcome such limitations, the approaches discussed in the rest of this section integrate a cyber network and/or device emulation on top of the power system simulators.

### 6.3.2 EPICTWIN (Digital Twin of EPIC)

Even though physical testbeds are helpful for research and training, there are limitations on the type of experiments that can be conducted in physical testbeds due to the set-up and maintenance costs involved and the probability of damaging the physical equipment. Therefore, Nandha et al. in [Kandasamy *et al.*, 2021] developed a digital twin of the EPIC testbed to conduct research and security training. The physical infrastructure of the testbed is designed using Matlab Simulink, and the simulation is running in real-time simulation hardware, namely 'Speedgoat'. The communication between the simulation and the rest of the system is handled by different Python programs through a UDP data bus. Virtual Machines are used to represent PLCs, IEDs, and Network switches controllers, while the run-time for the PLCs and IEDs is established using a web browser-based flow editor called 'Node-RED' [*Node-RED* n.d.]. The AMI meters in the testbed are simulated using Ubuntu docker containers. The network architecture of the testbed is implemented using Open Virtual Switches (OVS), which uses Generic Routing Encapsulation (GRS) to assign IPs and route networks. Similar to PLCs and IEDs, SCADA is also implemented using Node-RED. Three protocols are employed in the testbed digital twin: 1) Message Queuing Telemetry Transport (MQTT) – to achieve data-bus functions; 2) IEC 61850 MMS – to communicate control and measurements between PLCs, IEDs, and SCADA; 3) IEC 61850 GOOSE – to communicate between IEDs. The implementation of IEC protocols is achieved through libIEC61850. Finally, EPICTWIN has a built-in attack tool called EPICSPLOIT that allows users to configure basic man-in-the-middle attacks.

### 6.3.3 Smart Grid Cyber Range

While EPICTWIN was specifically developed to emulate EPIC testbed, an effort to develop a more generic model of a distribution substation is developed in [Roomi *et al.*, 2020]. In the study, the requirements to design any cyber range and conduct cyber-attack studies on the cyber range using open-source software are discussed. A 66/11kV sub-transmission level modernized substation is considered for the experiment. The physical layout of the substation is simulated using Pandapower software running in a virtual machine (VM). The protection logic are implemented through OpenPLC, and the graphical visualization is accomplished using ScadaBR and is implemented in separate VMs. The virtual sensors are deployed on the same VM as the simulator to read the measurements and store them in the database. Given the limitations with Pandapower, such as no direct measurement of frequency and transformer temperature, the frequency and temperature values are calculated based on Load Frequency Control (LFC) and the loading percent of the corresponding transformer, respectively. These values are stored along with the sensor values in the database. The communication to the PLC and the SCADA is established through the Modbus TCP library in Python. This paper mainly focuses on the FDI attack against the measurements and status that are communicated to OpenPLC. The impact of the FDI on the control device in the field is detailed with results in [Roomi *et al.*, 2020]. The architecture of the cyber range is shown in Figure 6.3.

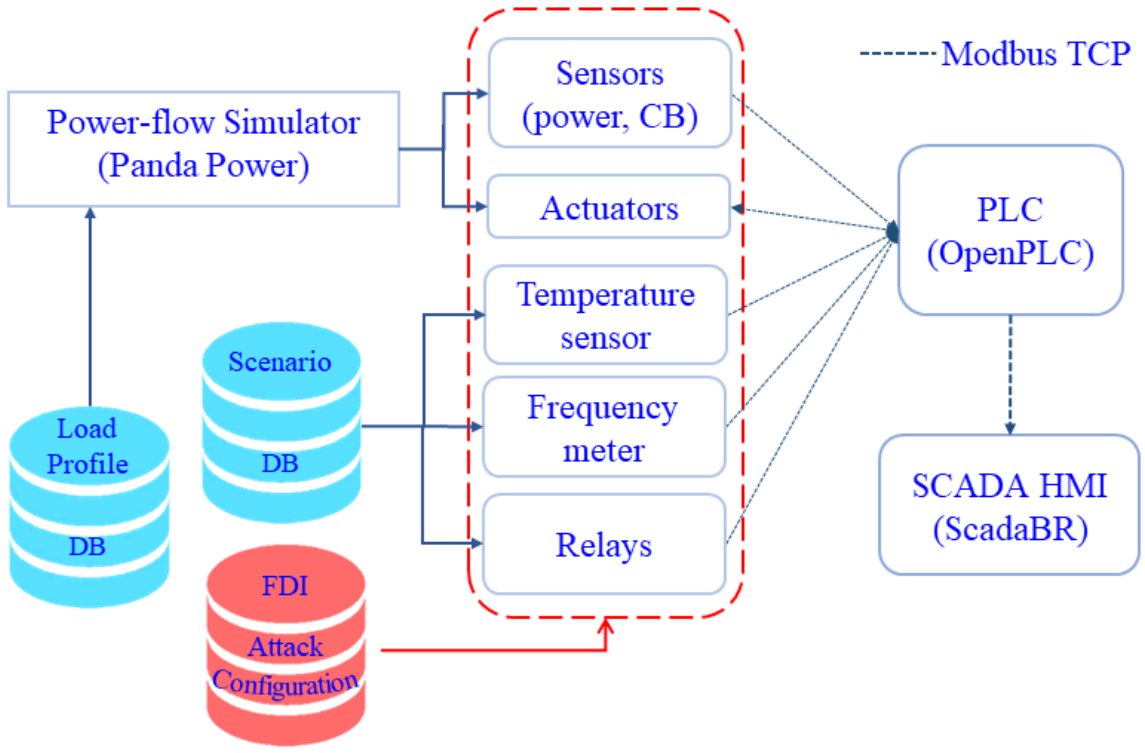

**Figure 6.3:** Smart Grid Cyber Range Architecture [Roomi *et al.*, 2020].

Though the aforementioned design strategies are sufficient to generate generic cyber ranges and also conduct cyber-attack studies, the process involved is not automated. Therefore, as an enhancement to the explained design strategy, a modeling framework was developed to automatically generate a cyber range of any smart grids [Mashima *et al.*, 2023]. This framework uses a set of XML schemas, which are defined in both machine- and human-readable format, to configure the virtualized version of a real smart grid. The cyber range is tested with single and multi-substation models. The communication protocols employed are MMS, GOOSE, Routable GOOSE (R-GOOSE), and Routable SV (R-SV). An open-source framework has been published to demonstrate inter-substation communication [Hussain *et al.*, 2023a; Hussain *et al.*, 2023b].

The architecture of the SG-ML (Smart Grid Modeling Language) framework is depicted in Figure 6.4. There are three main components in the framework: input components, processor components, and cyber range components. The bottom layer in the figure (highlighted in brown) defines the input files that are necessary for the framework. Two main XML files are needed: 1) the IEC 61850 System Configuration Language (SCL) files – which define the physical and cyber network of any smart grid; 2) Supplementary files – which define the missing components in the IEC 61850 SCL files. The next layer (highlighted in orange) is a sub-layer of the input files. These blocks show the details that are extracted from the input files for the SG-ML processor. The third layer (highlighted in blue) is the SG-ML processor layer. In this layer, the two XML files are integrated to create a full-fledged model file. For instance, the SCL processor for the physical network integrates the physical topology that is obtained from the SCL file with the parameter specification in the supplementary XML file. Therefore, in the top layer (highlighted in red in the figure), when the cyber range for the physical system is simulated to run the power flow, the input file contains the components, their specifications, and the connections between them. Similarly, the cyber network elements are also emulated. In this study, the following open-source software is used to create the cyber range based on the system configurations automatically. Pandapower [*Pandapower* n.d.] is used as a power system simulator, and the measurements are stored in MySQL [*MySQL* n.d.]. Virtual IEDs and PLCs are implemented using OpenPLC61850 [Roomi *et al.*, 2022b; Roomi *et al.*, 2022a]. Finally, the cyber network topology is configured through the Mininet emulator [*Mininet* n.d.].

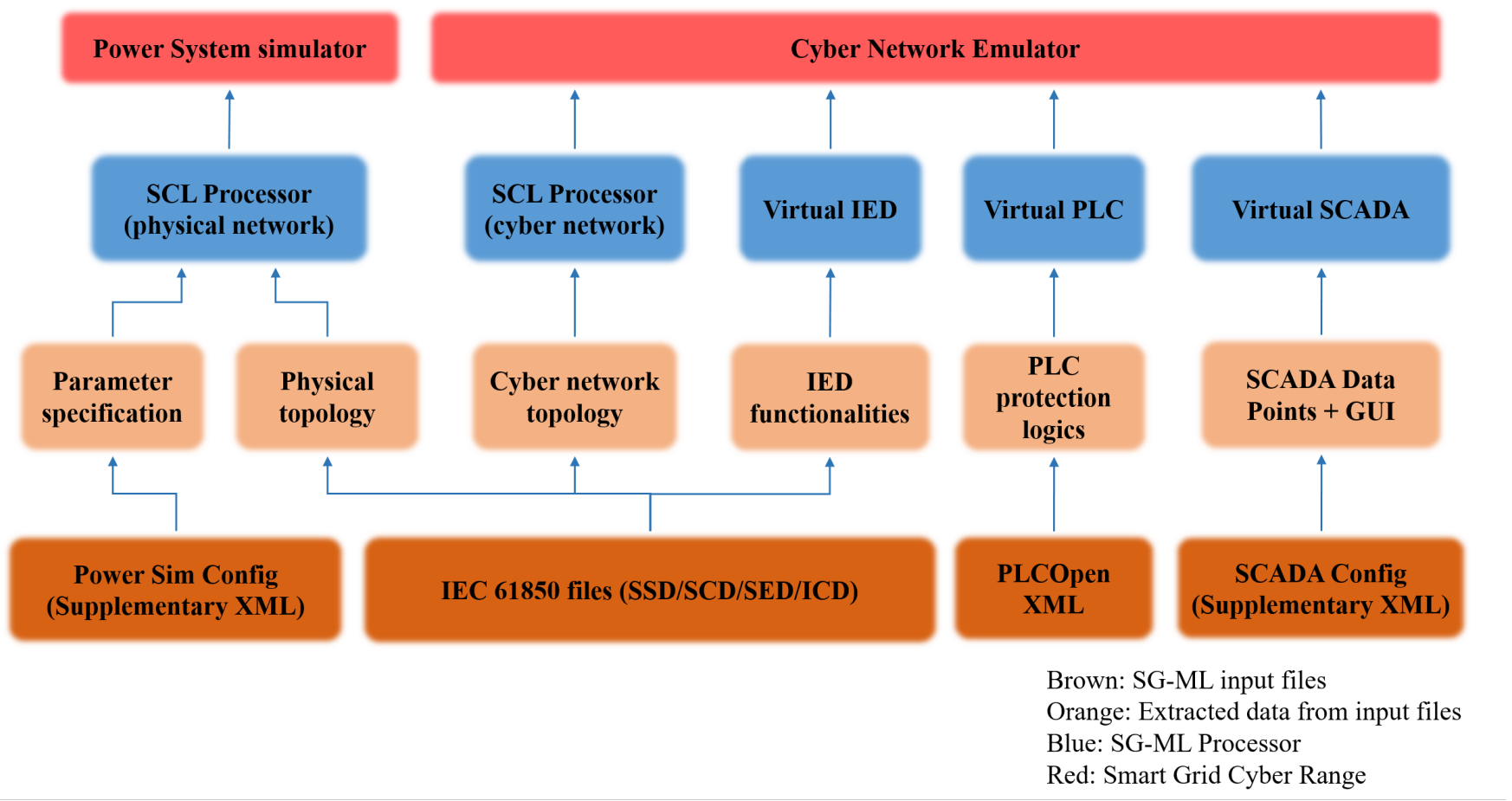

**Figure 6.4:** Architecture of SG-ML Modeling and Automated cyber Range Generation Framework

In order to demonstrate the capability of the automated SG-ML framework, a 66/11kV sub-transmission level substation model is generated. The physical topology and the cyber network topology are generated using the SCL files that are available for the substation. The power flow and measurement communication are carried out by utilizing supplementary XML files, as mentioned in the previous paragraph. The input and the output physical configuration of the generated cyber range through the SG-ML framework is depicted in Figures 6.5 and 6.6. Similarly, the cyber network topology generated is highlighted in Figures 6.7 and 6.8. The demo videos and some cyber attack studies using the cyber range are available in [*Smart Grid Cyber Range* n.d.]. Currently, the implementation of the cyber range using Matlab/Simulink as an alternative to Pandapower is explored [Roomi *et al.*, 2023b] and making this cyber range as a cloud based service is a part of the road map.

### 6.3.4 Smart Grid Honeypot

High-interaction smart grid honeypot can also be used as a cyber security testbed. One such examples is found in [Mashima *et al.*, 2020]

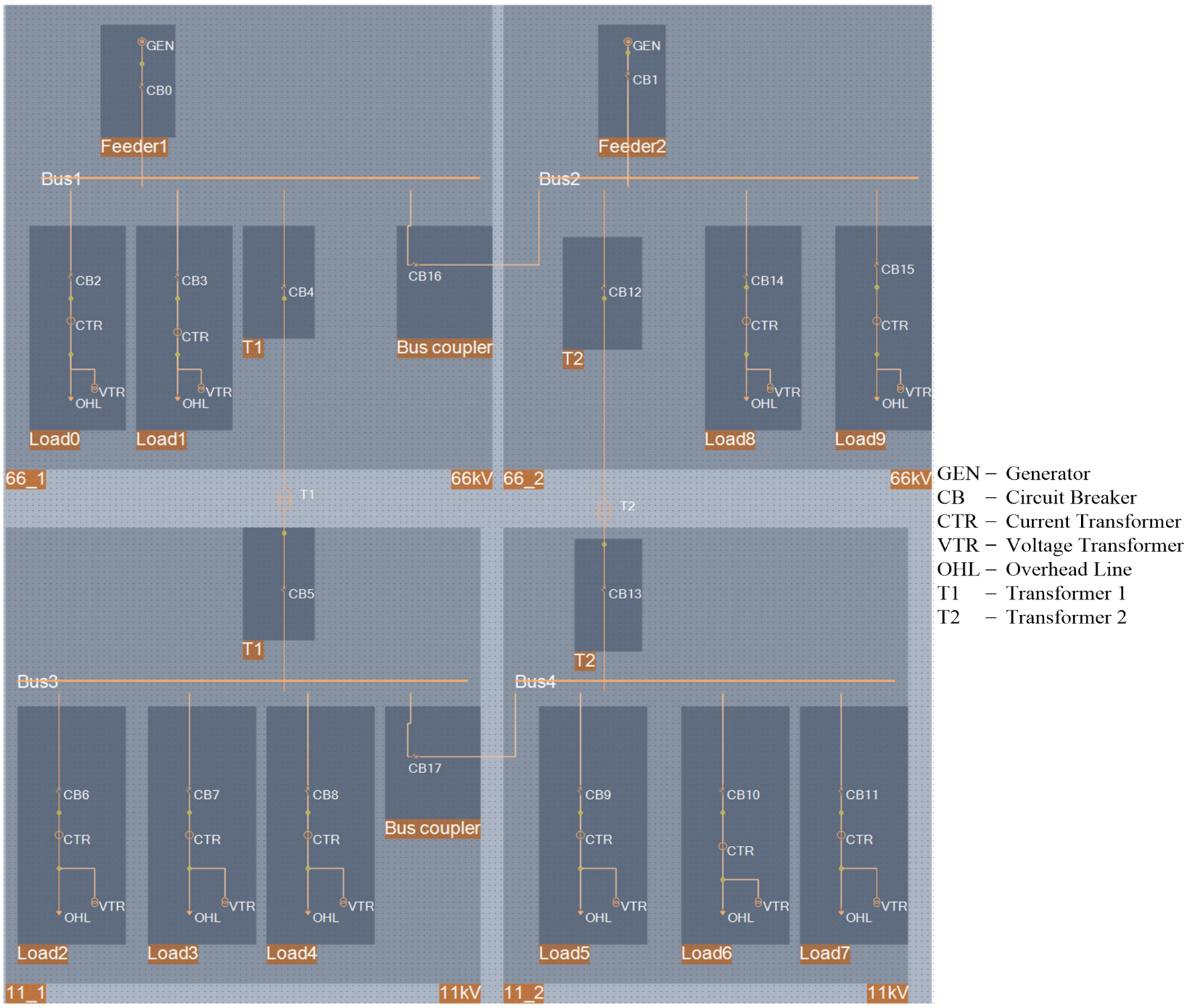

**Figure 6.5:** Graphical View of System Specification Description (SSD) Input to SG-ML Framework

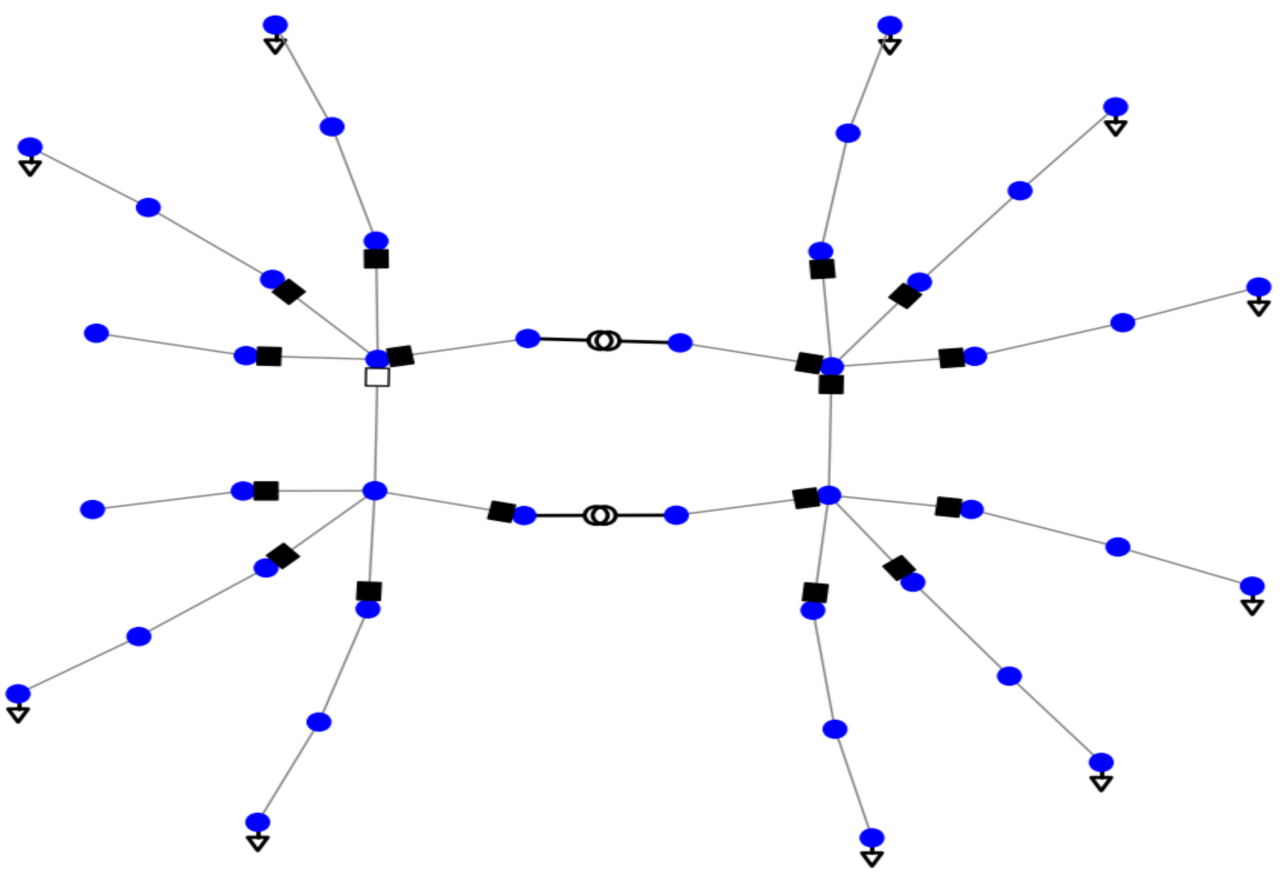

**Figure 6.6:** Power System Simulation Model Generated by SG-ML Framework

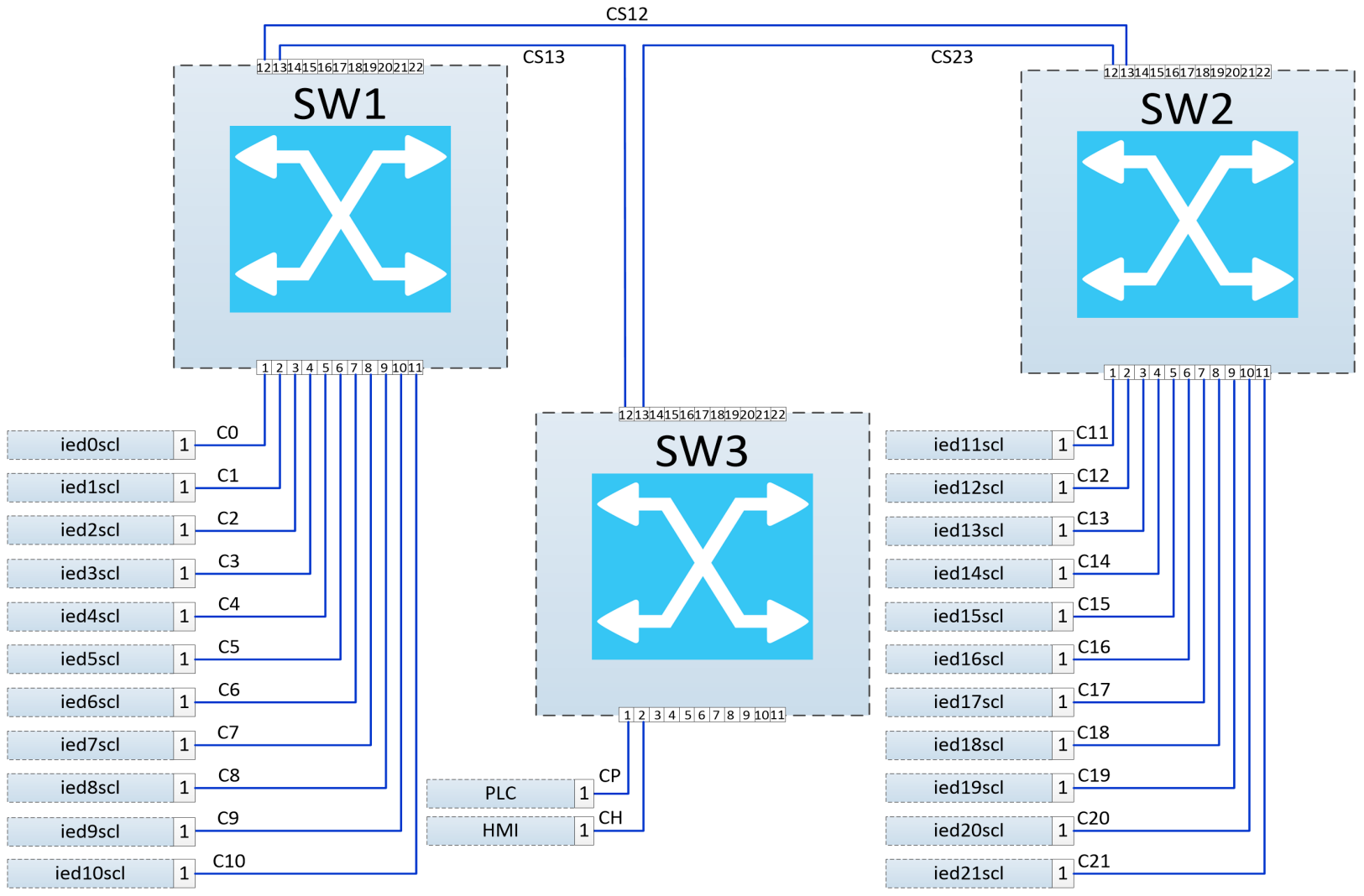

**Figure 6.7:** Graphical View of System Configuration Description (SCD) Input to SG-ML Framework

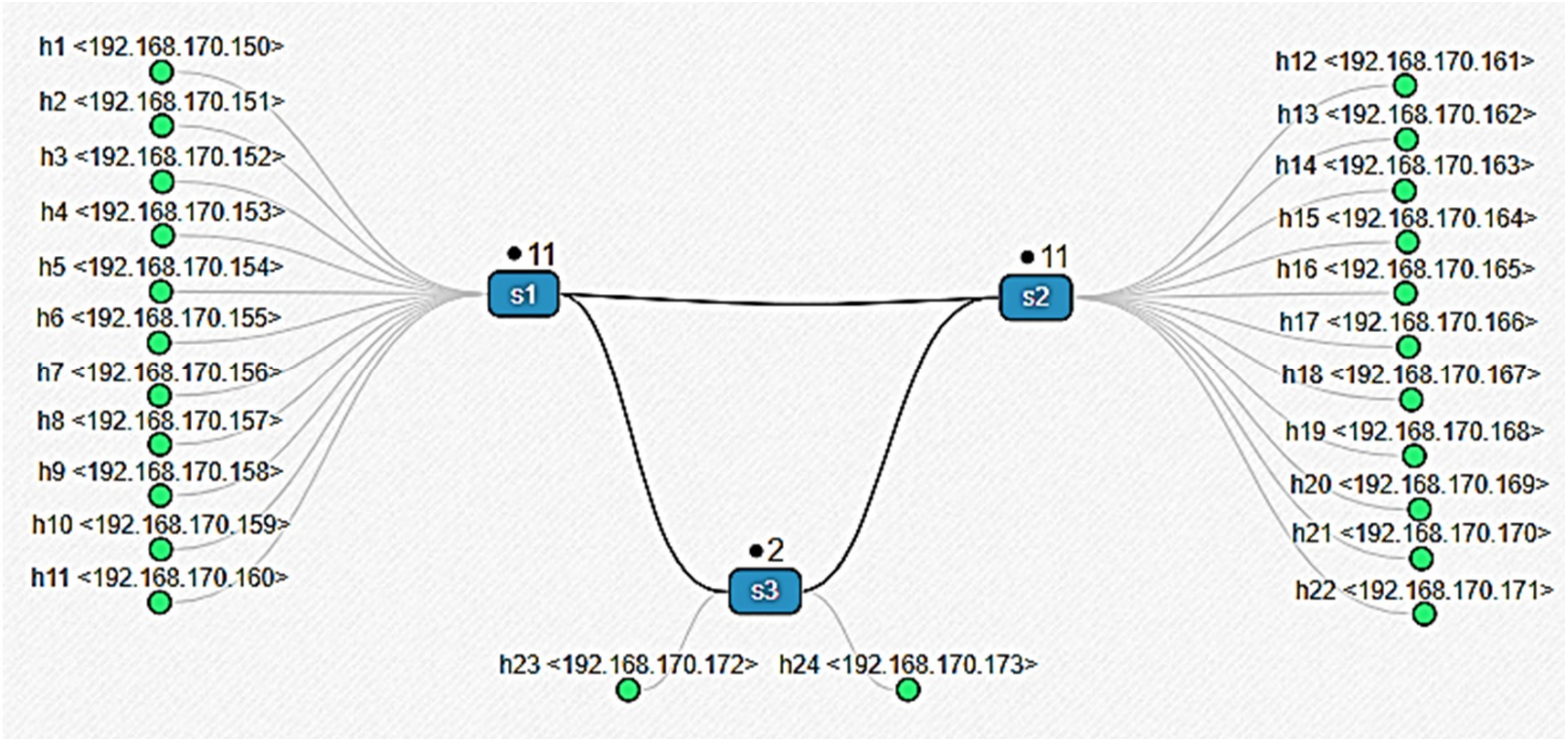

**Figure 6.8:** Network Emulation Model Generated by SG-ML Framework

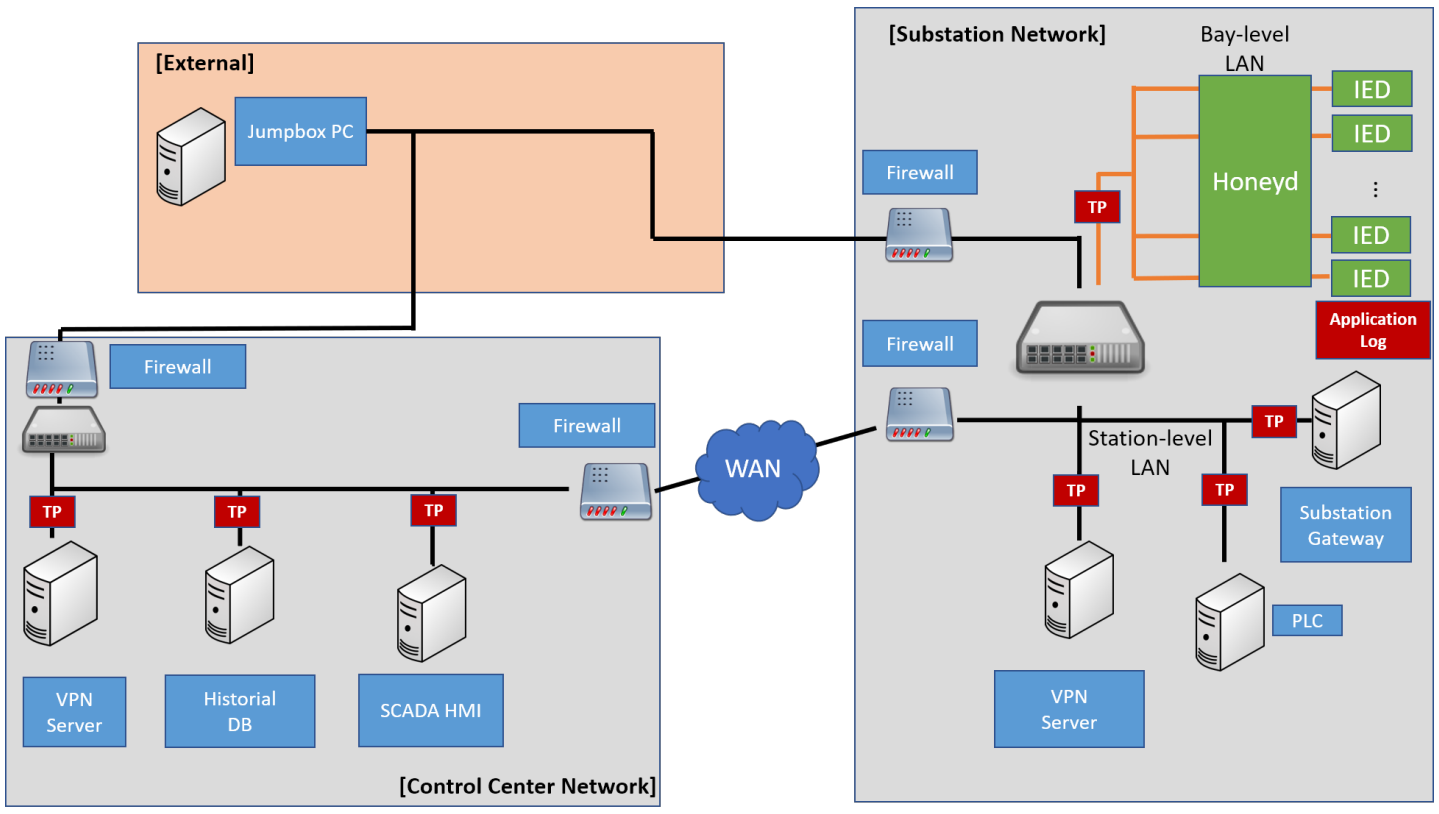

**Figure 6.9:** Smart Grid Honeypot Architecture [Mashima *et al.*, 2020].

(refer to Figure 6.9). The authors implemented a virtual smart grid infrastructure that imitates realistic systems in a control center and a field substation. The former consists of a SCADA HMI workstation, a Historian database, general-purpose workstations, a VPN server, and firewalls. The latter includes a substation gateway (which translates protocols between IEC 61850 and IEC 60870-5-104), IEDs, a VPN server, and firewalls. The servers and substation gateways are implemented as separate virtual machines with Windows or Linux OS, while IEDs are implemented using an open-source tool called Honeyd to imitate device characteristics and fingerprints. These are connected to SoftGrid, to emulate the physical behavior of power systems and are discussed later. Such a honeypot can be utilized for penetration testing exercises, as demonstrated by the paper.

### 6.3.5 SoftGrid

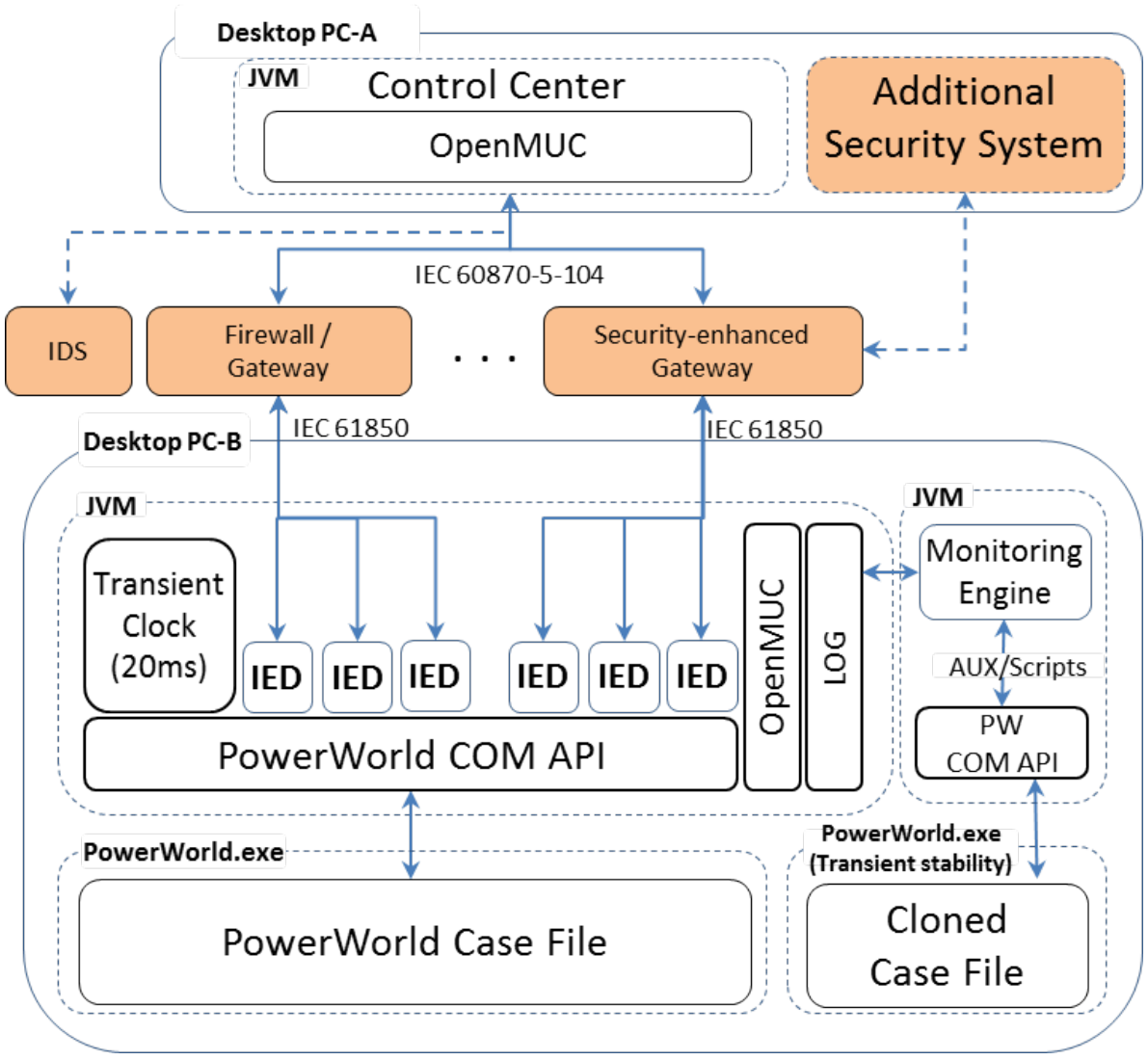

**Figure 6.10:** SoftGrid Architecture [Gunathilaka *et al.*, 2016a].

SoftGrid [Gunathilaka *et al.*, 2016a], shown in Figure 6.10, is developed as a venue for cybersecurity experiments and technology demonstration in IEC 61850 compliant substation system. SoftGrid consists of a virtual control center module, which communicates with field substations using IEC 60870-5-104 protocol for remote monitoring and control, and a virtual substation, which hosts a number of IEC 61850 compliant virtual IEDs and substation gateway (i.e., a protocol translator between IEC 60870-5-104 and IEC 61850). SoftGrid incorporates a PowerWorld simulator (commercial product) for steady-state as well as transient-state power flow simulation to evaluate the impact on power grid systems. On the other hand, it emulates minimal functionality on virtual IED and substation gateway as the communication endpoints and network topology are simplified.

## 6.4 Qualitative Comparison of Testbeds

Each type of smart grid security testbed has different sets of advantages and disadvantages. In this section, a qualitative comparison of different testbeds in Table. 6.1 is presented to guide users to find a suitable one for their needs. The comparison is made based on the qualitative metrics such as cost, accessibility, fidelity, scalability, and reproducibility to compare the aforementioned testbeds. Additionally, the difference in the configurations is included to provide details regarding the physical architecture.

From the table, as EPIC [iTrust, n.d.] is a hardware-based testbed with real hardware components, the overall cost associated with the testbed is high and accessibility is payment based and hence, it is limited. However, the fidelity will be high as no synthetic factor has to be added to the measurements. Given the fact the testbed is already setup, it requires intensive cost domain knowledge to reconfigure or redesign the architecture of the testbed. Therefore, the scalability and the reproducibility is low.

EPICTwin [Kandasamy *et al.*, 2021] is a virtual testbed that integrates a real-time plant simulator, which makes its cost to be medium while offering medium-high fidelity by offering real-time, dynamics simulation. On the other hand, scalability is bounded by the simulator hardware. This twin at the moment is closed accessible only to authorized personnel. Hence, the accessibility is closed-source. EPICTwin is specifically designed as virtual counterpart of EPIC, and thus simulation models are specifically designed for this purpose, based in part on proprietary data. Thus, its reproducibility is ranked low.

As smart grid cyber range (SGCR) [Roomi *et al.*, 2020] and smart grid honeypot [Mashima *et al.*, 2020] involves only open-source software to implement the technology, the cost is low compared to the other testbeds and it is open-sourced. However, the fidelity can vary from medium to high, mainly depending on the power system simulator utilized. The scalability and reproducibility of SGCR is high as the substation model can be scaled up or scaled down without much effort.

Although the reproducibility of Honeypot is also ranked high since it is also purely software based, scalability needs effort to reconfigure, and thereby, it is ranked low.

Softgrid [Gunathilaka *et al.*, 2016b] involves commercial software, namely PowerWorld [*PowerWorld* n.d.]. Hence, these can be classified as medium. Softgrid has medium scalability support with high reproducibility. The system can be used with publicly-available or shared PowerWorld simulation models, and then cyber components can be configured accordingly. On the other hand, use of computationally intensive transient-state simulation for the power system may limit scalability for real-time, interactive usage.

A detailed evaluation between the physical, hybrid and virtual testbeds to evaluate the testbeds' suitability for cybersecurity experimentation is found in [Mumrez *et al.*, 2023]. Furthermore, the authors have used the MITRE ATT&CK for ICS matrix to demonstrate the feasibility and coverage of tactics and techniques for assessing applicability of these testbeds.

Lastly, power system simulators, ranging from open-source ones to high-end simulator hardware, can be used to evaluate the impact of cyber attacks on a power grid. However, most of them offer no or limited support for the cyber side. Thus, they are not suitable for interactive cyber attack experiments, and fidelity is ranked low to (at most) medium. As any power grid models can be configured with the power system simulators, and the models can be easily shared, power system simulators offers high reproducibility.

## 6.5 Case Studies

In this section, cyber-attack case studies are demonstrated utilizing the cyber ranges described in the section 6.3.3. Common attacks such as false data injection (FDI) and man-in-the-middle (MITM) attacks are considered for this study, and the attack launch along with the impact is illustrated.

**Table 6.1:** Qualitative Comparison of Testbeds

| Testbed | Type | Cost | Accessibility | Fidelity | Configuration | Scalability | Reproducibility |
|---|---|---|---|---|---|---|---|
| EPIC [iTrust, n.d.] | Physical | High | Limited (on-site) | High | 4 sectors (generation, transmission, micro-grid, smart home) | None | None |
| EPIC Twin [Kandasamy *et al.*, 2021] | Virtual | Medium | Closed-source | Medium-high | Same as above | Low | None |
| Smart Grid Cyber Range [Roomi *et al.*, 2020] | Virtual | Low | Open-source | Medium | Substation | High | High |
| Smart Grid Honeypot [Mashima *et al.*, 2020] | Virtual | Low | Open-source | Medium-high | Comprehensive control center and substation | Low | High |
| SoftGrid [Gunathilaka *et al.*, 2016a] | Virtual | Medium | Open-source (with commercial simulator) | Low | SCADA HMI and substation | Medium | High |
| Power System Simulators | Virtual | Low - Medium | Open-source or commercial | Low - Medium | Power grid only | High | High |

### 6.5.1 FDI Attack on Cyber Range

In this section, the FDI attack on the smart grid cyber range explained in [Roomi *et al.*, 2020] is demonstrated. In a smart grid, protection logic is implemented to maintain the stability of the system. As such, four common protection strategies, which are over-current, automated transfer, under-frequency load shedding, and transformer over-temperature, are implemented in the smart grid cyber range. In order to demonstrate the FDI attack on this cyber range, one of the protection strategies is compromised, and the impact on the system is illustrated in this section. Figure 6.11 depicts the SCADA interface of the smart grid cyber range during normal operation. Each of the incoming generators (generator 1 and generator 2) is associated with a circuit breaker (CB) and a relay. These incoming lines are connected to each other through a bus coupler (named sensor 16 in the figure). This bus coupler is 'open' during the normal operation of the grid. When one of the generators goes out of service, then this bus coupler CB is 'closed' such that the active generator feeds the entire system. However, the closure of this bus coupler CB depends on the status of the CB and relay associated with the generator. For example, when one of the incoming generators is out of service, the associated CB will be 'open'. Subsequently, the PLC logic checks for the relay status. If there is no fault or discrepancy in the system, then the relay status will be good, and this condition is the criterion for the PLC to trigger the 'close' status for the bus coupler CB.

To offer more details for this case study, generator 1 is considered out of service as mentioned before, which is depicted by the 'open' CB in Figure 6.12. During this state, as the relay status in the disconnected line is good (indicated by green), PLC triggers the bus coupler CB to close. Now the FDI attack is launched by manipulating the relay status reported to the PLC. As such, when the false data is injected, the status of the CB in the disconnected line changes from good to bad (indicated in red in Figure 6.13). This tampered data misled the PLC not to initiate the control to open the bus coupler CB. During this condition, as loads 2 to 4 receive power from generator 2 through the bus coupler CB (sensor 17 in Figure 6.13), these loads are not disconnected. However, the primary loads that are connected to the main line (loads 0 and 1) are de-energized, resulting in a partial outage in the smart grid.

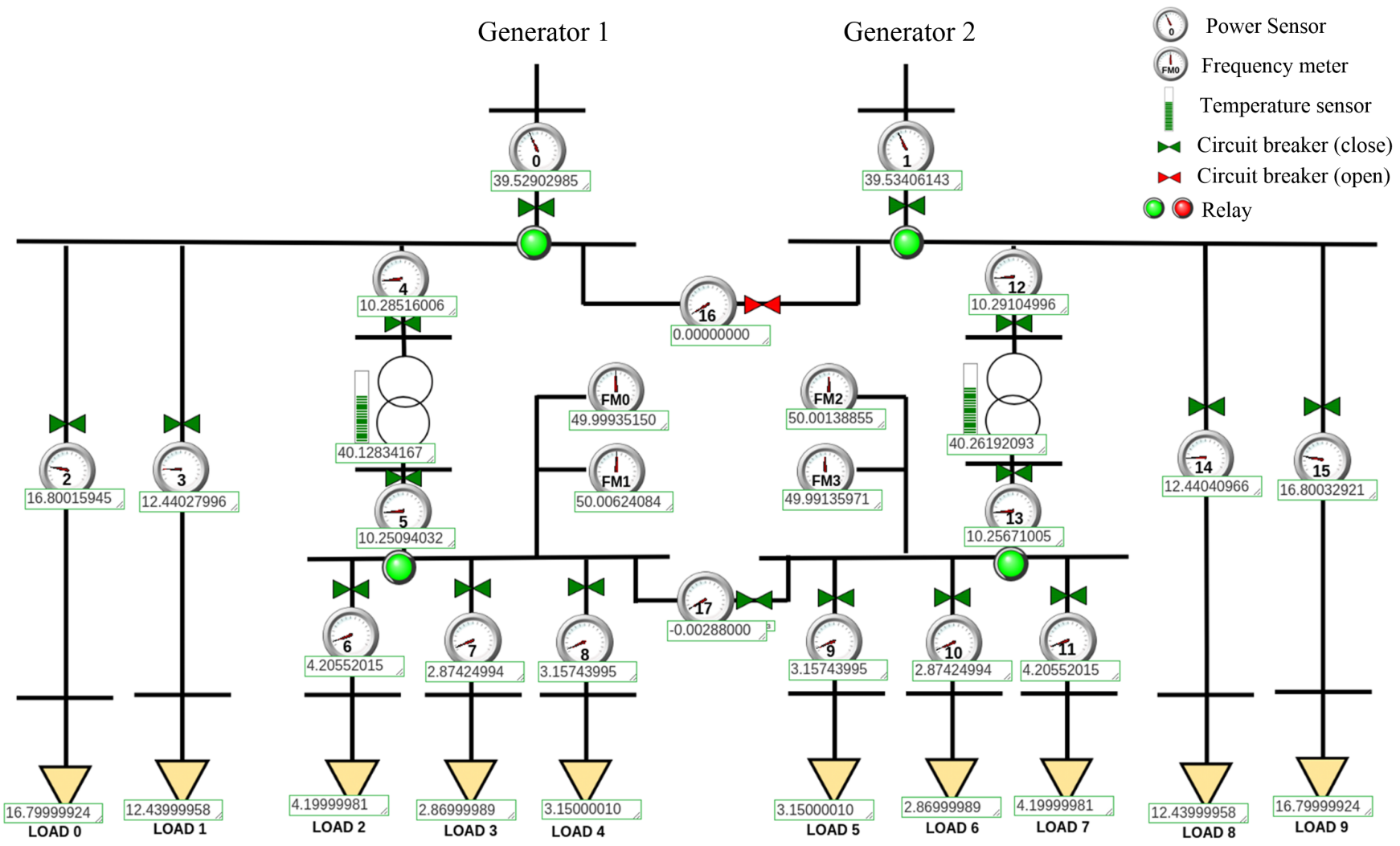

**Figure 6.11:** Normal State of Substation

### 6.5.2 MITM Attack on Cyber Range

Smart grids are vulnerable to MITM attacks as the grids contain communication devices. During a MITM attack, an attacker who gained access to the system can either eavesdrop on the communication between two devices or impersonate one of the real devices and carry out normal yet false exchanges of information. As such, the attacker can inject fake data or inappropriate commands to disrupt the normal operation of the grid. As MMS messages in an IEC-61850-based substation are using an application layer protocol, an attacker can mount the MITM attack on this communication using Address Resolution Protocol (ARP) spoofing. Therefore, this compromised device can modify the data and send false data to the destination. In Figure 6.14, the method to conduct a MITM attack on the cyber range is demonstrated. The cyber network topology for the 66/11kV model explained earlier is used in the figure. In this network, one of the IEDs located at node ‘h12’ is frequently communicating with node ‘h24’. The assumption is one of the network switches (sw) has been compromised, and the attacker is in the system. With the attacker gaining access to the network, the attacker eavesdrops on the communication that is happening between ‘h12’ and ‘h24’. Subsequently, in the following message exchange instead of following the regular network path h12 → s2 → s3 → h24 (highlighted in green), the network path is configured to h12 → s2 → sw → attacker VM → sw → s2 → s3 → h24. Through this strategy, the attacker can launch MITM attacks that can either inject false measurements into the destination device or alter a command so that the destination device malfunctions. The impact of these kinds of attacks can vary from a simple display of false data in SCADA HMI to a much more complex power outage in the grid.

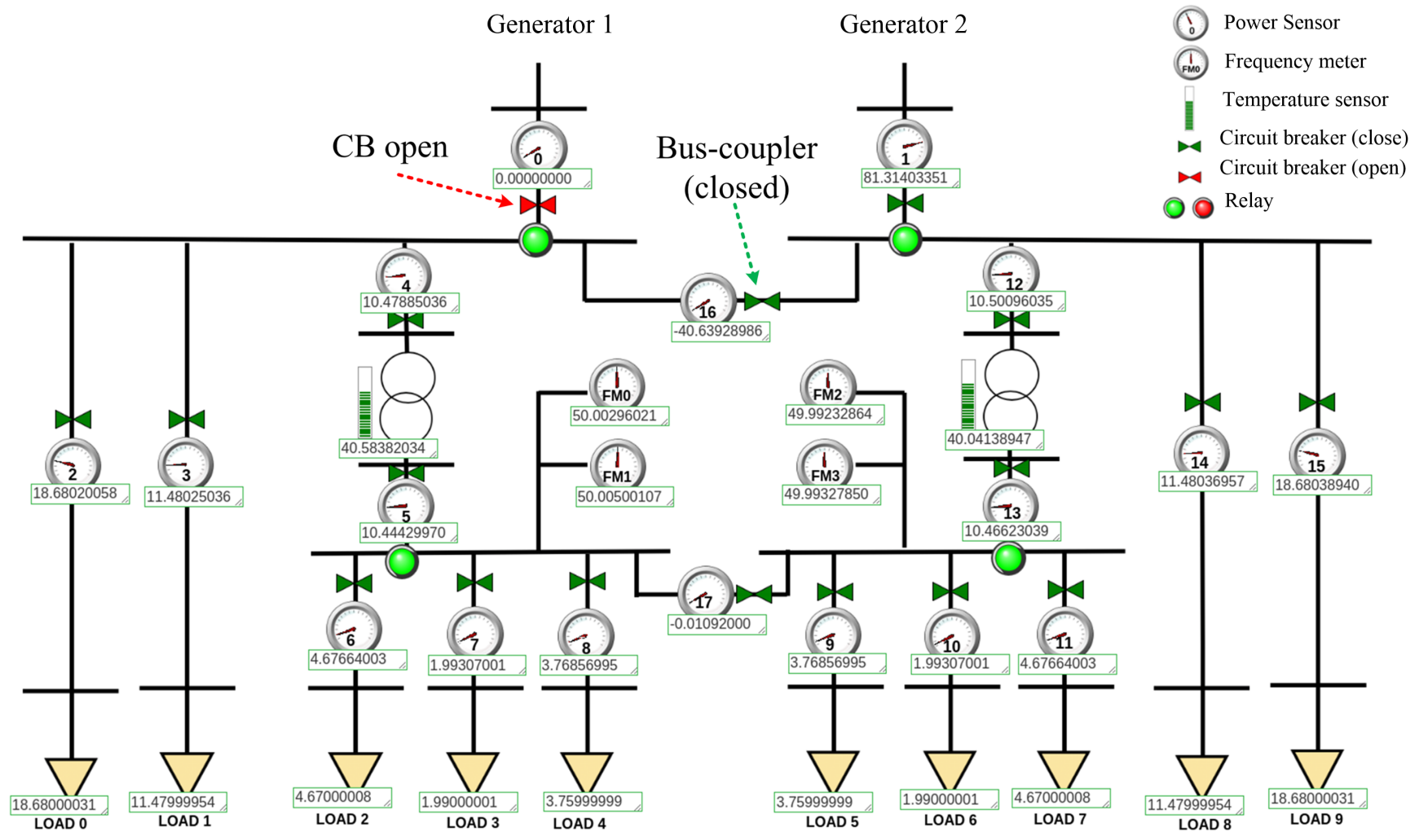

**Figure 6.12:** Incoming Line Out of Service

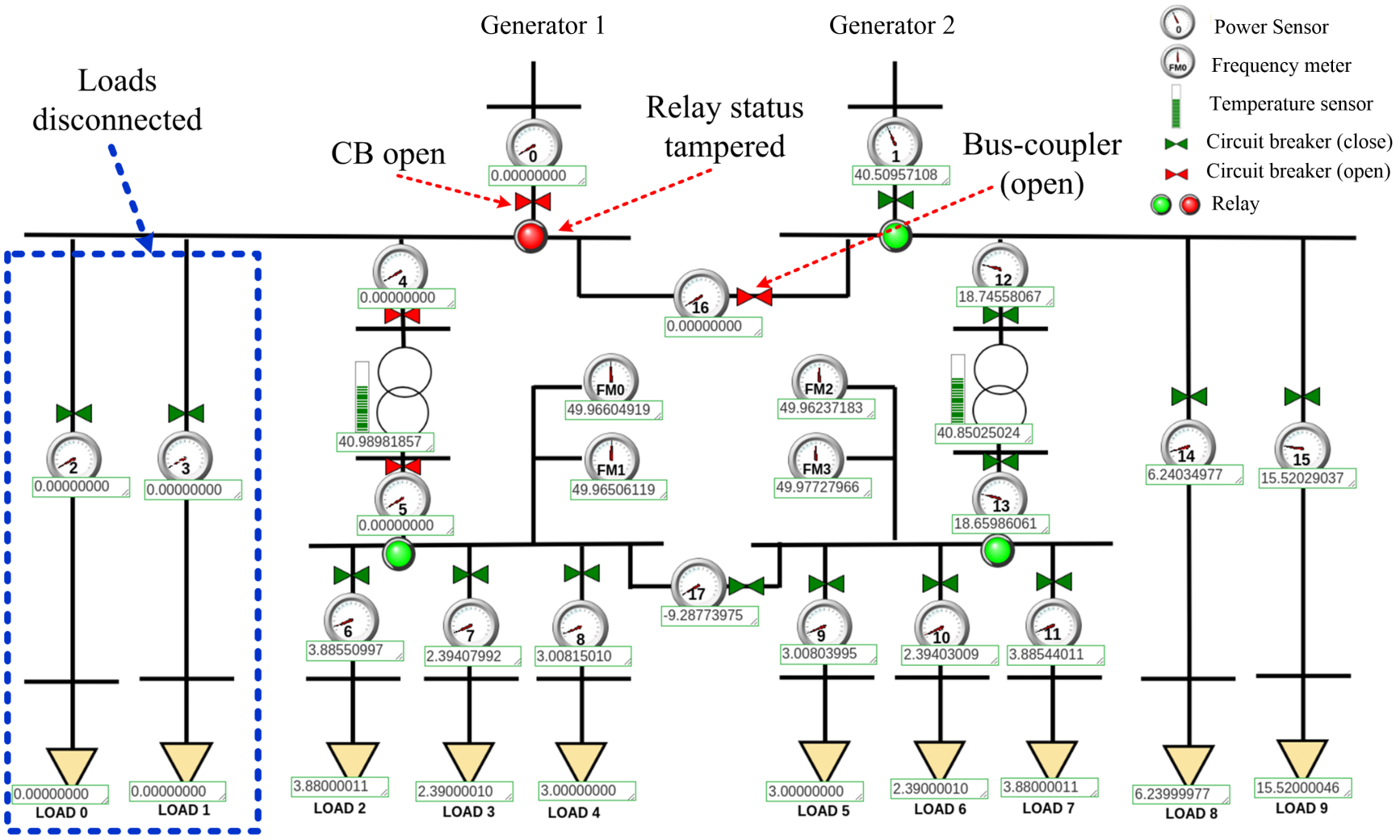

**Figure 6.13:** Impact of Tampered Relay Status

In this subsection, we discussed two different attacks as the case studies. The smart grid cyber range can be further utilized for experimenting cybersecurity solutions. For instance, by deploying cryptographic message authentication discussed in Chapter 4, we can effectively counter the discussed FDI attack. Network-based intrusion detection systems discussed in Chapter 5, which can flag the suspicious change in the mapping of IP address and MAC address and/or significant increase of ARP packets, are considered effective to counter the MITM attack.

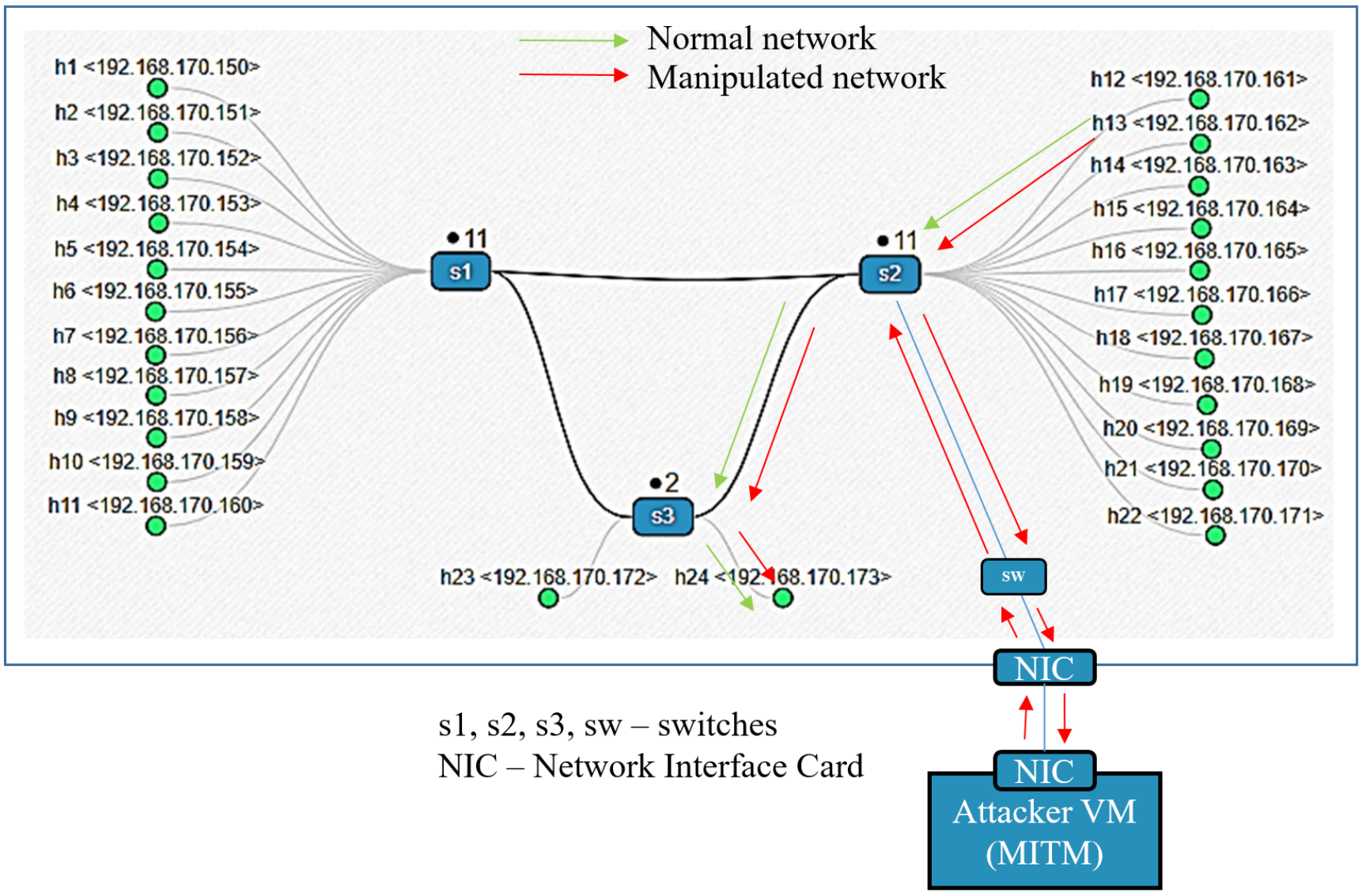

**Figure 6.14:** MITM Attack on Cyber Range

# 7

# Conclusions

A power grid system is arguably one of the most critical national infrastructures that affect the availability and stability of the other infrastructures. However, smart grid infrastructure often became a target of cyberattacks because of its importance. Cybersecurity for smart grid systems poses a set of requirements and constraints that are different from enterprise IT systems. For instance, the infrastructure is physically distributed and consists of a number of legacy devices for monitoring and control.

In this book, we covered a brief overview of state-of-the-art cybersecurity solutions for smart grid systems of different kinds. The coverage of the literature is not comprehensive, but we picked representative examples of each category to illustrate the concepts. The overall mitigation provided by the discussed smart grid cybersecurity solutions is summarized in Figure 7.1. As seen in the mapping to MITRE ATT&CK Matrix, each technology aims at countering different attack stages and tactics, and thus it is necessary to combine multiple cybersecurity measures of different categories, namely deterrence (Chapter 3), prevention (Chapter 4), and detection (Chapter 5). Some of the attack tactics are countered by multiple types of solutions (represented with darker blue color). For the tactics with darker blue color, we strongly recommend to combine as many types of cybersecurity measures as possible within the budget constraint for "defense in depth", instead of selecting only one. On the other hand, we can also see that the coverage is not yet comprehensive. Thus, further research and development efforts should be more focused on the tactics that are not colored yet in Figure 7.1.

| Initial Access | Execution | Persistence | Privilege Escalation | Evasion | Discovery | Lateral Movement | Collection | Command and Control | Inhibit Response Function | Impair Process Control | Impact |
|---|---|---|---|---|---|---|---|---|---|---|---|
| Drive-by compromise | Change operating mode | Modify program | Exploitation for Privilege Escalation | Change operating mode | Network connection enumeration | Default credentials | Automated collection | Commonly used port | Activate firmware update mode | Brute force I/O | Damage to property |
| Exploit public facing application | Command line interface | Module firmware | Hooking | Exploitation for evasion | Network sniffing | Exploitation of remote services | Data from information repositories | Connection proxy | Alarm suppression | Modify parameter | Denial of control |
| Exploitation of remote services | Execution through API | Project file infection | | Indicator removal on host | Remote system discovery | Lateral tool transfer | Detect operating mode | Standard application layer protocol | Block command message | Module firmware | Denial of view |
| External remote services | Graphical user interface | System firmware | | Masquerading | Remote system information discovery | Program download | I/O image | | Block reporting message | Spoof reporting message | Loss of availability |
| Internet accessible device | Hooking | Valid accounts | | Rootkit | Wireless sniffing | Remote services | Man in the middle | | Block serial COM | Unauthorized command message | Loss of control |
| Remote services | Modify controller tasking | | | Spoof reporting message | | Valid accounts | Monitor process state | | Data destruction | | Loss of productivity and revenue |
| Replication through removable media | Native API | | | | | | Point & tag identification | | Denial of service | | Loss of production |
| Rogue master | Scripting | | | | | | Program upload | | Device restart/shutdown | | Loss of safety |
| Spear-phishing attachment | User execution | | | | | | Screen capture | | Manipulate I/O image | | Loss of view |
| Supply chain compromise | | | | | | | Wireless sniffing | | Modify alarm settings | | Manipulation of control |
| Transient cyber asset | | | | | | | | | Rootkit | | Manipulation of view |
| Wireless compromise | | | | | | | | | Service stop | | Theft of operational information |
| | | | | | | | | | System firmware | | |

**Figure 7.1:** Overall Mitigation on MITRE ATT&CK Matrix for ICS.

Another universal challenge that the smart grid operators and practitioners are facing is the lack of an environment for evaluating cybersecurity measures through experiments and hands-on, interactive exercises. The evaluation of cyberattack impacts and the assessment of cybersecurity solutions are the key components in developing a secure smart grid. As such, hardware-based, virtual, and software-based testbeds are explored in this book. The hardware-based testbeds are good alternatives for real smart grid infrastructure to conduct impact study and security evaluation, while it is a viable, accessible option to everyone. Digital twins of smart grid for cybersecurity experiments (i.e., cyber ranges), which has better accessibility and flexibility, are also discussed along with state-of-the-art technology for automated generation of such virtual testbed. A qualitative comparison of these testbeds is conducted, and the findings are tabulated. Finally, various cyberattack case studies that are conducted on the cyber ranges are demonstrated. We hope such discussions on the testbeds provide practitioners and researchers useful guidelines and practical solutions to deal with cyberattacks.

Looking forward, developing robust IDS systems is becoming more important than ever before, given the growing number of attack surfaces in a power grid and the infeasibility of preventing cyber-attacks completely. Despite the increasing research and development efforts devoted in this field, still, a number of open issues remain, such as differentiating natural system oscillations from cyber attacks (especially with growing renewables, inverter-based resources, etc.) and developing intrusion detection systems where grid operators typically have limited sensing devices (such as distribution network/end-user sites). Addressing these issues requires operators to integrate different sources of information from various sensing sources, such as smart meters, distribution phasor measurement units, etc., and utilize edge computing.

While the attack vectors and tactics are evolving, there are promising technologies that can defend smart grid systems. For instance, AI technologies can contribute for such defenses. To name a few, reinforcement learning can be utilized to implement adaptive deception technologies. Prediction of the future state of the system and message content can help further accelerate message authentication by reducing the search space for pre-computation. Physics-informed neural network (PINN), which incorporates power system physical laws to guide the neural network model, could improve the accuracy of attack and anomaly detection.

The application of Large Language Models (LLMs) such as GPT-4 could play an instrumental role in addressing cybersecurity issues for smart grids. Leveraging LLMs can aid in pattern recognition, anomaly detection, and threat prediction by analyzing vast amount of data produced by smart grids and identifying patterns that suggest possible cyber attacks. LLMs could be trained to understand and predict malicious behavior based on historical data from cyber incidents. For instance, the notable 2020 SolarWinds cyber attack and the 2021 Colonial Pipeline ransomware attack provide excellent datasets for training these models. These attacks led to significant disruptions in critical infrastructures, but at the same time provided valuable insight into the tactics, techniques, and procedures used by attackers. Utilizing LLMs in this way can help develop proactive security measures, enabling swift detection and prevention of threats before they can inflict damage on smart grid systems. This novel use of LLMs could also aid in developing advanced Intrusion Detection Systems (IDS) and Security Information and Event Management (SIEM) solutions, enhancing overall smart grid resilience.

# Acknowledgements

This research is supported in part by the National Research Foundation, Prime Minister's Office, Singapore under its Campus for Research Excellence and Technological Enterprise (CREATE) programme and in part by the National Research Foundation, Singapore, Singapore University of Technology and Design under its National Satellite of Excellence in Design Science and Technology for Secure Critical Infrastructure Grant (NSoE_DeST-SCI2019-0005). This work is also supported by the National Research Foundation, Singapore under its AI Singapore Programme (AISG Award No: AISG2-TC-2021-002). Any opinions, findings and conclusions or recommendations expressed in this material are those of the authors and do not reflect the views of National Research Foundation, Singapore.

The authors also thank Dr. Ertem Esiner and Dr. Utku Tefek from Illinois Advanced Research Center at Singapore for their valuable input.